\documentclass[twocolumn,tighten,times,twocolappendix]{aastex631}

\usepackage{comment}

\newcounter{Rco}
\newcommand{\Ionst}[1]{\setcounter{Rco}{#1}\Roman{Rco}}
\newcommand{\Ion}[2]{\mbox{#1\,{\scriptsize\Ionst{#2}}}}
\newcommand{\Ionw}[3]{\mbox{#1\,{\scriptsize\Ionst{#2}}~$\lambda\,#3$\,\AA}}

\newcommand{\Rsol}{$R_\odot$}

\begin{document}

\title{Variability of Central Stars of Planetary Nebulae with the Zwicky Transient Facility. II. Long-Timescale Variables including Wide Binary and Late Thermal Pulse Candidates\footnote{Based in part on observations obtained with the Hobby-Eberly Telescope (HET), which is a joint project of the University of Texas at Austin, the Pennsylvania State University, Ludwig-Maximillians-Universit\"at M\"unchen, and Georg-August Universit\"at G\"ottingen. The HET is named in honor of its principal benefactors, William P. Hobby and Robert E. Eberly.}}

\author[0000-0003-2071-2956]{Soumyadeep Bhattacharjee}
\affiliation{Department of Astronomy, California Institute of Technology, 1216 E. California Blvd, Pasadena, CA, 91125, USA}

\author[0000-0002-0119-7883]{Nicole Reindl}
\affiliation{Landessternwarte Heidelberg, Zentrum für Astronomie, Ruprecht-Karls-Universität, Königstuhl 12, 69117 Heidelberg, Germany}

\author[0000-0003-1377-7145]{Howard E. Bond}
\affiliation{Department of Astronomy \& Astrophysics, Penn State University, University Park, PA 16802, USA}
\affiliation{Space Telescope Science Institute, 3700 San Martin Dr., Baltimore, MD 21218, USA}

\author[0000-0002-6428-2276]{Klaus Werner}
\affil{Institut f\"ur Astronomie und Astrophysik, Kepler Center for
  Astro and Particle Physics, Eberhard Karls Universit\"at, Sand~1, 72076
  T\"ubingen, Germany}

\author[0000-0003-2307-0629]{Gregory R. Zeimann}
\affil{Hobby-Eberly Telescope, University of Texas at Austin, Austin, TX 
78712, USA}

\author[0000-0003-3947-5946]{David Jones}
\affiliation{Instituto Astrofísico de Canarias, E-38205, La Laguna, Spain}
\affiliation{Departamento de Astrofísica, Universidad de la Laguna, E-38206 La Laguna, Tenerife, Spain}

\author[0000-0002-6871-1752]{Kareem El-Badry}
\affiliation{Department of Astronomy, California Institute of Technology, 1216 E. California Blvd, Pasadena, CA, 91125, USA}

\author[0009-0006-3864-7645]{Nina Mackensen}
\affiliation{Landessternwarte Heidelberg, Zentrum für Astronomie, Ruprecht-Karls-Universität, Königstuhl 12, 69117 Heidelberg, Germany}

\author[0000-0002-8767-3907]{Nicholas Chornay}
\affiliation{Institute of Astronomy, University of Cambridge, Madingley Road, Cambridge CB3 0HA, UK}
\affiliation{Department of Astronomy, University of Geneva, Chemin d'Ecogia 16, 1290 Versoix, Switzerland}

\author[0000-0001-5390-8563]{S. R. Kulkarni}
\affiliation{Department of Astronomy, California Institute of Technology, 1216 E. California Blvd, Pasadena, CA, 91125, USA}

\author[0000-0002-4770-5388]{Ilaria Caiazzo}
\affiliation{Institute of Science and Technology Austria, Am Campus 1, 3400 Klosterneuburg, Austria}

\author[0000-0002-2626-2872]{Jan van~Roestel}
\affiliation{Anton Pannekoek Institute for Astronomy, University of Amsterdam, 1090 GE Amsterdam, The Netherlands}

\author[0000-0003-4189-9668]{Antonio C. Rodriguez}
\affiliation{Department of Astronomy, California Institute of Technology, 1216 E. California Blvd, Pasadena, CA, 91125, USA}

\author{Thomas A. Prince}
\affil{Department of Astronomy, California Institute of Technology, 1216 E. California Blvd, Pasadena, CA, 91125, USA}

\author[0000-0001-7648-4142]{Ben Rusholme}
\affiliation{IPAC, California Institute of Technology, 1200 E. California
             Blvd, Pasadena, CA 91125, USA}

\author[0000-0003-2451-5482]{Russ R. Laher}
\affiliation{IPAC, California Institute of Technology, 1200 E. California
             Blvd, Pasadena, CA 91125, USA}

\author[0000-0001-7062-9726]{Roger Smith}
\affiliation{Caltech Optical Observatories, California Institute of Technology, Pasadena, CA 91125}

\correspondingauthor{Soumyadeep Bhattacharjee}
\email{sbhatta2@caltech.edu}



\begin{abstract}

In this second paper on our variability survey of central stars of planetary nebulae (CSPNe) using the Zwicky Transient Facility (ZTF), we report 11 long-timescale variables with variability timescales ranging from months to years. We also present preliminary analyses based on spectroscopic and\slash or photometric follow-up observations for six of them. Among them is NGC~6833, which shows a $980$~day periodic variability with strange characteristics: `triangle-shaped' brightening in $r$, $i$, and WISE bands but almost coincidental shallow dips in the $g$-band. The most plausible explanation is a wide binary with the photometric period being the orbital period. Long-period near-sinusoidal variability was detected in two other systems, NGC~6905 and Kn~26, with periods of $700$~days and $230$~days, respectively, making them additional wide-binary candidates. The latter also shows a short period at $1.18$~hours. We then present CTSS~2 and K~3-5, which show brightening and significant reddening over the whole ZTF baseline. A stellar model fit to the optical spectrum of CTSS~2 reveals it to be one of the youngest post-AGB CSPNe known. Both show high-density emission-line cores. We propose these to be late-thermal-pulse candidates, currently evolving towards the AGB phase. We then present recent HST/COS ultraviolet spectroscopy of the known wide-binary candidate LoTr~1, showing that the hot star is a spectroscopic twin of the extremely hot white dwarf in UCAC2~46706450. Similar to this object, LoTr~1 also has a fast rotating wide subgiant companion. We suggest that the long photometric period of 11 years is the binary orbital period. Finally, we briefly discuss the ZTF light curves of the remaining variables, namely Tan~2, K~3-20, WHTZ~3, Kn~J1857+3931, and IPHAS~J1927+0814. With these examples, we present the effectiveness of the von~Neumann statistics and Pearson Skew-based metric space in searching for long-timescale variables.

\end{abstract}



\section{Introduction} \label{sec:intro}

\defcitealias{Chornay20}{C20}
\defcitealias{Chornay21Distance}{CW21}
\defcitealias{Bhattacharjee24}{Paper I}
\defcitealias{Chen25}{C25}

Over the past decade, studies of photometric variability of central stars of planetary nebulae (CSPNe) have significantly improved our understanding of these systems. The majority of past works have been dedicated to the search for binarity in CSPNe in support of the common-envelope (CE) formation channel of planetary nebulae (PNe; see e.g.; \citealt{Bond90Binary,Balick02,demarco09,jones17,Boffin19}). Wide-field photometric surveys like the Optical Gravitational Lensing Experiment (OGLE), \emph{Kepler}-K2, Transiting Exoplanet Survey Satellite (TESS), and recently the Zwicky Transient Facility (ZTF) have together discovered more than a hundred binary CSPNe \citep{Miszalski09a, Miszalski09b, Jacoby21, Aller20, Aller24, Bhattacharjee24, Chen25}. Periodicity has also led to the characterization of other phenomena like rotational variability caused by stellar spots (see for example \citealt{Werner19, Bond24}), or pulsations of the hot CSPN \citep{Bond90Pulsator, Ciardullo96, Hajduk14, Corsico21}.

Though much less explored compared to periodic variables, CSPNe also exhibit aperiodic variability. One of the causes is strong winds in CSPNe, mostly causing `jittering' variability (see for example \citealt{Handler97,Arkhipova12,Arkhipova13,Corsico21}). Unusual and large-amplitude aperiodic photometric variability of CSPNe has also led to the discoveries of R Coronae Borealis variables inside PNe, providing significant insights into the late thermal pulse formation channel of this class of stars \citep{Gonzalez98,Asplund97,Clayton97,Jeffery19,Rosenbush15,Jeffery19}. Recently, transits of dust and debris discs was proposed as an explanation for the deep quasi-periodic dips seen in the nucleus of the planetary nebula WeSb~1 \citep{Bhattacharjee24,Budaj25}. These demonstrate the variety of causes that can induce photometric variability in CSPNe. 

Most of the aforementioned phenomena occur on short timescales, ranging from a few minutes to a few days. On the other hand, there have only been very few instances of long-timescale photometric modulation in CSPNe. One example is PN IC~4997 \citep{Aller66} which shows decade-long brightening and dimming periods, which may result from either a long-period binary \citep{Kostyakova09} or wind\slash mass loss variations \citep{Arkhipova20}. Long-period light curve modulation due to binarity is seen in LoTr~5 (period of 7.4~years, \citealt{Kovari+2019}). Long-term photometric monitoring also resulted in the discovery of the $\sim2600$~day binary in pre-PN V510~Pup \citep{Manick21}. Post-asymptotic giant branch (AGB) stars undergoing (very) late thermal pulses also show long-timescale photometric changes (\citealt{Genderen95} for FG~Sge and \citealt{Schaefer15} for Stingray Nebula). There are a few other similar examples in the literature. However, there has not been any systematic search for long-timescale variability in CSPNe.

This is the second paper of the series conducting a systematic study of CSPN variability with Zwicky Transient Facility (ZTF, \citealt{Graham19,Masci19}). Here, we focus on the CSPNe showing long-timescale photometric variability in ZTF (the first one focused on the short-timescale variables, \citealt{Bhattacharjee24}, hereafter \citetalias{Bhattacharjee24}). We find 11 objects showing prominent long-term variations in the light curve, making this by far the largest study of long-timescale variability in CSPNe. The rest of the paper is as follows. In section \ref{sec:method}, we briefly review the methodology employed in selecting the variable sources. In section \ref{sec:results} we present our results. This includes spectroscopic observations of some of the objects and related analyses. Finally, we present our conclusions in section \ref{sec:conclusion}.

\section{Method and Sample Selection}\label{sec:method}

The detailed procedure for the selection of variable CSPNe in ZTF data has been described in \citetalias{Bhattacharjee24}. We briefly review the methodology here. Our starting sample is the \emph{Gaia}-EDR3-selected list of CSPNe presented in \cite{Chornay21Distance}. We restrict ourselves to sources with CSPN identification reliability $>50\%$, resulting in 1812 CSPNe, of which ZTF data are recovered for 991 sources. Following this, we applied several photometric quality cuts, primarily to avoid sources where the extended nebula significantly affects the photometry. This reduced the number of sources to 490. To quantify variability, we calculate the Normalized Excess Variance metric (NEV) for the light curves. Using an appropriate cut on this metric value, we arrive at our final list of 94 highly variable CSPNe (HNEV sample). Among the HNEV sources, we classify 83 objects to be short-timescale variables (the focus of \citetalias{Bhattacharjee24}). The variability of the remaining 11 objects occurs on much longer timescales (months to years) and is well resolved with ZTF cadence. We classify these objects as long-timescale variables, which are the focus of this paper. The list of the 11 long-timescale variables is presented in Table \ref{tab:long_timescale_vars}. For convenience of readers, we also provide upfront the summary of our analyses in Table~\ref{tab:long_timescale_vars_summary}, the details being presented in the respective sections.

In \citetalias{Bhattacharjee24}, we had presented a two-dimensional metric space defined by the metrics von~Neumann statistic (vonN) and Pearson Skew (SkewP) as a simple and novel technique to identify ``exotic variables." Briefly, the former metric quantifies the orderliness in the variability (against random variability from photon noise), and the latter differentiates between dimming and brightening events. We discussed that objects with low vonN ($\lesssim$$1$) are often ``interesting," i.e., objects which can teach us new phenomena. We demonstrated that the debris-disk candidate WeSb~1 lies in the low vonN space and its large negative SkewP successfully represents the deep eclipses. In Appendix \ref{sec:vonn_skewp}, we identify the long-timescale variables in this space and show its effectiveness in identifying them from large data sets. Such a discovery tool is expected to come in handy with upcoming large datasets with the advent of new surveys like the Legacy Survey of Space and Time (LSST, \citealt{Ivezic19}).

\section{Results and Analyses}\label{sec:results}

\begin{deluxetable*}{cccccccccc}[t]
\tablenum{1}
\tablecaption{List of the long-timescale variable CSPNe}
\label{tab:long_timescale_vars}
\tabletypesize{\scriptsize}
\tablewidth{-1pt}
\tablehead{
    \colhead{PNG} &
    \colhead{Name} & 
    \colhead{RA (deg.)} &  
    \colhead{Dec (deg.)} & 
    \colhead{Gaia DR3 ID} & 
    \colhead{PNStat} &
    \colhead{PNRad ($''$)} &
    \colhead{Rel} &
    \colhead{Known Period} &
    \colhead{ZTF Period} 
}
\startdata
082.5+11.3 & NGC 6833                  & 297.4440 & 48.9611  & 2086764737566091136 & T      & 0.3   & 1.0  & --          & $\sim$980 d                                    \\
061.4-09.5 & NGC 6905                  & 305.5958 & 20.1045  & 1816547660416810880 & T      & 21.65 & 1.0  & --          & 700 d?  \\
084.6-07.9 & Kn 26                     & 320.7889 & 38.9700   & 1965268354805776896 & T      & 55.0  & 1.0  & --          & 230 d? 1.18~hr   \\
044.1+05.8 & CTSS 2                    & 282.6953 & 12.6249  & 4504643171308707072 & T      & --   & 0.99 & --          & --          \\
034.3+06.2 & K 3-5                     & 277.9409 & 4.0859   & 4283434037501295744 & T      & 5.0   & 0.99 & --          & --            \\
 228.2-22.1 & LoTr 1                    & 88.7775  & -22.9007 & 2917223705359238016 & T      & 71.0  & 0.78 & 11.28 yr (1), 6.48~d (2)          & 6.48~d          \\
069.6+15.7 & Kn J1857.7+3931           & 284.4260 & 39.5167  & 2103243049609815296 & P      & 75.0  & 1.0  & --          & --         \\
032.5-03.2 & K 3-20                    & 285.5423 & -1.8126  & 4262011019866089856 & T      & --   & 0.98 & --          & --                                      \\
014.0-02.5 & Tan 2                     & 276.3213 & -17.9413 & 4096278711670069248 & T      & 3.0   & 0.83 & --          & --                                   \\
045.0-12.4 & WHTZ 3                    & 299.4972 & 4.7921   & 4248349003725092608 & T      & 46.0  & 0.99 & --          & -- \\
044.3-04.1 & IPHAS J192717.94+081429.4 & 291.8248 & 8.2415   & 4296182638654046592 & P      & --   & 1.0  & --          & --   \\
\enddata
\tablecomments{Description of the last five column names. PNStat: the PN status as reported in HASH. T and P stand for True and Probable, respectively. PNRad: radius of the PN as reported in HASH. Rel: CSPN assignment reliability fraction from \citet{Chornay21Distance}. Known Period: previously reported photometric period. ZTF Period: period(s) detected in ZTF light curve.}
\tablerefs{1) \cite{Martinez22} 2) \cite{Tyndall13}}

\end{deluxetable*}

\begin{deluxetable*}{cc}[t]
\tablenum{2}
\tablecaption{Summary of our analyses for the long-timescale variables}
\label{tab:long_timescale_vars_summary}
\tabletypesize{\scriptsize}
\tablewidth{-1pt}
\tablehead{
    \colhead{Name} & 
    \colhead{Summary} 
}
\startdata
NGC~6833 & `Triangle-shaped' periodic brightening in ZTF $r$, $i$ and WISE bands associated with shallow dips in $g$. Eccentric binary with periastron interaction?   \\
NGC~6905 & Long periodic ($P$$\approx$$700$~days, also $P$$\approx$$245$~days?) near-sinusoidal variability. Wide binary? \\
Kn~26 & Long periodic ($P$$\approx$$630$~days, also $P$$\approx$$230$~days?) near-sinusoidal variability: wide binary? Short period ($P$$\approx$$1.18$~hours) arising from either pulsation or close binary. \\
CTSS~2 & Steady brightening associated with significant reddening. Very young pre-white dwarf CSPN ($T_{\rm eff}$$\approx$$90$~kK and $\log(g)$$=$$4.5$). Late thermal pulse? \\
K~3-5 & Very similar light curve behavior as CTSS~2, but no information about the CSPN. Late thermal pulse? Winds and dust? \\
LoTr~1 &  A very hot ($T_{\rm eff}$$\gtrsim$$100$~kK) white dwarf with a rapidly rotating cool giant ($9.7~R_{\odot}$ and $T_{\rm eff}$$=$$4590$~K) companion. Orbital period of 11 years?\\
Kn\,J1857+3931 & Long-timescale dimming over the whole ZTF baseline. \\
K~3-20 & Long-timescale brightening (symmetric in appearance) associated with reddening. Part of a longer-periodic binary light curve? \\
Tan~2 & Pointed dip in the beginning of ZTF light curve. Undergoing dip again after $\simeq$$6.5$ years. Wide-orbit debris eclipses or binary? \\
WHTZ~3 & Short amplitude likely periodic long-term modulation in ZTF. Wide binary? Noisy data prevents detailed analyses. \\
IPHAS\,J1927+0814 & Erratic high amplitude variability. Reason unclear: dust and debris? accretion? \\
\enddata

\end{deluxetable*}

Below we present the objects showing long-term photometric variability. We note up front that, except WHTZ~3, the ZTF variability of all the other objects below was recently reported in a completely independent work by \cite[][henceforth, \citetalias{Chen25}]{Chen25}. However, they provide no discussion on the origin of the variability or supplement their discovery with follow-up observations. In this work, we provide follow-up spectroscopic and photometric observations of six of the objects, shedding light on the nature of the objects and the origin of their variability. Additionally, we list the remaining five variables and urge further research by the community.

All the objects discussed here appear in the catalog of PN (candidates) detected by GALEX \citep{Munoz23}. Five among these objects (LoTr~1, Kn\,J1857+3931, NGC~6833, NGC~6905, and Kn~26) also appear in the catalog of hot subluminous stars by \citet{Culpan22}.

\subsection{NGC~6833}\label{subsec:ngc6833}

NGC~6833 is a well-known PN and a ``True PN'' in the HASH catalog,\footnote{Hong Kong\slash Australian Astronomical Observatory\slash Strasbourg Observatory H$\alpha$ Planetary Nebula (HASH, \citealt{Parker16}) catalog is the largest catalog of PNe, serving as an important reference for \citetalias{Bhattacharjee24} and this work. It ranks its PN candidate entries as ``True," ``Likely," and ``Probable" PNe, in decreasing order of their likeliness of actually being a PN\null. The True PNe are spectroscopically verified to be so.} with an apparent diameter of $0\farcs6$. Due to its compact angular size, current ground-based surveys cannot resolve the central star from the nebula. Thus, the recorded fluxes are a combination of both the star and the nebula. Images from the Hubble Space Telescope (see for example \citealt{Hyung10}) reveal a box-like shape of the PN\null. NGC~6833 is present in various catalogs over decades, and has been the subject of several spectroscopic studies. Its photometric variability, however, remained unnoticed. 

\subsubsection{Light Curve and Spectroscopy}\label{subsubsec:ngc_6833_lc}

\begin{figure*}[t]
    \centering
    \includegraphics[width=\linewidth]{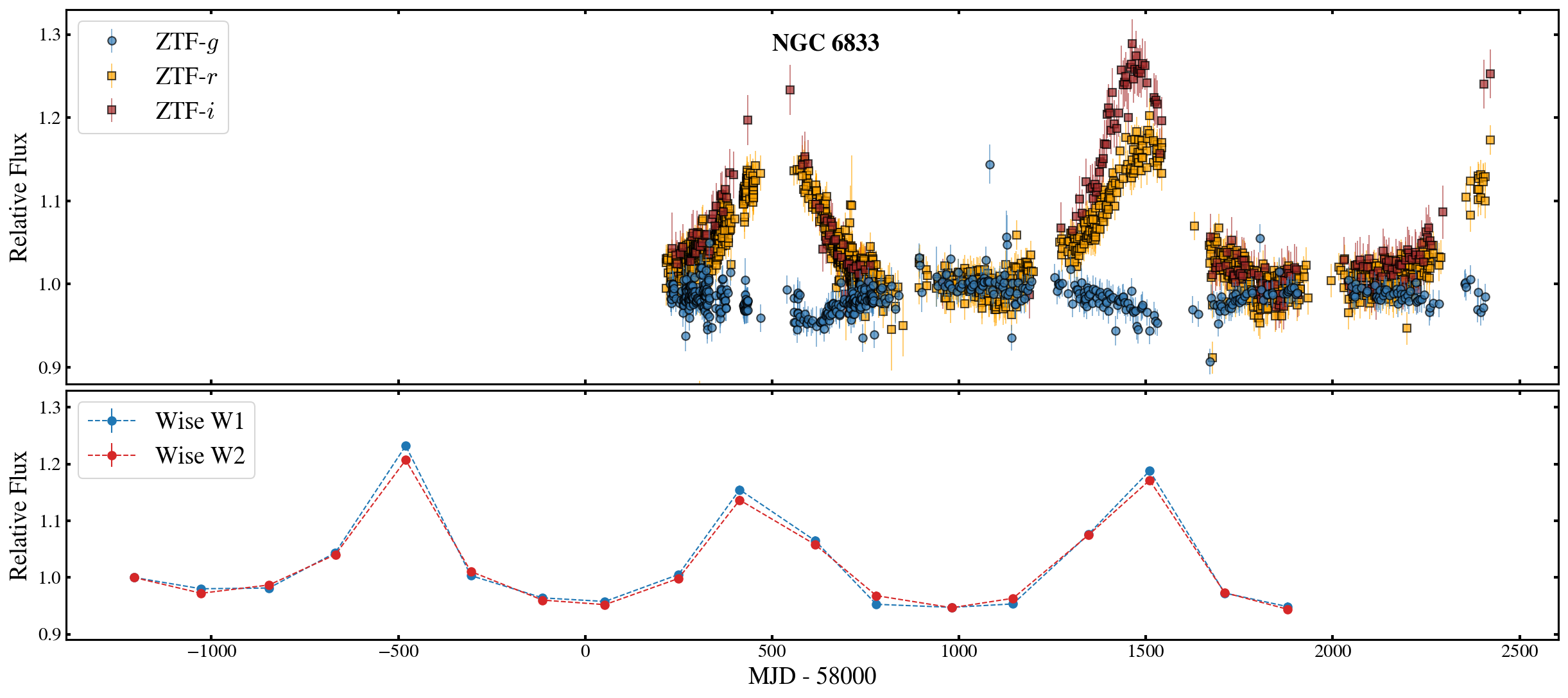}
    \caption{ZTF and neoWISE light curves for NGC~6833 (we do not show the two much older allWISE data points for ease of representation). The regular brightening events in ZTF-$r$ and redder bands, coincident in the optical and mid-IR are evident. The shallow dip in the ZTF-$g$ band corresponding to the peaks at the redder bands is also seen. For the details and discussion on this object, see Section \ref{subsec:ngc6833}.}
    \label{fig:NGC6833_ztf_wise}
\end{figure*}

\begin{figure}[t]
    \centering
    \includegraphics[width=\linewidth]{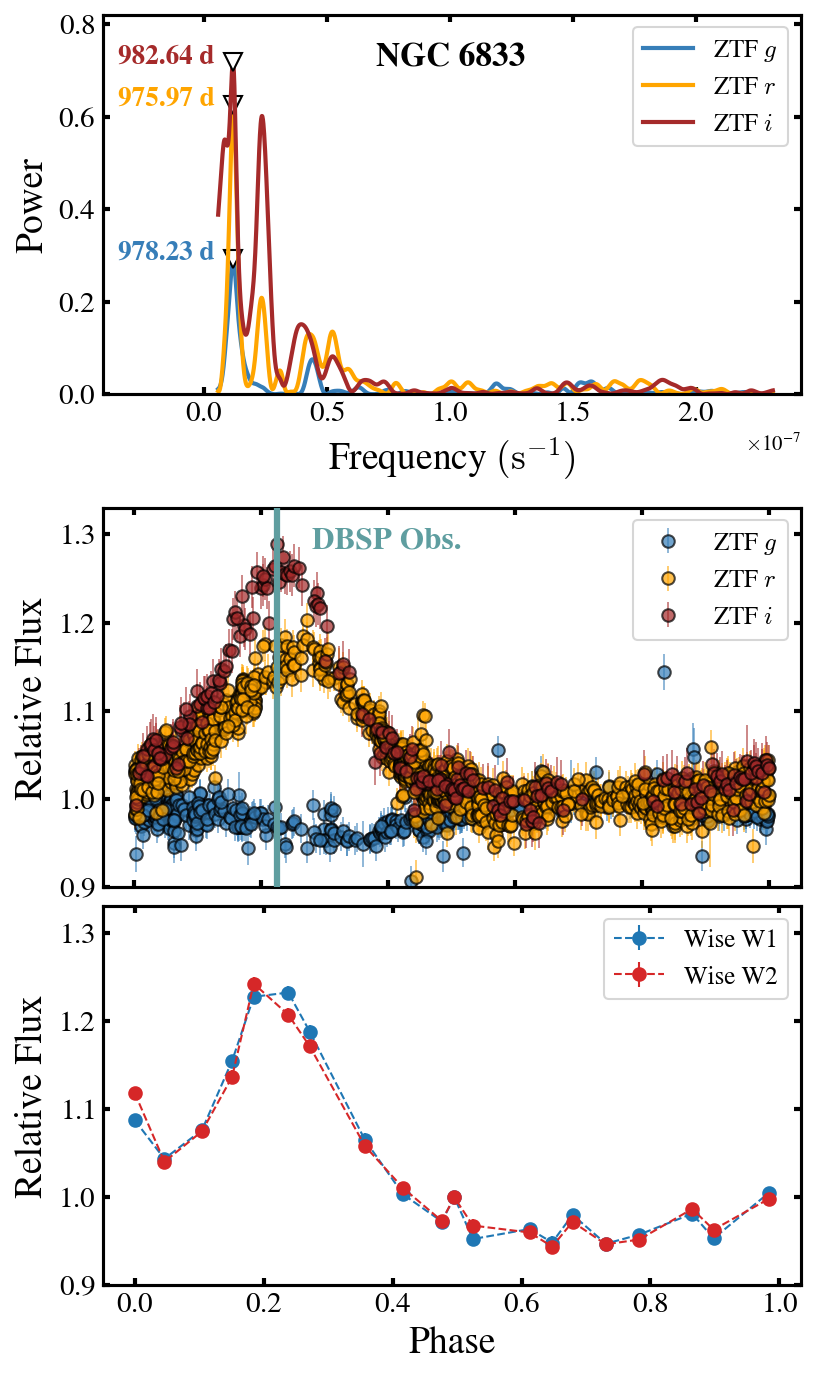}
    \caption{Top panel The Lomb-Scargle periodogram over for the ZTF light curve calculated over the period range of $50$~days ($2.3\times10^{-7}~{\rm Hz}$) to $2000$~days ($5.8\times10^{-9}$), with a resolution of $10^{-10}~{\rm Hz}$. Significant peaks are detected at ${\rm P\approx980}$~days. The power is strongest in the $i$-band, followed by $r$, and $g$. Middle panel ZTF light curve phase folded at the mean period of the three ZTF bands inferred from Lomb-Scargle of 978.95 days. The phase at the spectroscopic observation with DBSP has been marked. Bottom panel: WISE light curve (including the two allWISE data points) phase folded at the ZTF-inferred mean period. The first data points in each band are likely outliers.}
    \label{fig:NGC_6833_period_long}
\end{figure}

The ZTF light curve for NGC~6833 is shown in Figure \ref{fig:NGC6833_ztf_wise} (top panel). Fortunately, this target has good $i$-band coverage. The most prominent feature is the regular brightening events in the ZTF $r$ and the redder bands, with symmetric ingress and egress. The amplitude is in the range of $20-30\%$, with the largest amplitude in $i$-band. We note here though that the nebular flux can reduce the perceived variability, so the intrinsic amplitude of the source (likely the central star) can be much higher (see a brief discussion in Section \ref{subsubsec:ngc6833_discussion}). The less prominent but more intriguing feature appears in the ZTF $g$ band, which exhibits shallow (amplitude of ${\rm \lesssim10\%}$) dips nearly coincident with the peaks in the redder bands. The behavior is periodic and significant peaks appear in the Lomb-Scargle periodogram\footnote{We first use the \texttt{astropy.timeseries} implementation of the Lomb-Scargle periodogram \citep{Lomb76,Scargle82,VanderPlas18}} in all the three ZTF bands. The result is shown in the top panel of Figure \ref{fig:NGC_6833_period_long}. The detected periods in the three bands are marginally different (which is not surprising with long periods), with a mean value of $\sim980$ days. The phase-folded ZTF light curve at this mean period is shown in the middle panel of the same figure. Here, the activity appears to be slightly phase-delayed from the redder to the bluer bands. However, the current data are insufficient to establish this property with high confidence.

We also present the WISE light curve in Figure \ref{fig:NGC6833_ztf_wise}. For better presentation, we only show the neoWISE light curve in this Figure. The periodic brightening events are evident and coincident with the activity in ZTF. Unlike ZTF, the variability amplitude in both W1 and W2 appear similar. The lomb-Scargle periodogram applied to the WISE data shows a similar period of $\approx980$~days, as in ZTF. For consistency, we use the ZTF-inferred mean period to phase-fold the WISE data. The result is shown in the bottom panel of Figure \ref{fig:NGC_6833_period_long}. Except for the first data point (which happens to be the first allWISE reading and likely outlier affected by systematics relative to neoWISE), the phase folded light curves closely resemble that of the red bands in ZTF. 

Before any further discussion, we caution that the brightness of this object (${\rm \approx12.3}$ mag in $g$ and $\approx12.8$ mag in $r$) is very close to the saturation limit of ZTF ($\approx12-13$ in all bands). In fact, in one of the ZTF fields, the reference image is indeed saturated in the $g$-band (where the object is brightest, mainly due to the very strong [\ion{O}{3}] emission lines). Though such data points get rejected through the imposed quality cuts, the chance remains that the other flux readings are unreliable. To ascertain that this has not significantly affected the observed variability in the $g$-band, especially the dip, we checked the $g$-band light curve from the All Sky Automated Survey for SuperNovae (ASAS-SN) light curve provided at Sky Patrol V2.0\footnote{\url{http://asas-sn.ifa.hawaii.edu/skypatrol/}} (\citealt{Shappee14,Hart23}), which has a larger saturation brightness (${\approx11-12}$ magnitude). The ASAS-SN $g$-band light curve displays similar behavior as ZTF and the dips are visible. The data quality, however, is worse than ZTF thus not shown. For ZTF $r$ and $i$ bands, manual inspection shows that neither the reference nor the science images are saturated. We thus proceed with the assumption that the light curves are trustworthy. 

At the time of this work, the object was at its photometric extremum. Assuming the ZTF period to be stable, the best estimate places the previous optical observations of \cite{Wright05} and \cite{Hyung10} at the quiescent phase. This prompted us to obtain another spectrum, with the DouBle SPectrograph (DBSP; \citealt{Oke82}) attached to the Cassegrain focus of the Palomar 200-in Hale Telescope, to look for any significant changes. However, we could not find any notable difference between the spectra. The emission-line fluxes also agree well within error bars. The new spectrum is provided in Appendix \ref{app:ngc6833_opt_spec}.

We use the spectrum to determine if the observed photometric variability can arise from solely the variation in the emission lines. We consider the ZTF-$i$ band ($\approx7000-9000~{\rm \AA}$) which shows the largest amplitude of $\approx30\%$. We calculate the emission-line fluxes by subtracting the continuum. The emission line (including the Paschen series) contribution to the total flux is $\approx25\%$. With this spectrum taken at the $i$-band maximum, this would require a significant decrement in the line strengths of all the emission lines to contribute meaningfully to the variability, which is unlikely. This leads us to conclude that there is a significant variation in the continuum. However, given the compact nature of the PN, it is not possible to separate the stellar and nebular continuum with the current data.

\subsubsection{Discussion}\label{subsubsec:ngc6833_discussion}

The stable periodicity in the light curve strongly suggests binarity. The shape of the light curve with larger amplitude in redder bands\footnote{The strong emission lines, however, can significantly suppress any underlying continuum variability. To estimate the effect, we use the spectrum to subtract the emission lines and re-calculate the variability amplitudes in the ZTF bands. The amplitudes increased significantly. In fact, the ZTF $r$ band now has a higher amplitude than the $i$ band. This shows that the effect is significant and interpretation need to be made with sufficient care.} suggests an irradiation effect. This would mean that the photometric period is the binary orbital period. This would make it the fourth-longest known binary period inside a PN (after BD+33\,2642 \citealt{VanWinckel14, Jones17lotrngc}, LoTr~5, and NGC~1514 \citealt{Jones17lotrngc}). The large amplitude at such a long period suggests an eccentric binary. This is not a surprise as the other known long-period binaries are also eccentric. Irradiation, however, is not the only possible explanation. The light curves in the red bands resemble several RV Tauri systems showing long-term photometric modulation (known as the RVb phenomenon, see for example the light curves in \citealt{Kiss17}). In this case, the widely accepted explanation to the variability is the phase-dependent dust-obscuration of the flux-dominant star in the presence of a thick circumbinary disc \citep{VanWinckel99, Gezer15}. In this case, too, the photometric period represents the orbital period.

Neither of the two scenarios above can individually explain the anti-correlated behavior in the $g$ and the redder bands. A combination of both effects, however, could explain the variability. In an eccentric orbit, interaction (either directly between the stars or circumstellar disks) is enhanced near periastron. This can heat up the circumstellar dust leading to brightening in the red bands. Additionally, enhanced accretion can also lead to variability in emission lines, especially H$\alpha$. Such phenomena are sometimes observed during "pulsed accretion" processes in protostars (see for example \citealt{Muzerolle13}, though the corresponding period is much shorter) and AGB binaries \citep{Chen17,Bollen17}. Irradiation of the companion by the hot CSPN can also contribute to the brightening in the red bands. The dip in the blue band can then be caused by the heavy dust obscuration of the white dwarf, which is the dominant contributor in this wavelength range. The stochastic nature of the interaction may also explain any offset between the blue and red extrema.

We now consider possible cases where the photometric period does not relate to binarity. Rotational modulation is unlikely as, in most cases, the light curves in different bands are in phase. \citetalias{Chen25} discusses $\alpha^2$ Canum Venaticorum (ACV, a group of magnetic and chemically peculiar stars) a possibility as they also show anti-correlation between blue and red bands (see for example \citealt{Grobel17, Faltova21}). However, the periods are much shorter (all known objects having $P$$\lesssim$$10$~days, \citealt{Faltova21}) compared to NGC~6833. We also consider the possibility of it being Be stars losing mass periodically due to pulsational instabilities. In most such cases however, the ingress time is significantly shorter than the egress time and the infrared amplitudes are much larger than observed in NGC~6833 (see, for example, the light curves presented in \citealt{Froebrich23}). This makes this scenario, too, unlikely.

\subsection{NGC~6905 and Kn~26}\label{subsec:ngc6905_kn26}

\begin{figure*}[t]
    \centering
    \includegraphics[width=\linewidth]{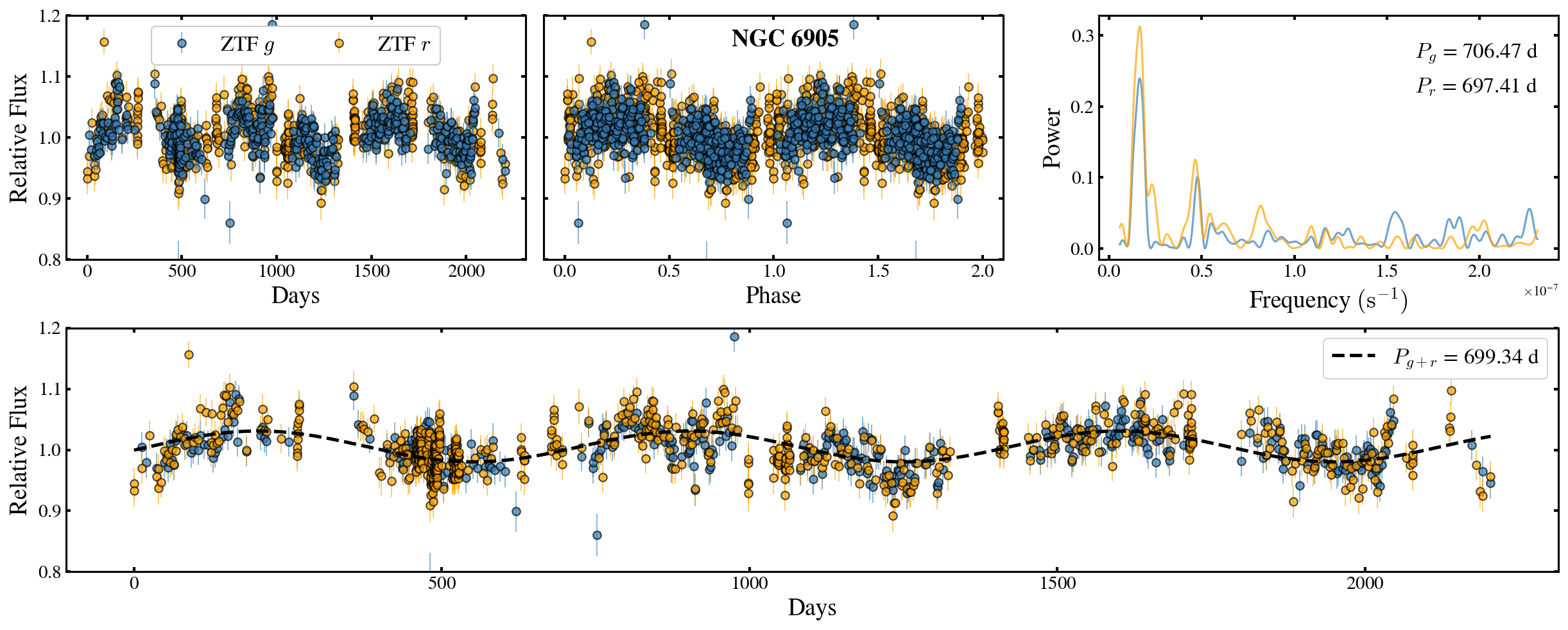}
    \caption{The long period variability in NGC~6905. Top Left: The raw ZTF light curve. Top Center: The ZTF light curve phased at the best period. Top Right: The periodogram for both the $g$ and $r$ bands. We also quote the respective periods in the panels. Bottom: Sinusoid at the best-fit period (in legend) of the combined $g+r$ data on the light curve for better visualization of the variability\slash periodicity.}
    \label{fig:ngc6905_period}
\end{figure*}

\begin{figure*}[t]
    \centering
    \includegraphics[width=\linewidth]{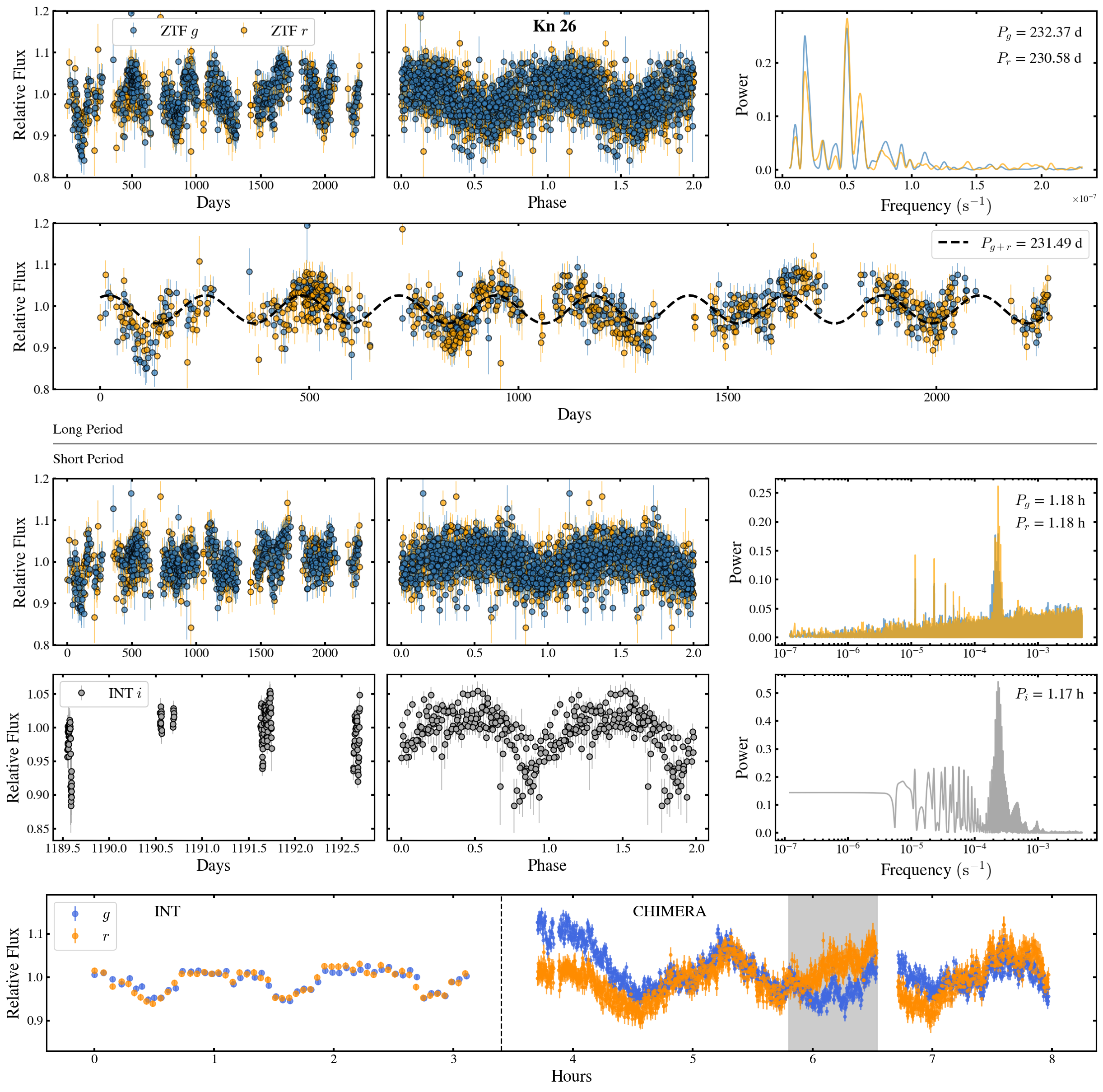}
    \caption{Periodicity in Kn~26. The long period is presented in the upper section in the same format as Figure \ref{fig:ngc6905_period}. The lower section presents the short period. The three panels in this section are as follows. Top panel, from the left: ZTF light curve with the long-period normalized, phase-folded at the best period (1.18~hr), and the corresponding Lomb-Scargle Periodogram. Middle panel: Same format as the middle panel with the INT $i$-band light curve. Bottom panel: The INT and CHIMERA $g$- and $r$-band light curves. For the latter, the shaded region represents the duration for which the instrument went out of focus. The reason for the between the light curves from the two instruments is unknown, but may very well be the result of suboptimal CHIMERA observations.}
    \label{fig:kn26_period}
\end{figure*}

NGC~6905 is a spectroscopically well-studied PN\null. Its central star is a [WR]-type star with a temperature of $\sim$150~kK \citep{Gomez22}. Spectral classifications of [WO] and [WC] have been given in the literature (e.g., \citealt{Gomez22} and \citealt{Corsico21}). The star also shows pulsations with periods in the range of $\sim$700--900 seconds \citep{Ciardullo96, Corsico21}. The nebula shows a bipolar morphology. 

Kn~26, on the other hand, is a known quadrupolar PN \citep{Guerrero13}, with Lanning~384 being its central star. \citet{Guerrero13} finds the SED to be consistent with a blackbody of $\sim70$~kK. Unlike NGC~6905, Kn~26 has not been spectroscopically classified before. We obtained follow-up optical spectroscopy with the Large Binocular Telescope (LBT) which reveals it to have a weak emission-line type ([WELS]) nucleus (see Appendix~\ref{app:kn26_spectrum}). 

\subsubsection{Light Curves}

Both NGC~6905 and Kn~26 show long-term modulation in the ZTF light curve. The Lomb-Scargle periodogram over the long-period range (same as mentioned in caption of Figure \ref{fig:NGC_6833_period_long}) shows a very prominent peak at $\approx700$~days and a weaker peak at $\approx245$~days for NGC~6905. For Kn~26, two peaks of similar powers at $\approx230$~days and $\approx630$~days (the former being slightly dominant) appear in the periodogram. The flux semi-amplitude corresponding to the most prominent periods for the two objects are $2.6\pm0.1\%$ and $3.4\pm0.2\%$, respectively. \footnote{\citetalias{Chen25} reports the 700~day period for NGC~6905 in their work. They, however, do not report these long periods in Kn~26. One might guess that this is due to the use of standard photometry against difference photometry. However, the long periods in both the objects are also recovered from the standard photometry data. This in turn shows that the photometric behavior is not an artifact of difference imaging.} The periods appear individually in the $g$ and $r$ bands (see Figures \ref{fig:ngc6905_period} and \ref{fig:kn26_period}, top panel). It is interesting to note that the inferred periods in the two objects are quite similar. This, and the closeness of the periods to possible systematic timescales (seasonal breaks and yearly observation cycles) prompted  several tests on the reality of the variability/periodicity. The details of the tests can be found in Appendix~\ref{appendix:ngc6905kn26periods}. We find no reason to reject the periods. 

In addition to the long periods, Kn~26 shows a short-period signal at $1.18$~hours. When a Lomb-Scargle analysis over the short-period range ($\sim$200 seconds to $\sim$100~days, same as used in \citetalias{Bhattacharjee24}) is applied on the raw ZTF light curve, the corresponding peak is visible, but it is subdominant to the $1$-day cadence peak. However, the peak appears dominantly when the long-period modulation is subtracted out from the light curve and the periodogram is re-run. This result is presented in the third panel in Figure \ref{fig:kn26_period}. The period is independently inferred in both the ZTF bands.

We used follow-up photometric observations to verify this period. Kn~26 was observed with the Wide Field Camera (WFC) at the Isaac Newton Telescope (INT) in 2021 July in the $i$-band with $120$~s exposure, and in 2022 August in $g$- and $r$-bands with $90$~s and $60$~s exposures, respectively. The data were debiased and flat-fielded, before performing differential photometry of the central stars against non-variable field stars, all using standard astropy routines \citep{Astropy18}. Recently, we re-observed Kn~26 with the Caltech HIgh-speed Multi-color camERA (CHIMERA) in $g$ and $i$ bands in 2024 September. The CHIMERA images were corrected using bias and flat images.\footnote{\url{https://github.com/caltech-chimera/PyChimera}} Photometry was extracted using aperture photometry (with a variable aperture size) using a modified version of the ULTRACAM pipeline \citep{Feline05}. We made a differential light curve using a single reference star. GPS timestamps were used to determine the mid-exposure time of each image.

A Lomb-Scargle periodogram applied to the INT $i$ band data yielded the same period as ZTF (fourth panel in the same figure). The morphology of the phase-folded light curve also resembles that of ZTF. Owing to fewer (and denser) observations with INT $g$ and $r$ bands and CHIMERA, we do not apply the period search. Rather, we show the full light curves in the bottom-most panel of the same figure. The INT light curves show a flat top with pointed periodic dips, in agreement with the INT $i$-band and ZTF light curve shapes. But the CHIMERA light curves appear markedly different. However, we note that there were some issues with the instrument which includes it going out of focus for a while (the shaded portion), which may have degraded the quality of the data. The periodicity, however, is visually evident in the light curves. 

\subsubsection{Discussion}

Ellipsoidal modulation or irradiation effect in a binary can result in the long-period variability. The systems can also be similar to LoTr~5, where the long-term photometric modulation is suspected to arise from stellar surface activity of a giant companion phased with the orbit \citep{Kovari+2019}. This is reinforced by the fact that the variability amplitudes of our objects are similar (slightly larger) to that of LoTr~5. In all of these cases, the orbital period is directly related to the photometric period (equal or double). These make both NGC~6905 and Kn~26 additional candidates for long-period wide binary CSPNe. 

The light curves also bear a resemblance to Long Secondary Period (LSP) photometric modulations observed in several RGB and post-AGB stars (for example, see \citealt{Kiss06}, \citealt{Nicholls09}, and \citealt{Soszynski21}, though the amplitudes are usually larger than seen in our targets.). This phenomenon is not yet well understood. But wide binarity and phase-dependent dust occultation (similar to RVb systems) is one proposed explanation. Here, too, the period corresponds to the binary orbital period. 

We now discuss two possibilities where the long period does not correspond to the orbital period, still requiring binarity. Magnetic activity cycles of giant stars in close binaries (RS~CVn systems) often result in long quasi-periodic photometric modulation. However, the associated periods are usually much longer ($\gtrsim$$5$~years, see \citealt{Martinez22}) than seen in NGC~6905 and Kn~26. Thus, this scenario is unlikely. The other possibility is stellar pulsations of giant stars, which is another widely accepted explanations for LSPs. \cite{Pawlak21} showed that LSPs tend to be stars in the RGB or AGB phase transitioning between period-luminosity sequences. Thus, with a twin-binary \citep{Elbadry19} setup it is possible that the companion of the CSPN is a pulsating LSP. High resolution radial velocity measurements is required to distinguish between the different scenarios.

The short period of 1.18~hours can arise from rotational modulation of the white dwarf \citep{Rosa24}. Another more exotic, but unlikely, possibility (also mentioned in \citetalias{Chen25}) is a close binary (this, with the long periods, may indicate a triple system!). The phase folded light curve is characterized by ``pointed" minimas, morphologically similar to ellipsoidal variability with grazing eclipses. In that case, the orbital period is 2.36~hr, making it the candidate for the shortest known CSPN binary period. At such short periods, for typical stellar parameters, we expect a large radial velocity variation of $\gtrsim$$100~{\rm km~s^{-1}}$. This should be detectable. The LBT spectra at different epochs, however, do not show any radial velocity shift.

\subsection{CTSS~2 and K~3-5}\label{subsec:ctss2_pnk35}

CTSS~2 (PN~G044.1+05.8) is a True PN in the HASH catalog. The nebula is faint and can be seen only in deep narrow-band ${\rm H\alpha}$ images (see Figure 12.8 in \citealt{Stanghellini16} for the HST image). There has been some uncertainty in the literature regarding the true nature of the central star of CTSS~2. \citet{Stanghellini16} note that the identified CSPN appears too bright to be associated with the faint and evolved PN, and considers the possibility of it being an unrelated field star. They also consider the possibility of it being a binary, as do \cite{MorenoIbanez16}. On the other hand, \cite{HernandezGarcia14} found emission signatures unusual for a PN in the {\em Spitzer} IR spectrum, and identified CTSS~2 to be a symbiotic system candidate. Confirmatory signatures of it being either a PN or symbiotic system are, however, lacking.

K~3-5 (PN~G034.3+06.2) is also a True PNe in HASH\null.  It is a resolved bipolar nebula, with an angular diameter of $10"$. The most recent spectroscopic study was performed in the spectral survey of central stars by \cite{Weidmann18}. They indicated the presence of emission lines both from the central stellar region and the extended nebula, thus making it a candidate double-envelope PN.  

\subsubsection{Light Curves}

\begin{figure*}[t]
    \centering
    \includegraphics[width=\linewidth]{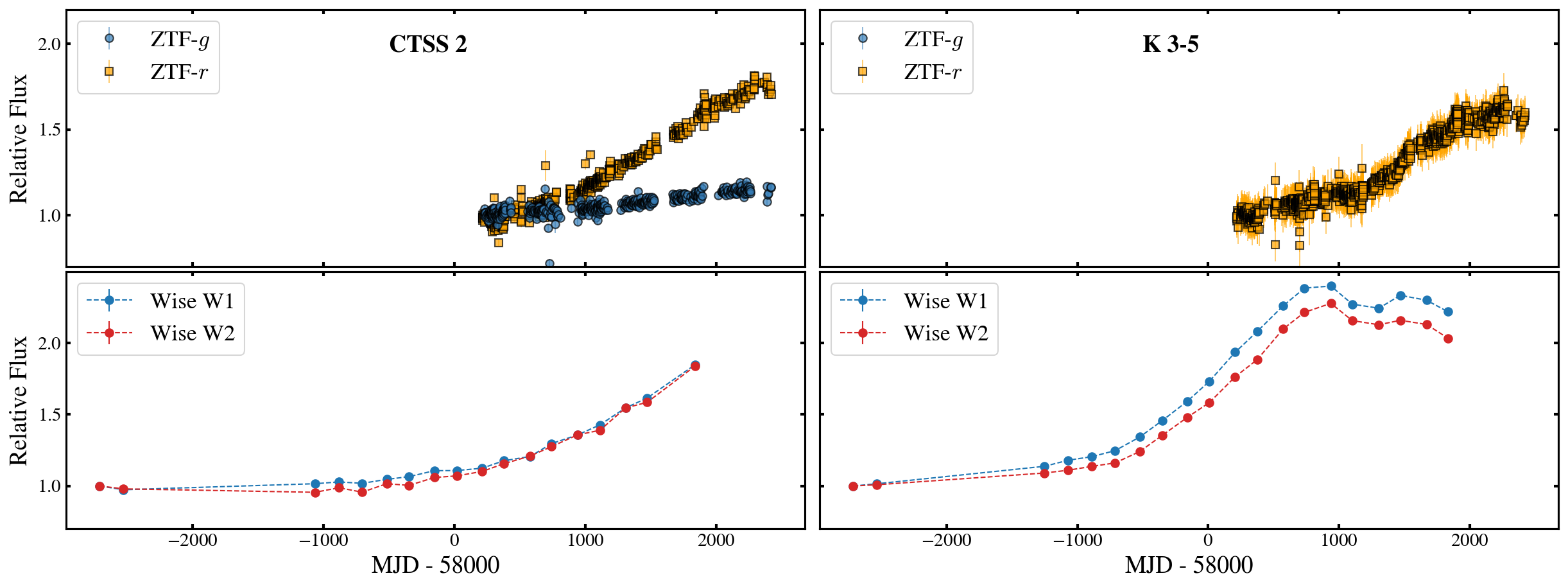}
    \caption{The ZTF and WISE light curves for CTSS~2 and K~3-5. Both show long-term photometric evolution, with the effect more pronounced in ZTF-$r$ and the redder bands. The $g$ band data for K~3-5 was rejected through the photometric quality cuts. However, manual inspection indicates a similar low amplitude trend as CTSS~2. For the details and discussion on these objects, see section \ref{subsec:ctss2_pnk35}.}
    \label{fig:CTSS2_PNK35_ztf_wise}
\end{figure*}

Figure \ref{fig:CTSS2_PNK35_ztf_wise} shows the ZTF and WISE light curves for CTSS~2 and K~3-5, with both objects showing very similar behavior. Over the past decade, the brightness of the objects in ZTF-$r$ and the redder bands have gradually increased by a factor of $\sim2$. An increase in brightness in ZTF-$g$ is also seen, albeit with much less prominence (flux increment by $\sim30\%$). The light curve for K~3-5 has two additional intriguing features. First is the turn-over of the WISE light curves at around ${\rm MJD\sim58700}$, unlike the optical light curve. Second, a significant bluing in the WISE bands (brightening of W1 compared to W2) is seen, contrary to the behavior in the optical. In CTSS~2, the WISE bands show no significant color dependence, unlike the ZTF light curves.

\subsubsection{Follow-up Optical Spectroscopy}

\begin{figure*}[t]
    \centering
    \includegraphics[width=\linewidth]{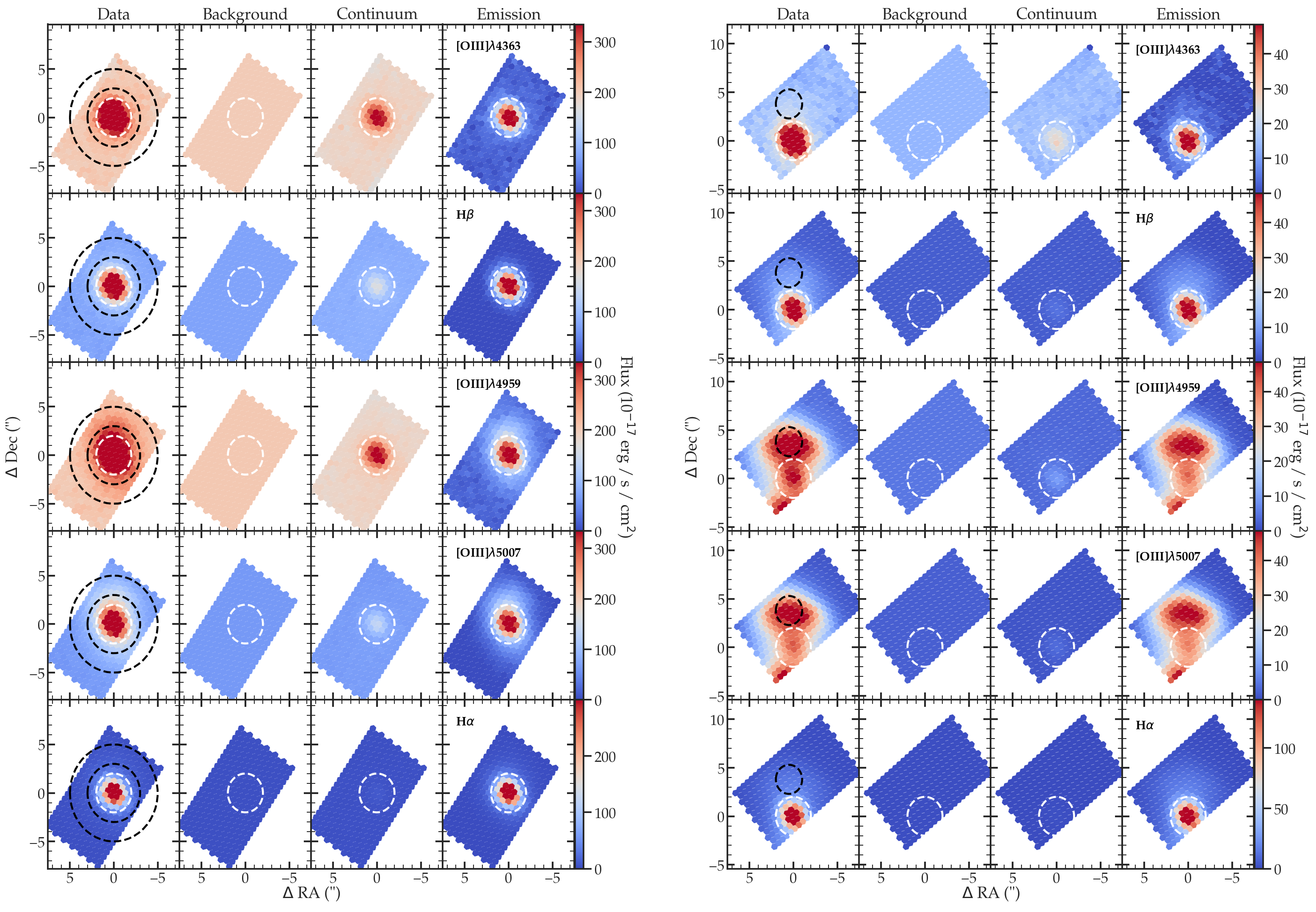}
    \caption{Synthetic narrow-band (NB, with a wavelength full-width of 16~\AA) images of CTSS~2 (left panel) and K~3-5 (right panel) created from the HET LRS2-B observations around the emission lines of [\ion{O}{3}]~4363,~4959,~5007~\AA, ${\rm H\alpha}$, and ${\rm H\beta}$. The `Data' column shows the NB readings without background\slash nebular subtraction. The second column shows the Background modeled from fibers beyond 6.5$''$ from the nucleus and sufficiently away from the primary emission regions. Care has been taken to exclude the nebula from the background as much as possible. The `Continuum' column shows the stellar continuum neighboring each emission line. The final `Emission' column shows the background and continuum subtracted image, showing the emissions. The dashed white circles show the 2$''$ radius extraction aperture centered on the nucleus for the one-dimensional spectrum. The black annulus (for CTSS~2, 3--5$''$ around the nucleus) and circle (for K~3-5, 1.5$''$ aperture) show the same for the corresponding nebulae. For further details on the synthesis of NB images from LRS2-B data, refer to sections 4.1 and 4.2 in \cite{Bond24egb6}. The central emitting core is distinctly visible against the nebula (which is diffused in case of CTSS~2 and bipolar for K~3-5). The prominence of the nebula is [\ion{O}{3}]~4959,~5007~\AA\ is also visible.}
    \label{fig:ctss2_k35_nb_images}
\end{figure*}

\begin{figure*}[t]
    \centering
    \includegraphics[width=\linewidth]{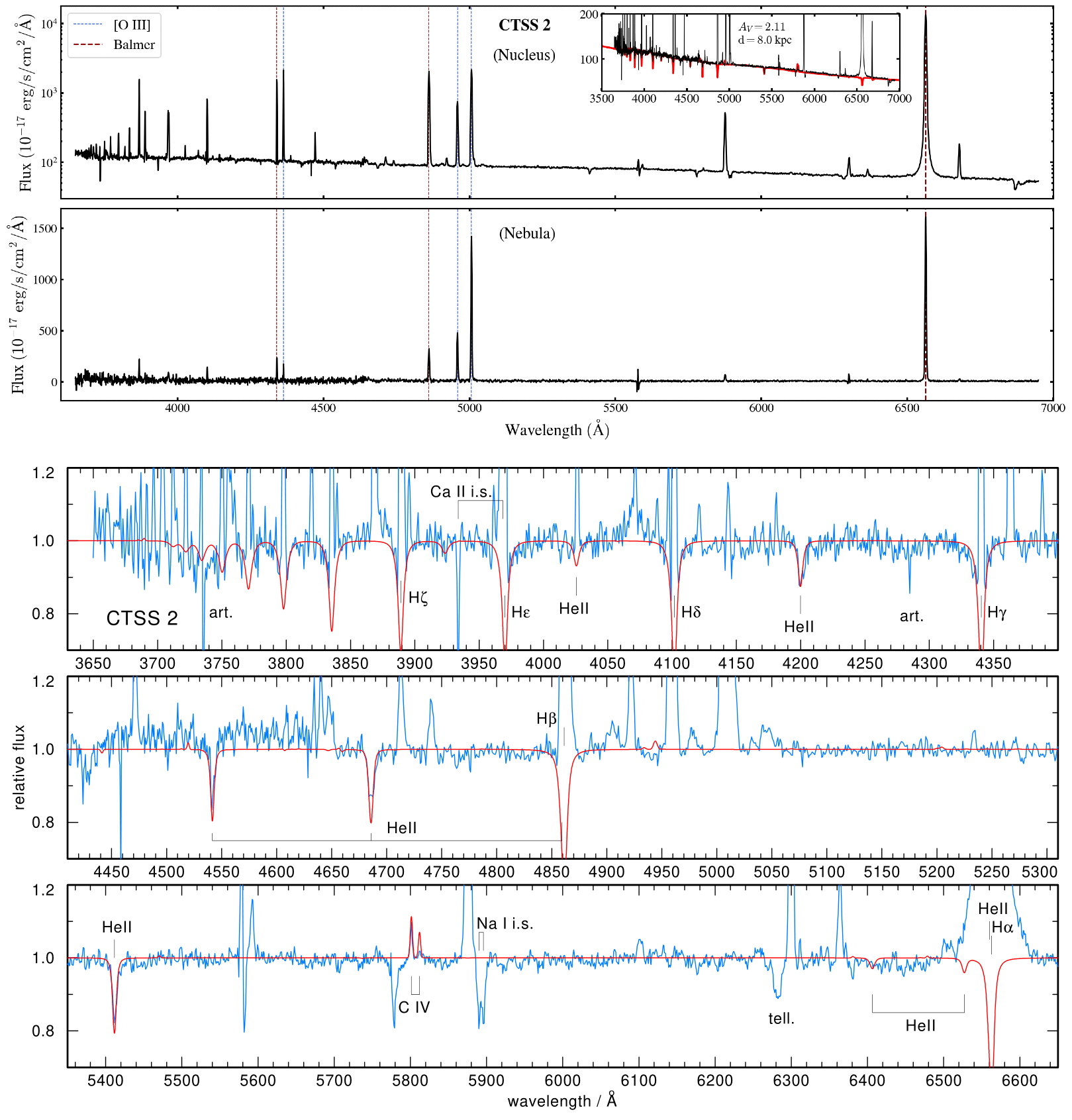}
    \caption{Upper section (two panels): One-dimensional HET spectra for the nucleus (top panel) and the nebula (bottom panel) of CTSS~2. The inset in the top panel shows the fitted stellar model compared to the un-normalized spectrum with assumed distance and extinction mentioned inside. Lower section: Model fit (red graph) to the HET spectrum of the CSPN of CTSS~2 (blue). Photospheric features are from \ion{He}{2}, \ion{C}{4}, and absorption wings in the higher-order Balmer lines.}
    \label{fig:ctss2_spectra_model}
\end{figure*}

\begin{figure*}[t]
    \centering
    \includegraphics[width=\linewidth]{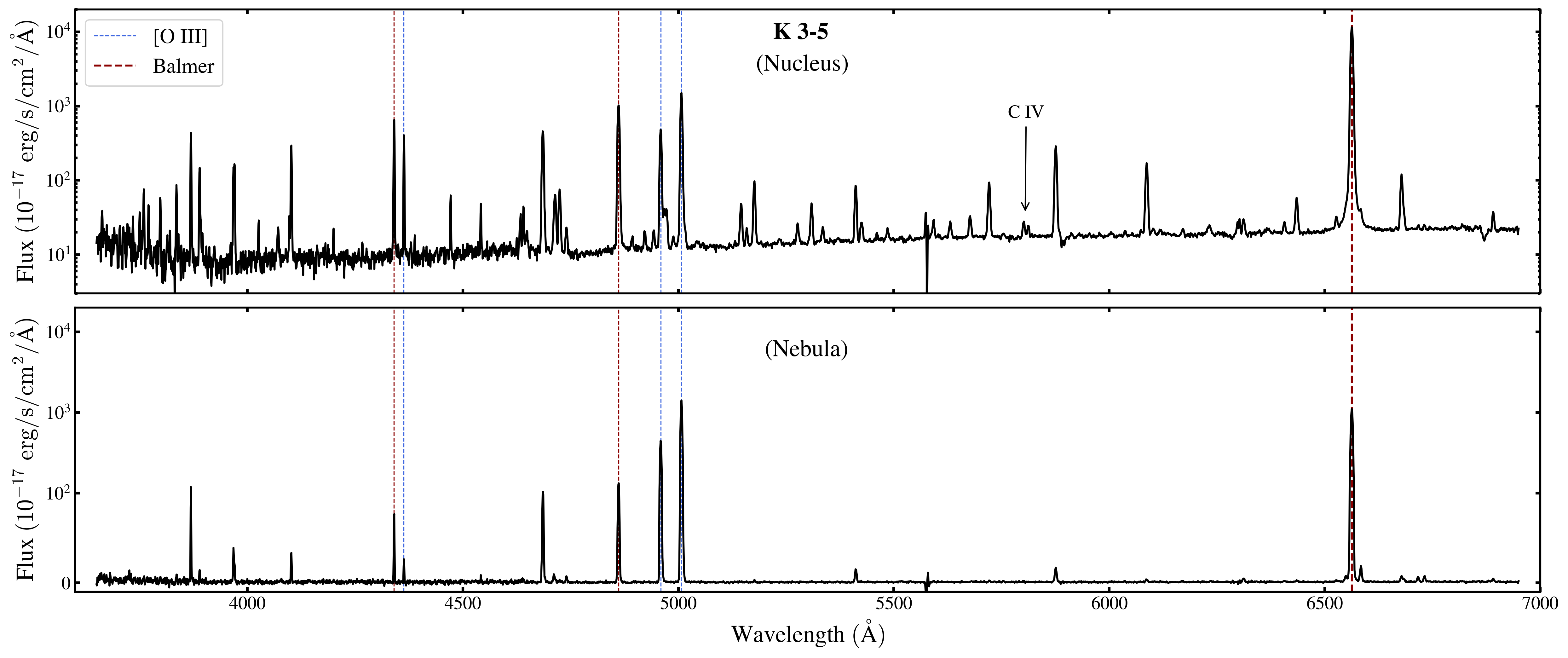}
    \caption{One-dimensional HET spectra for the nucleus (top panel) and the nebula (bottom panel) of K~3-5. The extraction fibers are mentioned in Figure \ref{fig:ctss2_k35_nb_images}. The \ion{C}{4}~5801,~5812~\AA\ emission from the nucleus is marked.}
    \label{fig:k35_spectrum_het}
\end{figure*}

\begin{figure}[t]
    \centering
    \includegraphics[width=\linewidth]{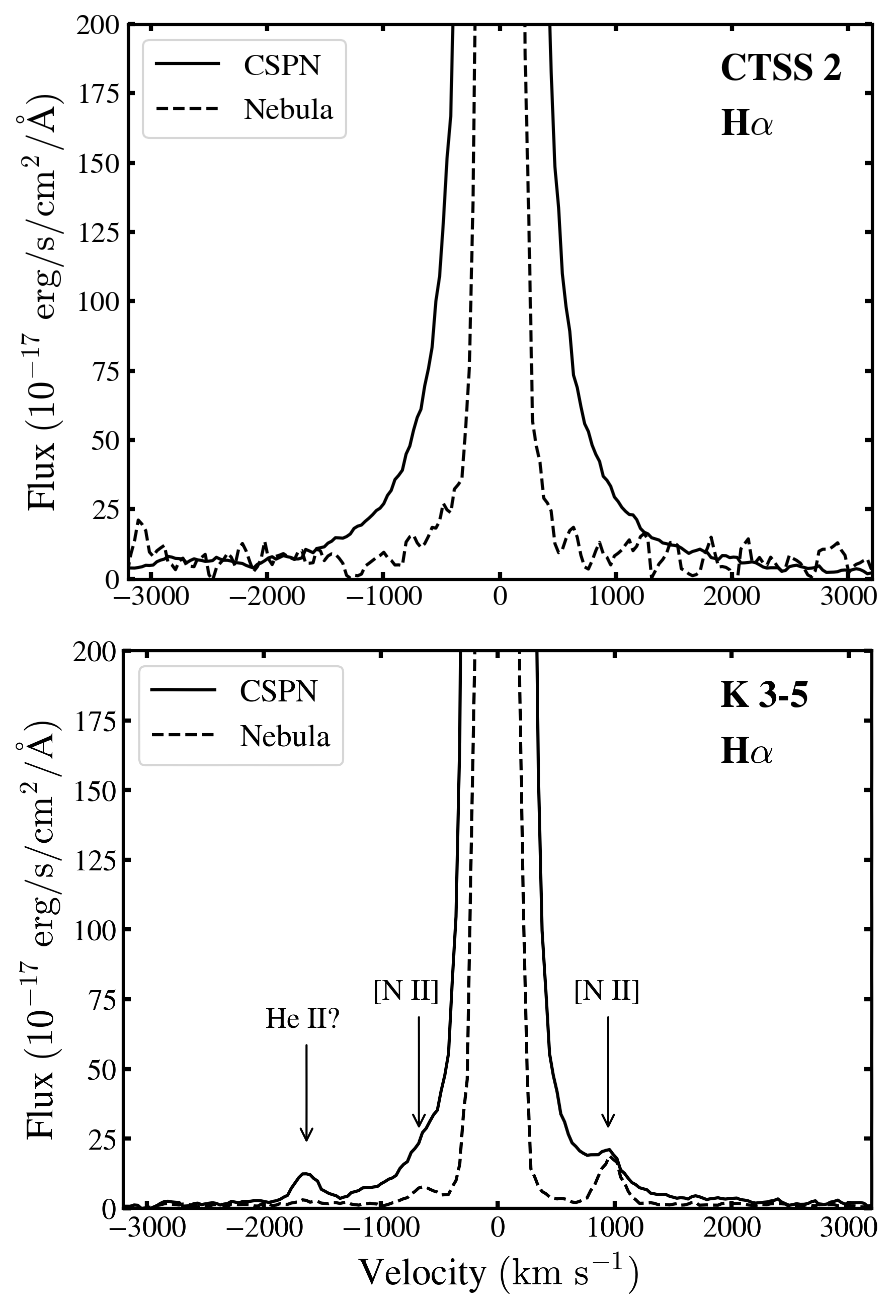}
    \caption{Zoom-in on the wings of the H$\alpha$ emission line in CTSS~2 (top panel) and K~3-5 (bottom panel) for both the nucleus (solid line) and nebula (dashed line). The extended wings up to $\gtrsim$$1000~{\rm km~s^{-1}}$ indicates wind\slash shocks. For K~3-5 the best guesses for the additional emission features are marked.}
    \label{fig:ctss2_k35_ha}
\end{figure}

CTSS~2 and K~3-5 were observed with the Low-Resolution Spectrograph (LRS2) on the 10-m Hobby-Eberly Telescope (HET; \citealt{Ramsey98, Hill21}) located at the McDonald Observatory in west Texas, USA, as a part of a spectroscopic survey of faint PN central stars (undertaken by H.E.B. and collaborators; see \citealt{Bond23} and subsequent papers). CTSS~2 was observed twice: 2024 September 15, $2\times250$~s, and 2024 October 15, $2\times300$~s. K~3-5 was observed once, on 2024 August 10, with a 900~s exposure.

The LRS2 instrument is described in detail by \cite{Chonis14,Chonis16}. Briefly, LRS2 provides integral-field-unit (IFU) spectroscopy with 280 $0\farcs6$-diameter lenslets that cover a $12'' \times 6''$ field of view (FOV) on the sky. LRS2 is composed of two arms: blue (LRS2-B) and red (LRS2-R). All of our observations are made with the target placed in the LRS2-B FOV\null. The LRS2-B arm employs a dichroic beamsplitter to send light simultaneously into two spectrograph units: the ``UV'' channel (covering 3640--4645 \AA\ with a resolving power of 1910), and the ``Orange'' channel (covering 4635--6950 \AA\ with a resolving power of 1140). Data reduction was performed using the pipelines developed by G.R.Z.\footnote{\url{https://github.com/grzeimann/Panacea}}$^{\rm ,}$\footnote{\url{https://github.com/grzeimann/LRS2Multi}} (Zeimann et al., in prep.).

The narrow-band (NB) images for both objects around the brightest emission lines in the spectrum: H$\alpha,~\beta$ and [\ion{O}{3}]~4363,~4959,~5007~\AA\ are shown in Figure \ref{fig:ctss2_k35_nb_images}. From the two-dimensional images, one-dimensional spectra were extracted for both the central core (nucleus) and the nebula. The corresponding apertures are shown within the NB images (white and black dashed apertures correspond to the nucleus and nebula, respectively). The extracted spectra for CTSS~2 and K~3-5 are shown in Figures \ref{fig:ctss2_spectra_model} (top panel) and \ref{fig:k35_spectrum_het} respectively.

Both CTSS~2 and K~3-5 were also observed with DBSP on 13th May, 2024 with exposures of 15~minutes. It is different from the HET spectrum only for its wider coverage in the red arm. However, the HET spectrum is of higher quality and, being an IFU, enables more accurate analyses. Thus, we base our discussion on the HET spectrum in the following sections. We provide the DBSP spectrum separately in Appendix \ref{appendix:ctss2_k35_dbsp}.

\subsubsection{Spectral Analysis: CTSS~2}

CTSS~2 consists of a bright nucleus and a faint surrounding nebula. The left panel of Figure \ref{fig:ctss2_k35_nb_images} shows the NB image of CTSS~2 around a few major emission lines. The bright core is very prominent and the faint nebula is also detected (especially in [\ion{O}{3}]~4959,~5007~\AA\ images). It is clear from these images that the nucleus is also very bright in the high-excitation emission lines, making it a double-envelope nebula.

We now turn to the one-dimensional spectra to understand the properties better (top panel in Figure \ref{fig:ctss2_spectra_model}). It is readily seen that the spectra (especially the relative strengths of the intensities of the major emission lines) of the nucleus and the nebula are markedly different. We now draw the attention of the readers to the strengths of H$\beta$ and [\ion{O}{3}]~4363~\AA\ relative to [\ion{O}{3}]~4959,~5007~\AA. In the nucleus, we find that the intensity of H$\beta$ is comparable to [\ion{O}{3}]~5007~\AA\, which is often attributed to very high electron densities. This is also supported by the high intensity of the [\ion{O}{3}]~4363~\AA\ emission line relative to the other [\ion{O}{3}] lines. 

To perform a more quantitative estimate, we define (as in \citetalias{Bhattacharjee24}) ${\rm O_{ratio}}=\frac{\rm [O~\sc{III}]~5007~\AA+[O~\sc{III}]~4959~\AA}{{\rm [O~\sc{III}]~4363~\AA}}$, which is related to the electron densities and temperatures (see Equation 3 in \citetalias{Bhattacharjee24}, taken from \citealt{Osterbrock04}). Overall, a lower value of ${\rm O_{ratio}}$ signifies very high densities and/or temperatures. We obtain ${\rm O_{ratio}}$$\approx$$3.5$ for CTSS~2 nucleus. This value is strikingly small for a PN. With this value, any reasonable temperature would estimate electron densities of $n_e$$\gtrsim$$10^6~{\rm cm^{-3}}$. Unlike the nucleus, the nebula as a much larger ${\rm O_{ratio}}$ of $\approx$$25.4$. This is consistent with the expectation that the diffuse nebula should be significantly less dense than the core. However, this is still lower than in most PNe and indicates moderately high electron densities of $n_e$$\gtrsim$$10^5~{\rm cm^{-3}}$.

We also find evidence of fast wind\slash shock activity in the nucleus. The top panel of Figure \ref{fig:ctss2_k35_ha} shows a zoom-in of the base of the H$\alpha$ emission lines for both the nuclear and the nebular spectra. Broad wings extending up to velocities of $\gtrsim$$1500~{\rm km~s^{-1}}$ are seen in the nuclear spectrum. In contrast, the nebular spectrum is much narrower, showing that the wind activity is mostly localized to the core. But the wings of the nebular H$\alpha$ still extend up to velocities of $\approx$$750~{\rm km~s^{-1}}$, which might indicate an extended influence of the nuclear activity. Note that such wind\slash shock may heat up the environment beyond that achievable through photoionization equilibrium, and contribute to lowering the inferred ${\rm O_{ratio}}$.

The continuum in the nuclear spectra contains absorption signatures (H and He), which we attribute to the central star. We also detect an emission doublet of \ion{C}{4}~5801,~5812\,\AA, typical of many hot stars. To infer the stellar parameters we performed a spectral fit. We computed a series of non-local thermodynamic equilibrium models of the type introduced by \citet{Werner19} using the T\"ubingen Model-Atmosphere Package \citep{2003ASPC..288...31W}. They are plane-parallel and in radiative and hydrostatic equilibrium.  They are composed of hydrogen and helium while carbon and nitrogen were included as trace elements keeping fixed the atmospheric structure. We assumed solar composition. Models were computed in the range $T_{\rm eff}$ = 50--90~kK and $\log(g/{\rm [cm~s^{-2}]})$ = 3.9--4.8. The best-fit parameters were determined by eye and found to be $T_{\rm eff}$$=$$70$$\pm$$10$~kK and $\log(g) = 4.5\pm0.2$ (bottom three panels in Figure \ref{fig:ctss2_spectra_model} shows the fit to the normalized spectrum). Using evolutionary tracks by \cite{M3B2016} this corresponds to stellar mass, radius, and luminosity of $M$$=$$0.65^{+0.06}_{-0.03}~M_{\odot}$, $R$$=$$0.75^{+0.24}_{-0.17}~M_{\odot}$, and $\log(L/L_{\odot})$$=$$4.08^{+0.48}_{-0.48}$. The corresponding age of the central star is inferred to be $\approx$$1500$$\pm$$500$~years. This makes it a candidate for one of the youngest post-AGB phase CSPN. We comment here on the possible ``contamination" from any nebular continuum. We do not expect significant contamination from the faint diffused nebula, but any continuum associated with the central emitting core can potentially dilute the absorption lines, rendering the spectral fit suboptimal. As this emission is mostly unresolved, appropriate modeling of the emission lines is needed to correct for any such contamination. Such a detailed study is beyond the scope of this paper.

The Gaia parallax distance to CTSS~2 is $d$$=$$8.77^{+6.97}_{-2.69}$~kpc. The analysis of \cite{Bailer-Jones} yields a slightly different distance estimate of $6.63^{+1.81}_{-1.47}$~kpc. We use the inferred stellar parameters to get an independent estimate of the distance to CTSS~2. We first employ a simple manual comparison of the model to the observed spectrum, by varying the reddening and the distance. A good visual fit was obtained with $E_{B-V}$$=$$0.68$ (i.e. $A_V$$=$$2.11$) and $d$$=$$8.0$~kpc (see inset in the top panel of Figure \ref{fig:ctss2_spectra_model}).\footnote{All dust extinction corrections in this work have been performed using the \texttt{dust\_extinction} \citep{dust_extinction, Gordon24_package_paper} implementation of the \cite{Karl23} galactic extinction models \citep{Decleir22, Gordon21, Fritzpatrick19, Gordon09}.} The distance is consistent with the Gaia distances. 

The extinction value used above is consistent with that inferred from the Balmer line ratios. We first use the nebular spectrum to estimate $A_V$. Following the same strategy as in \citetalias{Bhattacharjee24}, we get ${\rm BD_{\alpha\beta}}$$=$$6.13$ (the ratio of H$\alpha$ to H$\beta$) and ${\rm BD_{\beta\gamma}}$$=$$2.85$ which leads to extinction estimate of $A_V$$=$$2.12$$\pm$$0.02$. For the nuclear spectrum, we obtain ${\rm BD_{\alpha\beta}}$$=$$8.71$ and ${\rm BD_{\beta\gamma}}$$=$$2.98$ which translates to $A_V$$=$$2.7$$\pm$$0.3$. This is slightly higher than the nebular estimate but broadly consistent.

From the HST image in \cite{Stanghellini16}, we estimate the angular diameter of the nebula to be around $7''$. Assuming a distance to the nebula of $8$$\pm$$2$~kpc, gives the physical radius of the nebula is $\approx$$0.13$$\pm$$0.03$~pc. This gives an average expansion speed of $\approx$$88^{+78}_{-40}$$~{\rm km~s^{-1}}$. This is marginally higher than the average expansion velocities of PNe ($\approx$$40~{\rm km~s^{-1}}$, \citealt{Pereyra13, Jacob13}), but consistent within the error limits. 

\subsubsection{Spectral Analysis: K~3-5}

The right panel in Figure \ref{fig:ctss2_k35_nb_images} shows the narrow band images on K~3-5. It is quite intriguing to observe that the nucleus is brighter than the nebula in Balmer and [\ion{O}{3}]~4363~\AA\ emissions, however, the reverse is true for [\ion{O}{3}]~4959,~5007~\AA\ images.

We proceed to perform a similar analysis as with CTSS~2 with the one-dimensional spectra (Figure \ref{fig:k35_spectrum_het}): comparing the nuclear and nebular spectra and inferring the physical properties. The strengths of the emission lines are different between the nucleus and the nebula in a very similar way as in CTSS~2. The increased strength of the H$\beta$ and [\ion{O}{3}]~4363~\AA\ emission lines indicate a double envelope nature of the PN with a localized high-density emission core. This is in line with the inferences in \citet{Weidmann18}.

For the nebular spectrum, we get ${\rm O_{ratio}}$$\approx$$158$. This is at par with the typical values for evolved PNe. Additionally, [\ion{S}{2}]~6716,~6731~\AA\ lines are prominently detected in the spectrum. We obtain an intensity ratio of [\ion{S}{2}]~6716/[\ion{S}{2}]~6731$=$$0.91$. These two line ratios combined (from \citealt{Osterbrock04}) yield electron densities of $n_e$$\approx$$700~{\rm cm^{-3}}$ and temperature of $T_e$$\approx$$11$~kK. These properties are consistent with that of a PN. 

For the nucleus, however, we derive a significantly lower ${\rm O_{ratio}}$ of $\approx$$11$. For any temperature, this yields $n_e$$>$$10^5~{cm^{-3}}$, significantly higher than the inferred nebular density, in line with it being a high-density emission core. We note here that weak [\ion{S}{2}] lines are also detected in the nuclear spectrum, bearing a similar intensity ratio as the nebula. This is inconsistent with the low ${\rm O_{ratio}}$ value. We thus conclude that this is just a contamination in the nuclear aperture from the overlapping nebula. Unlike CTSS~2, no absorption signature from the central star is detected in the nuclear spectrum preventing any parameter estimation of the CSPN. However, certain emission lines might arise from the central stellar system. For example, we detect \ion{C}{4}~5801,~5812~\AA\ emission line in the nuclear spectrum (absent in the nebular spectrum). This doublet is often associated with either the photosphere of a hot CSPN (like CTSS~2) or an irradiated companion.

Like CTSS~2, K~3-5 also has broad H$\alpha$ emission wings in the nuclear spectrum, extending up to velocities of $\gtrsim$$1000~{\rm km~s^{-1}}$ (bottom panel of Figure \ref{fig:ctss2_k35_ha}). On the other hand, the nebula profile is much narrower, it's width probably being just limited by the instrument resolution. Note that faint [\ion{N}{2}]~6548,~6583~\AA\ are detected in both the nuclear and the nebular spectra (marked in the figure, though we suspect the former to be just nebular contamination like the [\ion{S}{2}] lines). In the same figure, we see an emission signature at $\approx$$-1500~{\rm km~s^{-1}}$ in the nuclear spectrum. This can be the \ion{He}{2}~6527~\AA\ line corresponding to the $n=14$ to $n=5$ transition (see for example \citealt{Lee01}).

We infer a very high reddening for this object. We infer ${\rm BD_{\alpha\beta}}$$=$$13.7$ and ${\rm BD_{\beta\gamma}}$$=$$3.4$ from the nuclear spectrum. The nebular spectrum also yields consistent values of $9.96$ and $3.67$ respectively. This translates to $A_V$$\approx$$3.7$$\pm$$0.4$. This explains the apparent red continuum of the nuclear spectrum. We note that such a high reddening revises the inferred values of ${\rm O_{ratio}}$ for the nucleus and the nebula to $\approx$$6$ and $\approx$$82$. None of our conclusion changes because of this, apart from increasing the density estimation for the nucleus to $n_e$$\gtrsim$$10^6~{\rm cm^{-3}}$.

\subsubsection{Discussion}

CTSS~2 is a good candidate for a late thermal pulse (LTP) scenario, in which the star experiences a late He-shell flash shortly after leaving the AGB (e.g. \citealt{Lawlor23}). LTPs are expected to occur frequently, and lead to an accelerated evolution of the star. Thus far, only two such stars have been "caught in the act": SAO\,244567 was observed during the initial rapid heating and contraction, which was followed around 2002 by a rapid cooling and expansion of the star \cite{Reindl+2014a, Reindl+2017}. The other example is FG~Sge which was observed only during the cooling an expansion phase, returning back to the AGB, where it turned into a H-deficient now an RCB star \citep{Jeffery06}. We stress, that the brightening rate of CTSS~2 is very similar to that of FG~Sge (see, for example, Figure~3 in \citealt{Genderen95}), and the strong reddening observed in the light curves of CTSS~2 suggests that the star is currently cooling. \citep{Jeffery06} explained the evolution of FG~Sge with a 0.625~$M_{\odot}$ LTP model, which agrees very well with our inferred mass of the CSPN of CTSS~2. We also note that the H-rich surface composition of CTSS~2 is what is expected from LTP models before the star manages to evolve back to the AGB (e.g. \citealt{Lawlor23}) and what was observed for SAO\,244567 and FG~Sge. K~3-5 might also be a LTP candidate, given the similarity in light curve behavior. But without any spectroscopic identification of the CSPN, the inference is more uncertain.

We now consider a few other possible scenarios. Firstly, we deem it unlikely that the observed light curves are part of a periodic binary signal. The lack of turnover in the light curves implies $P$$\gtrsim$$15$~years. At such a wide orbit, neither irradiation nor ellipsoidal variability can be strong enough to cause the observed brightening. The former might be significant for a highly eccentric orbit, but the duration of the periastron (thus the irradiation effect) then would be much smaller. Thus, we deem this scenario highly~unlikely.

The two systems can be similar to IC~4997 which shows decades-long photometric evolution comprising of both dimming and brightening episodes \citep{Aller66, Kostyakova09, Arkhipova20}. The cause for the variability is yet unclear. The other possibility is long term dust activity, as witnessed in M~2-29 \citep{Hajduk08,Miszalski11pnm229}. The spectral energy distributions of both CTSS~2 and K~3-5 indeed indicate presence of a large amount of dust (see Appendix \ref{subsubsec:ctss2_k35_ir_excess}). Long term photometric and spectroscopic\footnote{Comparison of the current and past spectra indeed indicates spectroscopic evolution of both CTSS~2 and K~3-5. The spectra of both of them on HASH (and in \cite{Stanghellini16} for CTSS~2) has H$\beta$ strength much less than the neighboring [\ion{O}{3}] lines, and a weak [\ion{O}{3}]~4363~\AA. This might indicate an increase in $n_e$ associated with the brightening. But we do not establish it with confidence because of a lack of knowledge about the details of the archival spectra (trace selection, nucleus vs nebula).} monitoring is needed for a more concrete understanding of these systems.

The high-density emission cores in both the systems warrant a separate mention. This is a property of the so-called EGB~6-type PNe \citep{Liebert13} bearing Compact Emission Knots (CEKs), hypothesized to be an irradiated dense accreted envelope around a wide-orbit companion to the CSPN white dwarf. Thus, this may be an indirect indication of the binarity of the CSPNe. For K~3-5, the bipolar nature of the nebula also supports multiplicity of the CSPN. Another explanation involving a companion might be a symbiotic interaction, where such high density cores are common. The lack of the spectral signatures of a Mira giant in the red end (see the DBSP spectrum with a better red coverage, Figure \ref{fig:ctss2_k35_dbsp}), however, speaks against them being the more common D-type symbiotic system. They can be the less common S or D'-type symbiotics. Presence of these systems inside PNe are rare. Confirmatory signatures of symbiotic behavior, like the Raman-scattered \ion{O}{6} emissions, are not seen in the spectra. But this does not rule out symbiotic interaction as such signatures are highly variable and not present in many systems.

\subsection{LoTr 1}\label{subsec:lotr1}

\begin{figure*}[t]
    \centering
    \includegraphics[width=\linewidth]{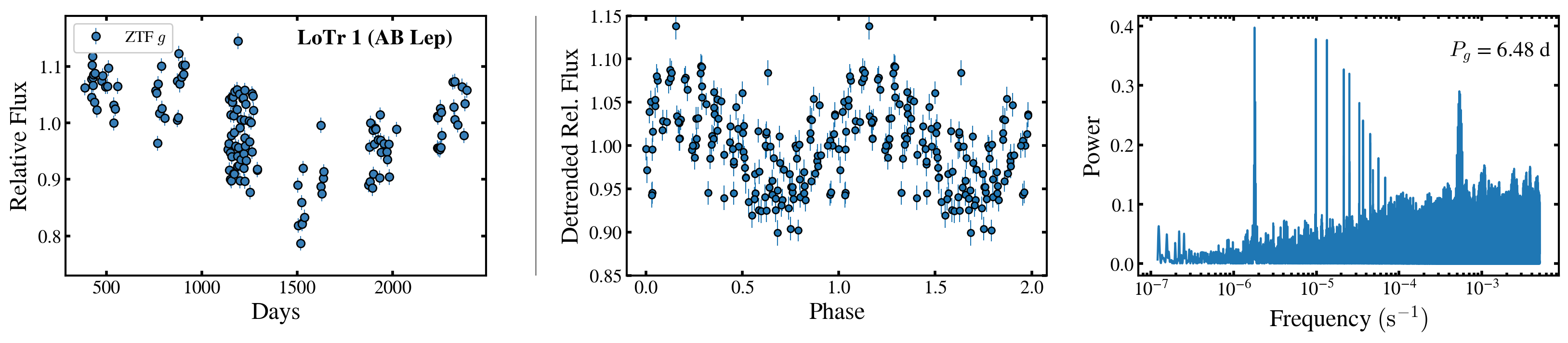}
    \caption{Variability in LoTr~1. Left: The raw ZTF light curve. Middle: Season-wise de-trended ZTF light curve phase folded to its best period. Right: The periodogram for the de-trended ZTF data and the best period.}
    \label{fig:lotr1_ztf}
\end{figure*}

LoTr~1 is a True PN in HASH and has a diameter of 145$''$. Its CSPN, AB~Lep, is a known variable. 

\subsubsection{Light Curves}

Long-term photometric modulation was reported in \citet{Martinez22}, with a period of 11.28 years. They present LoTr~1 as a RS~CVn system: a close binary with a giant companion where the magnetic activity cycle of the giant star leads to long-period photometric variability. On short timescales, it shows a period of 6.4~days, which has been attributed to the rotation of the giant star \citep{Tyndall13}.

The ZTF light curve shows a long-lasting `dip' (left panel in Figure \ref{fig:lotr1_ztf}). This is likely a segment around the minimum of the longer periodic variability (the time of the minima and the amplitude are consistent with that expected from \citealt{Martinez22}). We successfully recover the signal in the season-wise de-trended ZTF light curve (the right two panels in the same figure) as the dominant periodogram peak. Note that \citetalias{Chen25} also detected this period, but, unlike this work, it did not appear as their dominant peak (their Figure 14). 

We note that LoTr\,1 is similar to LoTr\,5, which consists of an unresolved hot (pre-)white dwarf and rapidly rotating magnetically active G5 giant pair. For the latter system \citet{Kovari+2019} uncovered a long-term photometric variability which appears in phase with the orbital radial velocity of the binary ($P_{\mathrm{orbit}}\approx 7.4$\,yr). Thus, one can likewise speculate that the long-term variability in LoTr\,1 might reflect the orbital period of the binary system. Unlike \cite{Martinez22}, we propose that this is a wide binary with a rapidly rotating giant companion.

\subsubsection{HST/COS Spectroscopy and Spectral-Energy Distribution}

The central star of LoTr\,1 was observed for 2299\,s on March 29, 2023, with the Cosmic Origins Spectrograph (COS) aboard the Hubble Space Telescope (HST), using the G130M grating centered on 1291\AA\ (dataset LEZCI2020, PI: N. Reindl). Another 1640\,s exposure was taken with HST/COS on July 30, 2023, using the G140L grating ($R\approx 1000$) that covers $\approx 1150-1700$\,\AA\ (dataset LEZCI5010, PI: N. Reindl). The spectra were retrieved from the MAST archive.

\begin{figure*}[t]
    \centering
    \includegraphics[width=\linewidth]{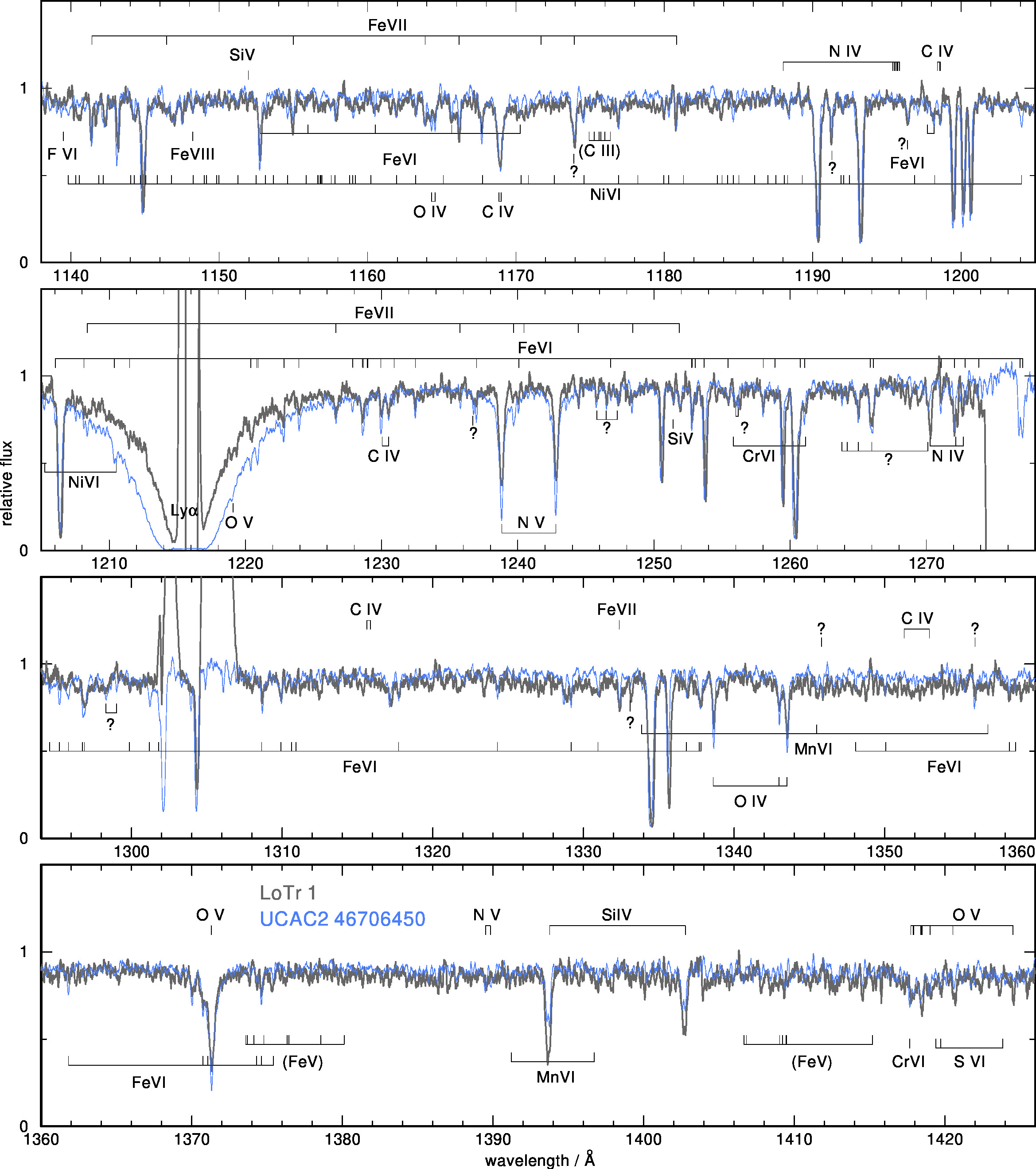}
    \caption{HST/COS G130M spectrum of the LoTr\,1 (gray) compared to the HST/COS G130M spectrum of its spectroscopic twin, the very hot white dwarf in UCAC2\,46706450 (blue graph). The spectra were smoothed with a 0.1\,\AA\ wide boxcar to increase the signal-to-noise ratio.
    Prominent photospheric lines are identified. Identifications in brackets denote uncertain detections. Question marks indicate unidentified photospheric lines.}
    \label{fig:cos}
\end{figure*}

The hot central star of LoTr\,1 turns out to be a spectroscopic twin of the extremely hot white dwarf in UCAC2\,46706450 \citep{Werner+2020} as it becomes obvious when looking at Fig.~\ref{fig:cos} where we compare the UV spectra of both stars. The lack of observed \Ion{Fe}{5} lines as well as absence of the \Ion{C}{3} multiplet at 1175\,\AA\ imposes a lower $T_{\rm eff}$\ limit of 100\,000\,K \citep{Werner+2020}. In the G140L spectrum a weak \Ionw{He}{2}{1640} can be detected which suggests that the central star of LoTr\,1 is either a very hot DAO white dwarf or a O(H)-type pre-white dwarf. Interestingly, the companion of UCAC2\,46706450 is also a rapidly rotating giant star.

In order to estimate the radii of both stars we carried out a two component fit to the spectral energy distribution (SED) of LoTr~1, employing the $\chi ^2$ SED fitting routine described in \cite{Heber+2018} and \cite{Irrgang+2021}. For the cool star we used PHOENIX models calculated by \cite{Husser+2013} and for the hot central star the metal free model grid computed by \cite{Reindl+2023}. We fixed the $T_{\rm eff}$ of the central star to 105\,000\,K (i.e. to the value that was derived for UCAC2\,46706450 by \citealt{Werner+2020}), and then let the let the angular diameter, reddening, surface ratio of both stars, as well as the $T_{\rm eff}$ of the cool companion vary freely. By that we derive a reddening of $E(44-55)=0.08$\,mag\footnote{\cite{Fitzpatrick+2019} employs $E(44-55)$, which is the monochromatic equivalent of usual $E(B-V)$, using the wavelengths 4400\,\AA\ and 5500\,\AA, respectively. For high effective temperatures such as for the stars in our sample $E(44-55)$ is identical to $E(B-V)$.} and a radius of 0.44\,\Rsol\ for the hot central star (identical to UCAC2\,46706450), indicating that it is just about to enter the white dwarf cooling sequence. For the cool giant star we derive $T_{\rm eff}$$=4590$\,K and a radius of 9.7\,\Rsol. These parameters are broadly consistent with a subgiant star of mass in the range of $\approx$$1-2~M_{\odot}$,\footnote{We compared the stellar parameters to Modules for Experiments in Stellar Astrophysics (MESA, \citealt{Paxton11,Paxton13,Paxton15}) evolutionary tracks queried from MESA Isochrones and Stellar Tracks (MIST, \citealt{Dotter16, Choi16}). Owing to the preliminary nature of the analysis we do not show the comparison. This will be presented elsewhere with spectroscopic observations.} which is also very similar to the companion of UCAC2\,46706450. With the 6.48~day rotational period, this implies a surface velocity of $75~{\rm km~s^{-1}}$.

\begin{figure}[t]
    \centering
    \includegraphics[width=\linewidth]{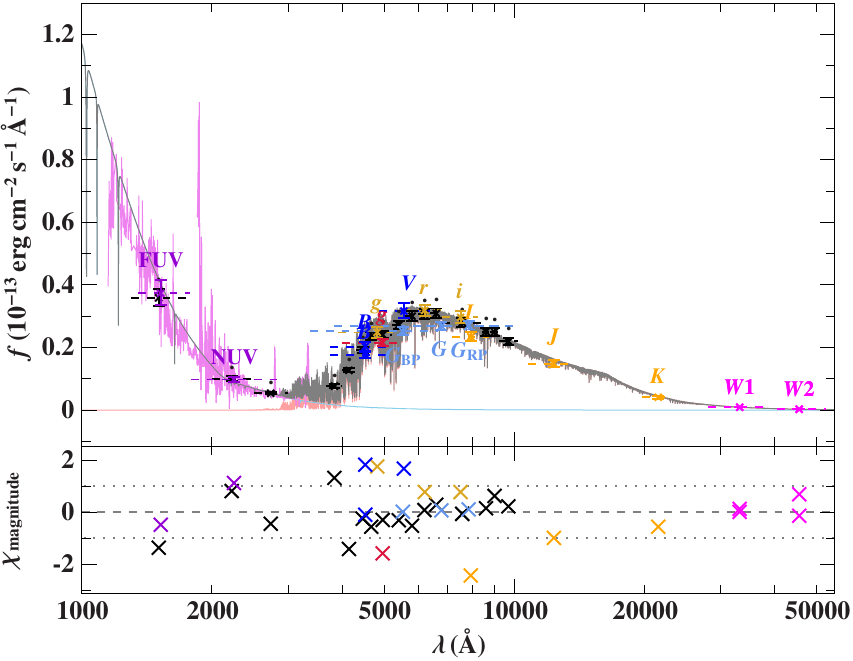}
    \caption{Top panel: SED fit for LoTr\,1. Filter-averaged fluxes converted from observed magnitudes are shown in different colors. The respective full width at tenth maximum are shown as dashed horizontal lines. In pink the spectrum from the International Ultraviolet Explorer is shown. The combined best-fitting model, degraded to a spectral resolution of 6\,\AA\ is plotted in gray. The blue graph represents the model for the hot central star and the salmon graph is the model for the cool giant. Bottom panel: the difference between synthetic and observed magnitudes.}
    \label{fig:lotr1_sed}
\end{figure}

\begin{figure}[t]
    \centering
    \includegraphics[width=\linewidth]{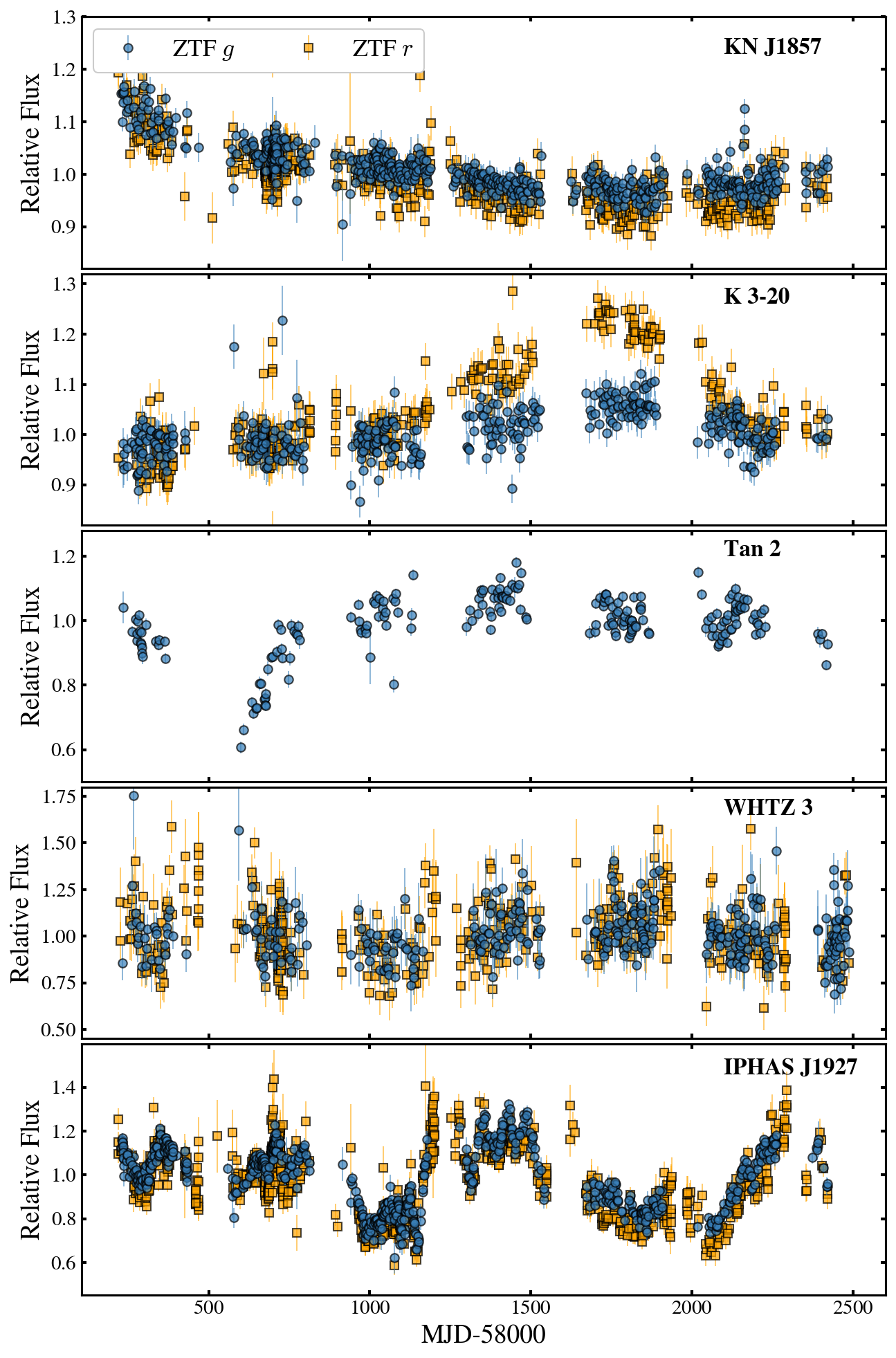}
    \caption{The light curves for the remaining five long-timescale variables within the HNEV sample. A brief discussion on these objects is presented in section \ref{subsec:long_vars}. }
    \label{fig:long_timescale_vars}
\end{figure}

\subsection{The other long-timescale variables}\label{subsec:long_vars}

In this section, we briefly introduce the remaining five long-timescale variables. Further spectroscopic and photometric studies of these objects are underway and will be presented in future works. The ZTF light curves are shown in Figure \ref{fig:long_timescale_vars}. Except for Kn\,J1857.7+3931 (henceforth, Kn\,J1857) and IPHAS\,J192717.94+081429 (henceforth, IPHAS\,J1927), all of the long-timescale variables are True PN on HASH with resolved nebulae. Kn\,J1857 has now been reclassified to `object of unknown nature' on HASH and IPHAS\,J1927 is a Possible PN. Neither of the two has a resolved nebula (thus either very young PNe or some stellar PN mimic). 

We first discuss the True PNe. K~3-20 shows a brightening event, with a larger amplitude in the ZTF $r$ band. Tan~2 shows a singular eclipse-like dip in the $g$ band. Unfortunately, the $r$ band data for this object did not pass the photometric quality cuts. The light curve of WHTZ~3, albeit noisier, shows low amplitude long-term modulation. The possible causes for variability of these objects are uncertain. The observed light curves can be a part of even longer timescale periodic variability associated with binarity. For example, the behavior in K~3-20 may result from irradiation effect in a long-period wide binary (similar to NGC~6833). Further long-term photometric monitoring is needed to test these scenarios. Concerning the remaining two objects, Kn\,J1857 shows a gradual dimming, by $\sim25\%$ in flux, with similar amplitude in both the ZTF bands. The variability in IPHAS\,J1927, on the other hand, is more irregular. Similar kinds of photometric behaviors are often seen in symbiotic systems or cataclysmic variables (CVs). Being only a ``Possible PN,'' spectroscopic data are necessary to determine the true nature of this object.

\section{Summary and Conclusions}\label{sec:conclusion}

This is the second in a series of papers undertaking a systematic study of the variability of central stars of planetary nebulae (CSPNe), using optical light curves from the Zwicky Transient Facility (ZTF). In \citetalias{Bhattacharjee24}, we applied appropriate variability metrics to identify significantly variable sources. The present paper reports the 11 objects which show long-timescale variability. For six of these objects, we present follow-up spectroscopic\slash photometric observations and provide extensive discussion, which is expected to serve as the first step in understanding the objects. We also present the remaining six objects and encourage follow-up studies by the community.

The first object we discuss is NGC~6833 (Section \ref{subsec:ngc6833}). It shows `triangle-shaped' brightening in the ZTF $r$, $i$, and WISE W1 and W2 bands, occurring at a period of $\sim$$980$~days. The more intriguing aspect is the anti-correlated behavior in the ZTF $g$-band, which shows shallow dips nearly at the same epochs as the peaks in the redder bands. We think that this is a long-period eccentric binary, with the photometric period corresponding to the orbital period. This makes it the candidate for the fourth-longest binary orbital period inside a PN\null. Though the irradiation effect can possibly explain the brightening, the simultaneous dips in $g$-band is puzzling. Dust occultation and extinction can potentially result in such a feature.

Long-periodic modulation was also detected in the light curves of NGC~6905 and Kn~26 (Section \ref{subsec:ngc6905_kn26}). The former showed a significant peak at $\sim$$700$~days and the latter at $~$$230$~days. Secondary significant peaks were also detected at $\sim$$245$~days and $\sim$$630$~days. The modulation appears broadly sinusoidal. This makes both the objects candidates for long-period binary CSPNe. Kn~26 displays an additional periodicity at 1.18~hr, which we verify through follow-up photometry. The origin of this period is uncertain. It can be due to a close binary (unlikely) or rotational origin.

We present CTSS~2 and K~3-5, which show slow and long-term brightening (Section \ref{subsec:ctss2_pnk35}). The objects brightened by almost a factor of $2$ in the ZTF $r$-band and WISE bands, but much less so (factor of $\lesssim$$1.25$) in the $g$-band, thus undergoing significant reddening. Spectroscopy of both the nucleus and the nebula shows that both objects host high-density ($n_e$$\gtrsim$$10^6~{\rm cm^{-3}}$) emission-line cores. Thus they are either EGB~6-type PNe with compact emission knots, or symbiotic systems. The optical spectrum of CTSS~2 displays stellar absorption signatures. A spectral fit yields a temperature of $70$~kK and a stellar mass of $0.65~M_{\odot}$. This translates to a post-AGB age of $\approx$$1500$~years, making it one of the youngest CSPNe known. Unfortunately, the spectrum of K~3-5 does not show adequate stellar features, thus preventing any spectroscopic analysis of the (pre-)white dwarf. We think that these might be late thermal pulse events. In this case, the light curves suggest that the stars are cooling down and expanding, moving back to another AGB phase. This is a very rare occurrence and would indicate the success of long-baseline surveys like ZTF to discover these objects. Other possible explanations include wind\slash dust activity of the CSPN system. 

We present new results on LoTr~1 (Section \ref{subsec:lotr1}), which is a known variable showing a long-period modulation at $P=11.28$~years and a shorter period at $6.5$~days. The latter is attributed to the rotational modulation of a fast-rotating cool giant companion. The longer period might be the orbital period. Ultraviolet spectroscopy with HST shows that the hot component in the system is a spectroscopic twin of the extremely hot white dwarf in UCAC2~46706450, with a temperature of $105$~kK. A fit to the SED reveals the stellar parameters of the giant to be $T_{\rm eff}=4590$~K and radius of $R=9.7~R_{\odot}$.

Finally, we briefly discuss the observed variability in the remaining five objects (Section \ref{subsec:long_vars}). They display a variety of photometric behaviors: long-timescale dimming, single brightening event, single dip event, or more erratic variability. Follow-up studies are necessary to understand these systems.

Overall, this study shows the effectiveness of long-baseline surveys in uncovering the variety of photometric behaviors observed in CSPNe. Efforts need to be made not only to understand the objects individually through follow-up observations, but, if possible, to establish evolutionary links to different classes of objects. For example, the peculiar photometric behavior of NGC~6833 may shed significant light on the formation and evolution of post-CE wide binary systems. Note here that ZTF can only access the northern sky. The majority of PNe are in the southern sky, near the Galactic center. Upcoming surveys like LSST will observe these targets, and we expect to find several more such intriguing systems. Simple variability metrics and metric spaces (as demonstrated in Appendix \ref{sec:vonn_skewp}) will significantly boost such discoveries. 

\section{Acknowledgements}
    This work is based on observations obtained with the Samuel Oschin Telescope 48-inch and the 60-inch Telescope at the Palomar Observatory as part of the Zwicky Transient Facility project. ZTF is supported by the National Science Foundation under Grants No. AST-1440341 and AST-2034437 and a collaboration including current partners Caltech, IPAC, the Oskar Klein Center at Stockholm University, the University of Maryland, University of California, Berkeley, the University of Wisconsin at Milwaukee, University of Warwick, Ruhr University Bochum, Cornell University, Northwestern University, and Drexel University. Operations are conducted by COO, IPAC, and UW.

    This work has made use of data from the European Space Agency (ESA) mission \emph{Gaia} (\url{https://www.cosmos.esa.int/gaia}), processed by the \emph{Gaia} Data Processing and Analysis Consortium (DPAC; \url{https://www.cosmos.esa.int/web/gaia/dpac/consortium}). Funding for the DPAC has been provided by national institutions, in particular, the institutions participating in the \emph{Gaia} Multilateral Agreement.

    We are grateful to the staffs of Palomar Observatory and the Hobby-Eberly Telescope for assistance with the observations and data management. The Liverpool Telescope is operated on the island of La Palma by Liverpool John Moores University in the Spanish Observatorio del Roque de los Muchachos of the Instituto de Astrofisica de Canarias with financial support from the UK Science and Technology Facilities Council.

    The Low-Resolution Spectrograph 2 (LRS2) on HET was developed and funded by the University of Texas at Austin McDonald Observatory and Department of Astronomy, and by Pennsylvania State University. We thank the Leibniz-Institut f\"ur Astrophysik Potsdam (AIP) and the Institut f\"ur Astrophysik G\"ottingen (IAG) for their contributions to the construction of the integral field units.
We acknowledge the Texas Advanced Computing Center (TACC) at The University of Texas at Austin for providing high performance computing, visualization, and storage resources that have contributed to the results reported within this paper.

The Isaac Newton Telescope is operated on the island of La Palma by the Isaac Newton Group of Telescopes in the Spanish Observatorio del Roque de los Muchachos of the Instituto de Astrof\'isica de Canarias

    S.B. thanks Frank J. Masci and Zachary P. Vanderbosch for useful discussions and suggestions regarding solving the issues with ZTF forced photometry on extended sources. S.B. also thanks Jim Fuller, Charles C. Steidel, Lynne Hillenbrand, and Adolfo Carvalho for useful discussions on methods and science. S.B. acknowledges  financial support from the Wallace L. W. Sargent Graduate Fellowship during the first year of his graduate studies at Caltech. N.C. was supported through the Cancer Research UK grant A24042.  

    N.R. is supported by the Deutsche Forschungsgemeinschaft (DFG) through grant RE3915/2-1.

    DJ acknowledges support from the Agencia Estatal de Investigaci\'on del Ministerio de Ciencia, Innovaci\'on y Universidades (MICIU/AEI) under grant ``Nebulosas planetarias como clave para comprender la evoluci\'on de estrellas binarias'' and the European Regional Development Fund (ERDF) with reference PID-2022-136653NA-I00 (DOI:10.13039/501100011033). DJ also acknowledges support from the Agencia Estatal de Investigaci\'on del Ministerio de Ciencia, Innovaci\'on y Universidades (MICIU/AEI) under grant ``Revolucionando el conocimiento de la evoluci\'on de estrellas poco masivas'' and the the European Union NextGenerationEU/PRTR with reference CNS2023-143910 (DOI:10.13039/501100011033).

    We have used \texttt{Python} packages Numpy \citep{harris2020array}, SciPy \citep{2020SciPy-NMeth}, Matplotlib \citep{Hunter:2007}, Pandas \citep{reback2020pandas}, Astropy \citep{Astropy13, Astropy18}, and Astroquery \citep{astroquery19} at various stages of this research.

\appendix

\section{The Von-Neumann and Pearson Skew metric space}\label{sec:vonn_skewp}

\begin{figure}[t]
\figurenum{A1}
    \centering
    \includegraphics[width=\linewidth]{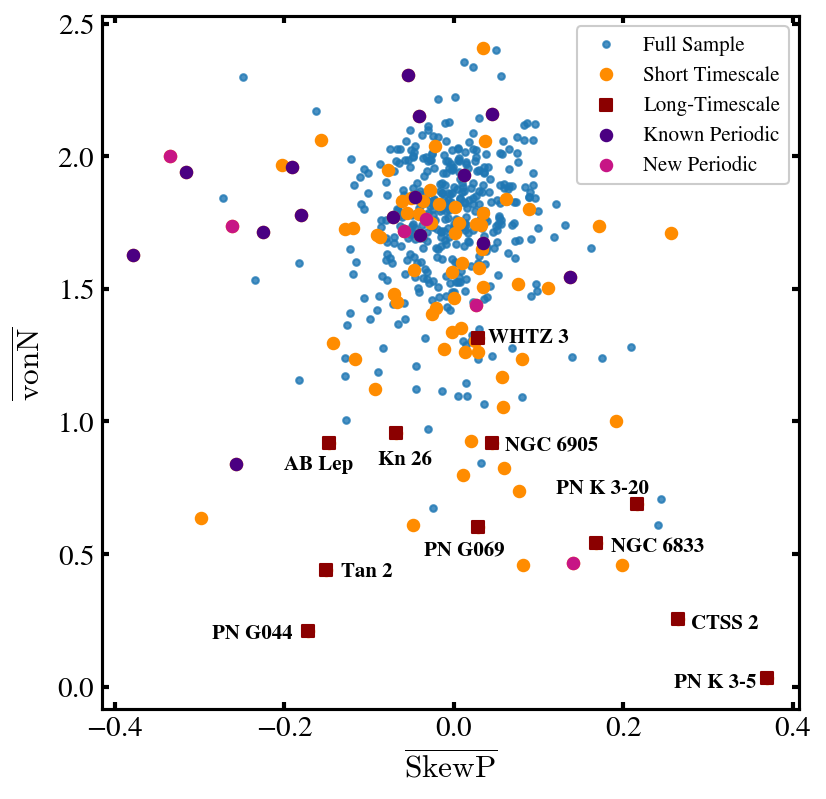}
    \caption{The position of the ZTF-variable CSPNe in the two-dimensional variability metric space defined by the Von-Neumann statistics (vonN) and Pearson-Skew (SkewP) metrics (see equations B1 and B2 in \citetalias{Bhattacharjee24}). This is the same as Figure 8 in \citetalias{Bhattacharjee24}, with the long-timescale variables labeled individually. All of them are characterized by a low value of Vonn ($<1$), making them easily identifiable. The value of SkewP provides further information on the nature of the variability (brightening/outburst vs dimming/transits).}
    \label{fig:vonn_skew}
\end{figure}

In Appendix B of \citetalias{Bhattacharjee24}, we presented a two-dimensional metric space defined by the variability metrics von Neumann statistics and Pearson Skew (vonN and SkewP, equations B1 and B2 in \citetalias{Bhattacharjee24} respectively). We demonstrated the ability of the metric space to identify `uniquely' variable light curves, like WeSb~1, from a larger pool of objects. We briefly review the metric space and its functionality. The Vonn metric quantifies the randomness in the time-series data. Theoretically, an adequately sampled completely random data should have ${\rm vonN\sim2}$ \citep{VonNeumann41}. SkewP, on the other hand, quantifies the directionality of the variability ($<0$ signifies dips/eclipses, and $>0$ indicates outbursts). Non-variable sources with just photon shot noise, thus, cluster around ${\rm vonN\sim2}$ (slightly lower at $\sim1.8$ in practice) and ${\rm SkewP\sim0}$.

In \citetalias{Bhattacharjee24}, we discussed the position of the short-timescale variables in this metric space. We now focus on the long-timescale variables. Figure \ref{fig:vonn_skew} is the same as Figure B1 in \citetalias{Bhattacharjee24} with all the 11 long-timescale variables marked individually. The notable feature is that almost all the sources (except WHTZ~3, where the light curve is noisier) have low values of the Vonn metric ($<1$). This is because the variability in these sources is well resolved in ZTF cadence and, thus, non-random. Based on the light curves presented in the main paper, the values of SkewP are easy to interpret. Sources showing brightening events, like NGC~6833, CTSS~2, K~3-5, and K~3-20, have ${\rm SkewP>0}$. Whereas objects showing dips like AB~Lep or Tan~2 have ${\rm SkewP<0}$. The variability in NGC~6905 and Kn~26 is symmetric, thus they don't have significant positive or negative SkewP values. The SkewP values of the more complex variables like PN~G044.3+04.1 can be difficult to interpret. Nevertheless, most of the long-timescale variables stand out from the other objects in the space and are easily identifiable. This thus potentially serves as a great tool to filter interesting sources from large data sets of light curves.

\section{Optical spectrum of NGC~6833}\label{app:ngc6833_opt_spec}

\begin{figure*}[t]
\figurenum{B1}
    \centering
    \includegraphics[width=\linewidth]{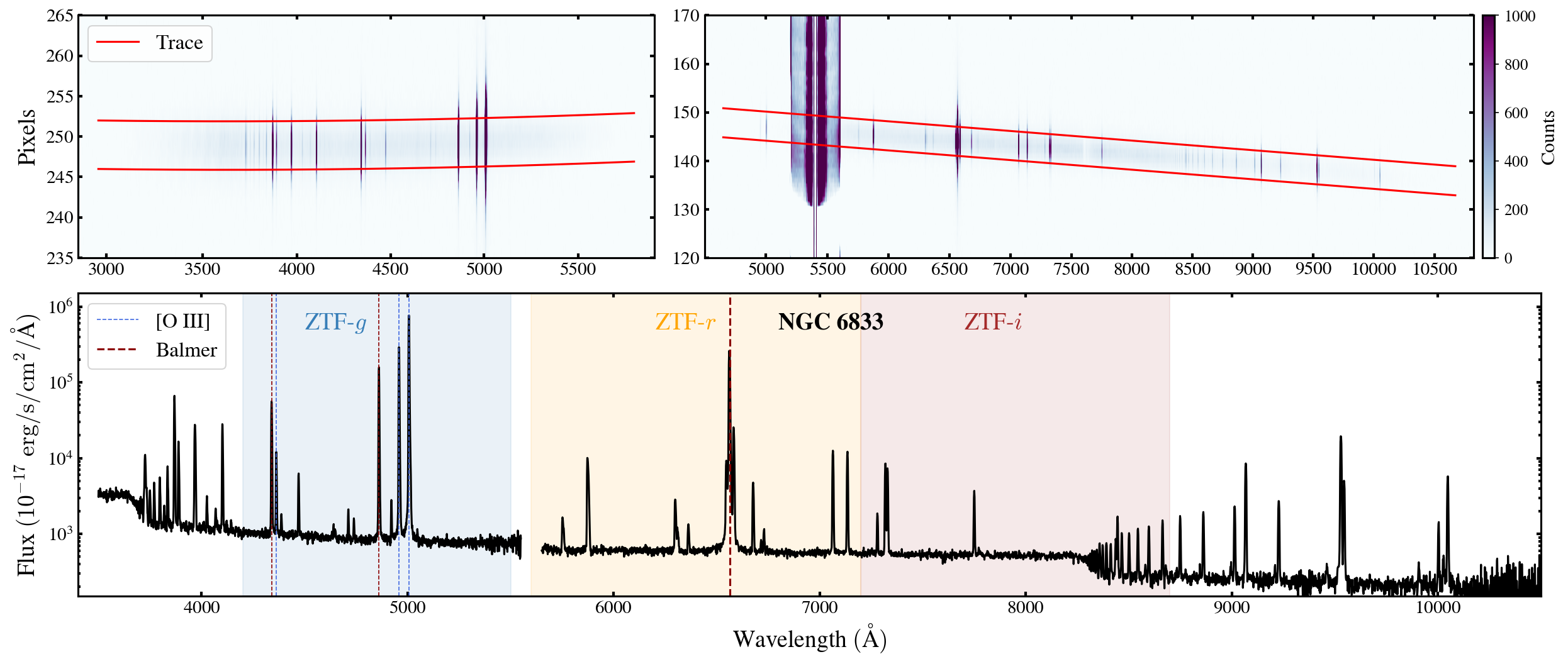}
    \caption{The DBSP spectrum for NGC~6833. The upper left and right panels show the two-dimensional spectra for the blue and the red arms respectively. The bottom panel shows the one-dimensional spectrum derived from the marked trace. We mark the first three Balmer and [\ion{O}{3}] emission lines in the spectrum. We also shade the approximate ZTF band passes.}
    \label{fig:ngc6833_trace_spectrum}
\end{figure*}

NGC~6833 was observed with the DouBle SPectrograph (DBSP; \citealt{Oke82}) attached to the Cassegrain focus of the Palomar 200-in Hale Telescope. We used the D55 dichroic, the 600 line~mm$^{-1}$ grating for the blue arm blazed at 3780~\AA, and the 316 line~mm$^{-1}$ grating for the red arm blazed at 7150~\AA. We used grating angles of 27$^{\circ}$17$'$ and 24$^{\circ}$38$'$ for the blue and red sides, respectively. With these setups and a slit width of 1.5$''$ long-slit, we achieve resolving powers of $R=1600$ in the blue arm and $R=1400$ in the red arm. These setups provide continuous spectral coverage across 2900--10800~\AA, with the division between blue and red arms typically occurring around 5650~\AA. We followed the same reduction procedure as described in Section 4.2.1 in \citetalias{Bhattacharjee24}. Briefly, we used PypeIt \citep{PypeIt_1, PypeIt_2} with the options \texttt{use\_2dmodel\_mask = False} and \texttt{no\_local\_sky = True} to avoid masking and local subtraction of the strong emission lines. Following this, we perform manual trace identification on the two-dimensional spectrum. 

Figure \ref{fig:ngc6833_trace_spectrum} shows the optical spectrum for NGC~6833. Given the very compact nature of the nebula, the spectrum is completely dominated by the nebular continuum and the emission lines. No stellar signature can be observed in the current spectrum. 

\section{Optical spectrum of Kn~26}\label{app:kn26_spectrum}

\begin{figure*}[t]
\figurenum{C1}
    \centering
    \includegraphics[width=\linewidth]{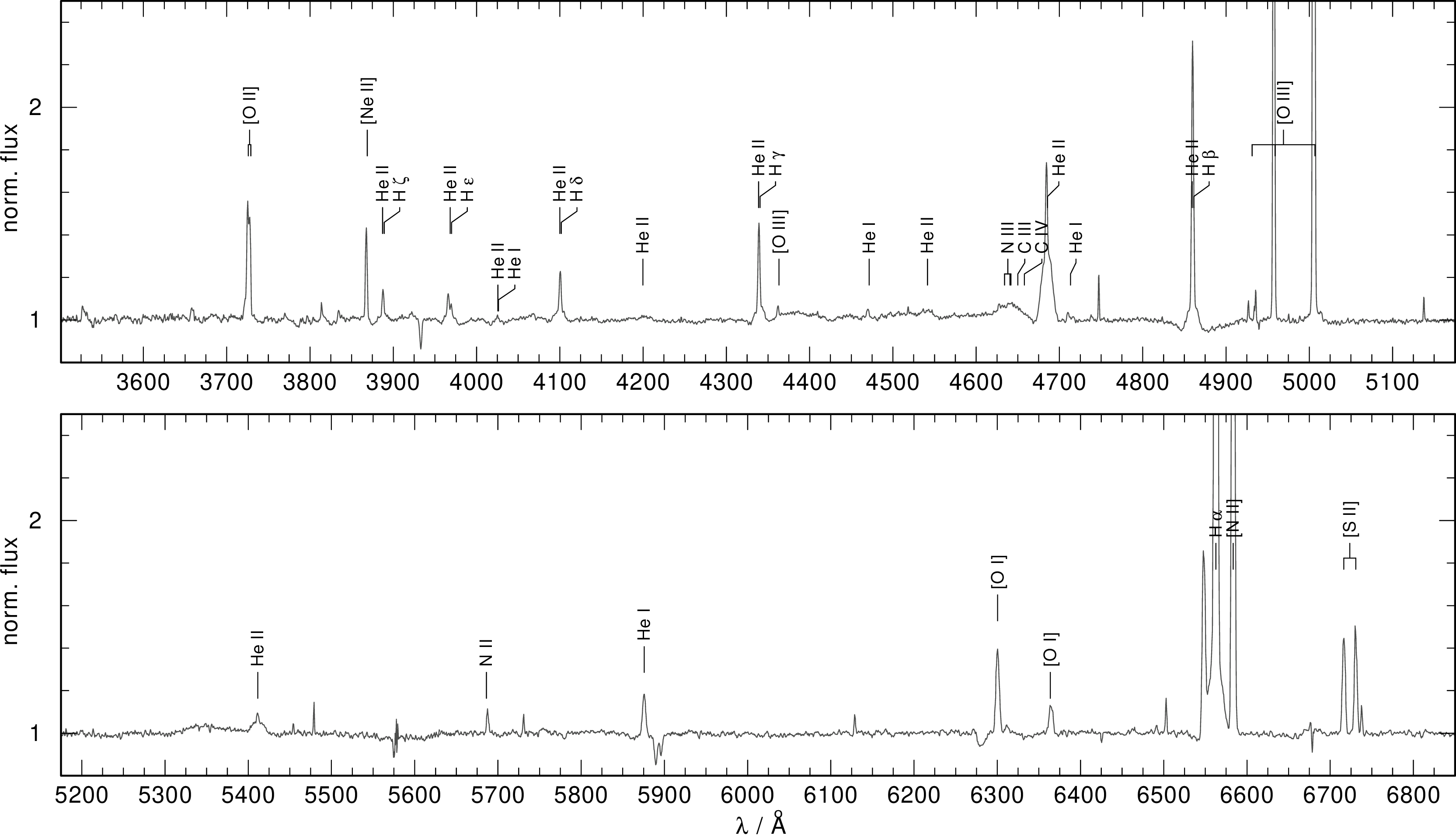}
    \caption{LBT/MODS spectrum of the WELS type central star of Kn\,26.}
    \label{fig:mods}
\end{figure*}

We took four 900\,s exposures of Kn~26 at the twin 8.4m Large Binocular Telescope (LBT) using the Multi-Object Double Spectrographs (MODS, \citealt{Pogge+2010}) on December, 2024. MODS provides two-channel grating spectroscopy by using a dichroic that splits the light at $\approx5650$\,\AA\ into separately optimized red and blue channels. The spectra cover the wavelength region $3330 - 5800$\,\AA\ and $5500 - 10\,000$\,\AA\ with a resolving power of $R\approx 1850$ and $2300$, respectively. The spectra were reduced using the 
modsccdred\footnote{\url{https://github.com/rwpogge/modsCCDRed}} \textsc{PYTHON} package \citep{Pogge2019} for basic 2d CCD reductions, and the modsidl\footnote{\url{https://github.com/rwpogge/modsIDL}} pipeline \citep{Croxall+2019} to extract 1d spectra and apply wavelength and flux calibrations.

The coadded MODS spectrum is shown in Fig.~\ref{fig:mods}. We note that the Balmer and \Ion{He}{2} lines appear to be a super composition of a narrow emission line (likely resulting from the PN) and weaker but broader emission line (possibly originating from the photosphere). We identify the \Ion{N}{3} \ \Ion{C}{3} \ \Ion{C}{4} emission line complex around 4650\,\AA\, but note that the the \ion{C}{4}~5801,~5812~\AA\ doublet is not visible. Therefore, Kn~26 can be considered as a weak emission-line type ([WELS]) central star \citep{Marcolino+2003}. We note here that the emission lines can also arise from the irradiated companion (as often the case with WELS type PN nuclei, see for example \citealt{Miszalski11}), and can be an indirect confirmation of the binarity of the system.

\section{NGC~6905 and Kn~26 Periodicity Tests}\label{appendix:ngc6905kn26periods}

\begin{figure*}[t]
\figurenum{D1}
    \centering
    \includegraphics[width=\linewidth]{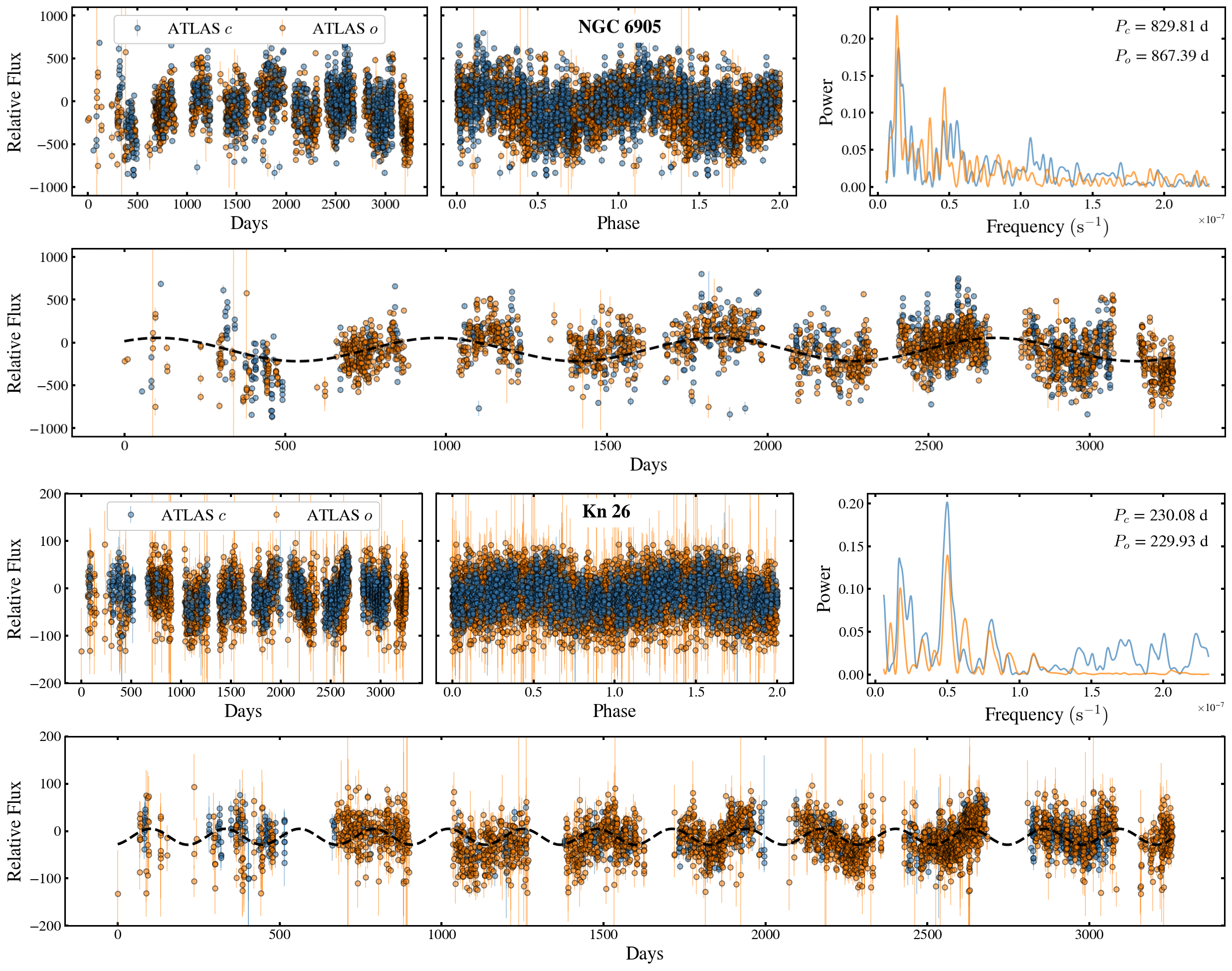}
    \caption{Same as Figure \ref{fig:ngc6905_period}, but with ATLAS light curve, for both NGC~6905 (upper section) and Kn~26 (lower section). Similar periods (same for Kn~26) as in ZTF are recovered, ruling out any ZTF-specific systematics.}
    \label{fig:ngc6905_kn26_period_atlas_merged}
\end{figure*}

We briefly describe here the various tests performed to confirm the reality of the inferred periods of NGC~6905 and Kn~26. Though not shown, the periods were recovered from standard photometry data too, ruling out any difference image artifact. To check if the observed variability is due to seasonal variation in atmospheric conditions, we compared the atmospheric seeing to the observed flux. No correlation was found. To test if the periods arise from the seasonal breaks or yearly cycles in the survey, we randomized the flux values keeping the time stamps constant, and manually investigated the periodogram in several iterations. No significant peak was obtained and the periodograms were as good as noise. This is also reflected as a negligible false alarm probability (FAP) calculated using the inbuilt bootstrap method in \texttt{astropy.timeseries}. 

We then query ZTF standard photometry light curves of all \emph{Gaia} sources within a couple of arcminutes from the targets (${\rm\sim200}$ objects for each source). Visually, none of the neighboring objects showed the variability seen in the target sources. These include sources behind the extended nebulae of the PNe, thus ruling out any spurious nebular activity resulting in the observed modulation. We applied the same period search procedure on all neighboring objects' light curves and manually inspected the periodograms. For the most significant period in each of the resultant periodograms, we investigate the distribution of the FAPs. None of the objects showed the periods inferred in our targets with any reasonable FAP or periodogram signal to noise. The only two spurious periods with low FAP in some of the objects are one year and $\sim2000$~days (the ZTF baseline), but nowhere close to the inferred periods in the two targets. 

For a final test, we queried the difference-image forced photometry data from the Asteroid Terrestrial-impact Last Alarm System (ATLAS) survey\footnote{We queried both the difference and science-image forced photometry data from the online service at \url{https://fallingstar-data.com/forcedphot/} \citep{Shingles21}. Owing to their warnings about the latter, we primarily use the former but use the science-image data for estimation of the reference flux and normalization. We reject all data where \texttt{err}$\neq0$.} for both the targets and perform period search. The results are presented in Figure \ref{fig:ngc6905_kn26_period_atlas_merged}. Qualitatively, similar variability as in ZTF is recovered in ATLAS. This rules out with confidence both any ZTF-specific systematics and region-specific atmospheric seeing variation inducing the variability. Significant peak in the periodogram appears for NGC~6905, albeit at a slightly higher period of $\sim850$~days. The disagreement may result from the ATLAS data being significantly noisier than ZTF. A secondary peak at a similar period as in ZTF is recovered. For Kn~26, we successfully recovered the dominant ZTF period of $\sim230$~days. The secondary peak in ATLAS has a lower signal-to-noise, but the corresponding period is broadly in agreement with that in ZTF. We briefly note here that ATLAS data is too noisy to detect the short period in Kn~26.

\section{DBSP Spectra of CTSS~2 and K~3-5}\label{appendix:ctss2_k35_dbsp}

\begin{figure*}[t]
\figurenum{E1}
    \centering
    \includegraphics[width=\linewidth]{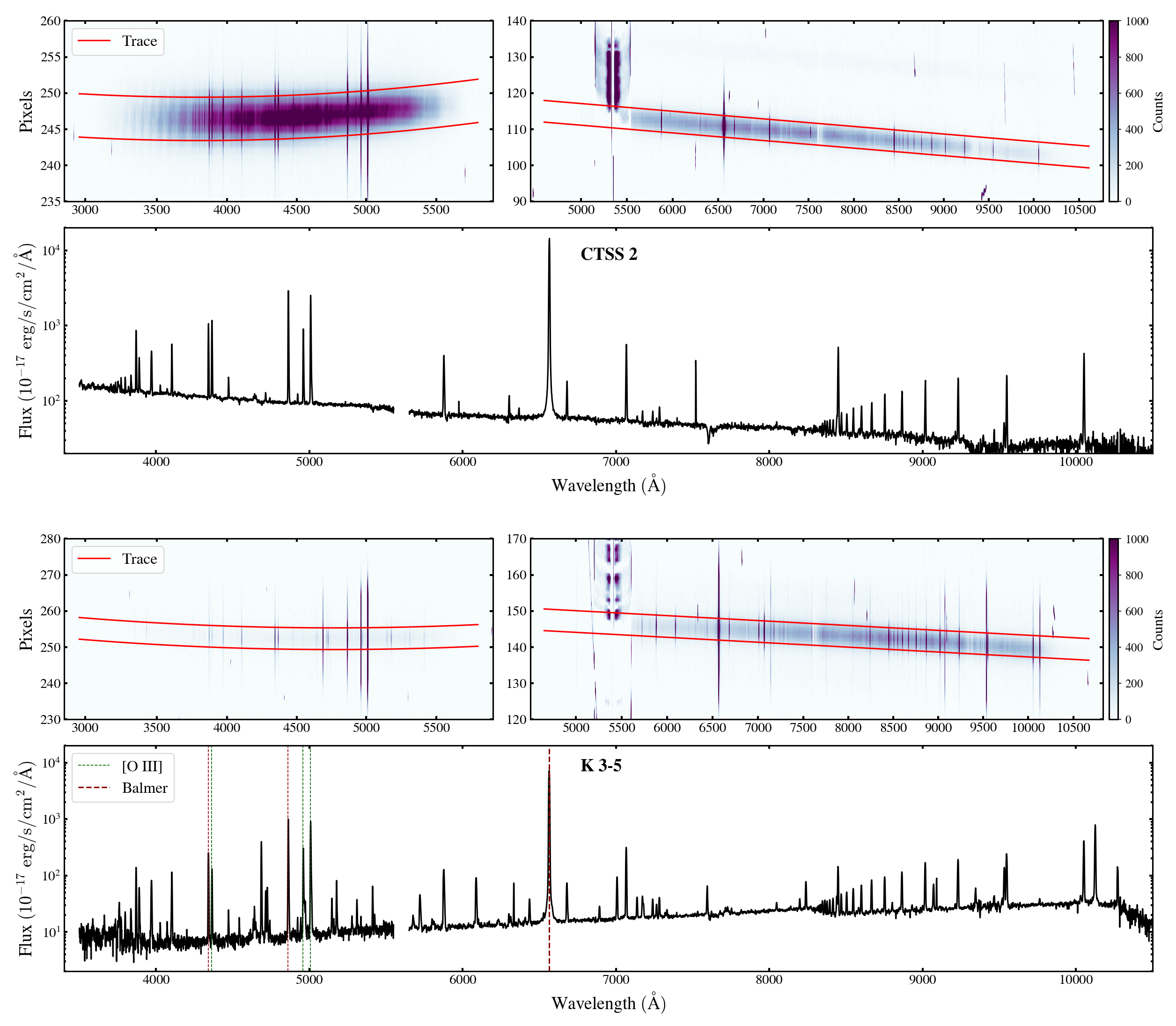}
    \caption{DBSP spectra of CTSS~2 (upper section) and K~3-5 (lower section) in the same format as Figure \ref{fig:ngc6833_trace_spectrum}.}
    \label{fig:ctss2_k35_dbsp}
\end{figure*}

Figure \ref{fig:ctss2_k35_dbsp} shows the DBSP spectrum for CTSS~2 and K~3-5. The same reduction was procedure as with WeSb~1 in \citetalias{Bhattacharjee24} and NGC~6833 was followed. With a slit spectrograph, it is difficult to separately identify the nucleus and the nebula. Thus, we choose a fiducial trace containing the main emission core and extract the corresponding one-dimensional spectra. The only advantage over the previously presented HET spectrum is its better spectral coverage in the red. This enables us to check for obvious indication of a giant star, which may indicate a symbiotic system. Such signatures are not present in the spectrum.

\section{Infrared Excess in CTSS~2 and K~3-5}\label{subsubsec:ctss2_k35_ir_excess}

\begin{figure}[t]
\figurenum{F1}
    \centering
    \includegraphics[width=\linewidth]{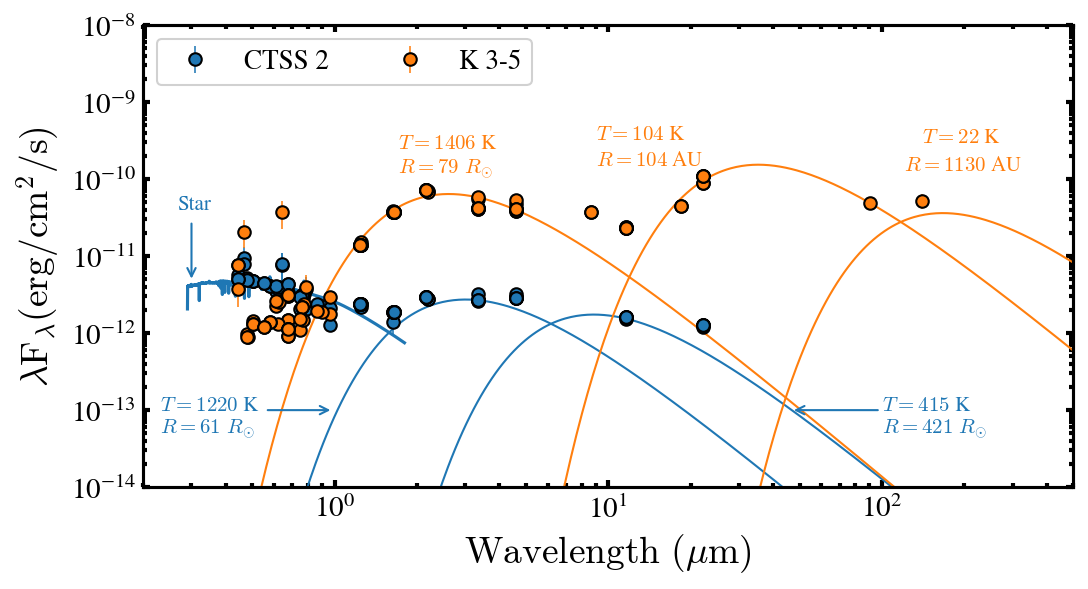}
    \caption{Spectral energy distribution (SED) of CTSS~2 (blue) and K~3-5 (orange). Significant infrared excess is evident. The blackbody fits the long-wavelength ($>$$1~{\rm \mu m}$) data are shown along with the inferred parameters. We interpret them as warm and cold dust structures. The Vizier query also yields fluxes at radio wavelengths (1.4~GHz and 5~GHz). They likely do not relate to our analysis (as the radio emissions are often of non-thermal origin) and thus we have disregarded them.}
    \label{fig:ctss2_k35_sed_ir_excess}
\end{figure}

Long term dust production is a possible explanation to the reddened brightening of CTSS~2 and K~3-5. Thus, to look for the presence of dust in these systems, we investigate their spectral energy distributions (SED). The result is presented in Figure \ref{fig:ctss2_k35_sed_ir_excess}. Some scatter is evident in the data, especially for K~3-5, which is possibly due to the extended nature of the objects leading to suboptimal photometry. Nevertheless, the presence of significant infrared excess in both the objects is evident.

For a more quantitative estimate, we perform multi-component blackbody fit to the data points with wavelengths $>$$1~\mu m$. For CTSS~2, a reasonable fit was obtained with a two-component fit. The stellar component was subtracted and distance of $8$~kpc was assumed. For K~3-5, we assumed the Gaia eDR3 parallax distance of $2481$~pc, and a reasonable fit was only observed with three components. The fit parameters of the different components are presented are annotations in the same Figure \ref{fig:ctss2_k35_sed_ir_excess}. 

We interpret the various blackbody components as representative of extended dust structures. We find that both CTSS~2 and K~3-5 appear to have warm dust with temperature in the range of $T$$=$$1000-1500$~K and radius $R$$=$$50-100~R_{\odot}$. We also find evidence of extended colder dust features. For K~3-5 it appears to be extended to around a thousand AU, with $T$$\lesssim$$100$~K. For CTSS~2, the second component appears hotter and more localized to around $400~R_{\odot}$. But this might be due to inadequate coverage in the redder wavelengths, thus missing out on the more extended cooler dust components.

\bibliography{sample631}{}

\begin{thebibliography}{}
\expandafter\ifx\csname natexlab\endcsname\relax\def\natexlab#1{#1}\fi
\providecommand{\url}[1]{\href{#1}{#1}}
\providecommand{\dodoi}[1]{doi:~\href{http://doi.org/#1}{\nolinkurl{#1}}}
\providecommand{\doeprint}[1]{\href{http://ascl.net/#1}{\nolinkurl{http://ascl.net/#1}}}
\providecommand{\doarXiv}[1]{\href{https://arxiv.org/abs/#1}{\nolinkurl{https://arxiv.org/abs/#1}}}

\bibitem[{{Aller} {et~al.}(2024){Aller}, {Lillo-Box}, \& {Jones}}]{Aller24}
{Aller}, A., {Lillo-Box}, J., \& {Jones}, D. 2024, arXiv e-prints, arXiv:2409.06332, \dodoi{10.48550/arXiv.2409.06332}

\bibitem[{{Aller} {et~al.}(2020){Aller}, {Lillo-Box}, {Jones}, {Miranda}, \& {Barcel{\'o} Forteza}}]{Aller20}
{Aller}, A., {Lillo-Box}, J., {Jones}, D., {Miranda}, L.~F., \& {Barcel{\'o} Forteza}, S. 2020, \aap, 635, A128, \dodoi{10.1051/0004-6361/201937118}

\bibitem[{{Aller} \& {Liller}(1966)}]{Aller66}
{Aller}, L.~H., \& {Liller}, W. 1966, \mnras, 132, 337, \dodoi{10.1093/mnras/132.2.337}

\bibitem[{{Arkhipova} {et~al.}(2012){Arkhipova}, {Burlak}, {Esipov}, {Ikonnikova}, \& {Komissarova}}]{Arkhipova12}
{Arkhipova}, V.~P., {Burlak}, M.~A., {Esipov}, V.~F., {Ikonnikova}, N.~P., \& {Komissarova}, G.~V. 2012, Astronomy Letters, 38, 157, \dodoi{10.1134/S1063773712020016}

\bibitem[{{Arkhipova} {et~al.}(2013){Arkhipova}, {Burlak}, {Esipov}, {Ikonnikova}, \& {Komissarova}}]{Arkhipova13}
---. 2013, Astronomy Letters, 39, 619, \dodoi{10.1134/S1063773713090016}

\bibitem[{{Arkhipova} {et~al.}(2020){Arkhipova}, {Burlak}, {Ikonnikova}, {Komissarova}, {Esipov}, \& {Shenavrin}}]{Arkhipova20}
{Arkhipova}, V.~P., {Burlak}, M.~A., {Ikonnikova}, N.~P., {et~al.} 2020, Astronomy Letters, 46, 100, \dodoi{10.1134/S1063773720020012}

\bibitem[{{Asplund} {et~al.}(1997){Asplund}, {Gustafsson}, {Lambert}, \& {Kameswara Rao}}]{Asplund97}
{Asplund}, M., {Gustafsson}, B., {Lambert}, D.~L., \& {Kameswara Rao}, N. 1997, \aap, 321, L17, \dodoi{10.48550/arXiv.astro-ph/9704005}

\bibitem[{{Astropy Collaboration} {et~al.}(2013){Astropy Collaboration}, {Robitaille}, {Tollerud}, {Greenfield}, {Droettboom}, {Bray}, {Aldcroft}, {Davis}, {Ginsburg}, {Price-Whelan}, {Kerzendorf}, {Conley}, {Crighton}, {Barbary}, {Muna}, {Ferguson}, {Grollier}, {Parikh}, {Nair}, {Unther}, {Deil}, {Woillez}, {Conseil}, {Kramer}, {Turner}, {Singer}, {Fox}, {Weaver}, {Zabalza}, {Edwards}, {Azalee Bostroem}, {Burke}, {Casey}, {Crawford}, {Dencheva}, {Ely}, {Jenness}, {Labrie}, {Lim}, {Pierfederici}, {Pontzen}, {Ptak}, {Refsdal}, {Servillat}, \& {Streicher}}]{Astropy13}
{Astropy Collaboration}, {Robitaille}, T.~P., {Tollerud}, E.~J., {et~al.} 2013, \aap, 558, A33, \dodoi{10.1051/0004-6361/201322068}

\bibitem[{{Astropy Collaboration} {et~al.}(2018){Astropy Collaboration}, {Price-Whelan}, {Sip{\H{o}}cz}, {G{\"u}nther}, {Lim}, {Crawford}, {Conseil}, {Shupe}, {Craig}, {Dencheva}, {Ginsburg}, {VanderPlas}, {Bradley}, {P{\'e}rez-Su{\'a}rez}, {de Val-Borro}, {Aldcroft}, {Cruz}, {Robitaille}, {Tollerud}, {Ardelean}, {Babej}, {Bach}, {Bachetti}, {Bakanov}, {Bamford}, {Barentsen}, {Barmby}, {Baumbach}, {Berry}, {Biscani}, {Boquien}, {Bostroem}, {Bouma}, {Brammer}, {Bray}, {Breytenbach}, {Buddelmeijer}, {Burke}, {Calderone}, {Cano Rodr{\'\i}guez}, {Cara}, {Cardoso}, {Cheedella}, {Copin}, {Corrales}, {Crichton}, {D'Avella}, {Deil}, {Depagne}, {Dietrich}, {Donath}, {Droettboom}, {Earl}, {Erben}, {Fabbro}, {Ferreira}, {Finethy}, {Fox}, {Garrison}, {Gibbons}, {Goldstein}, {Gommers}, {Greco}, {Greenfield}, {Groener}, {Grollier}, {Hagen}, {Hirst}, {Homeier}, {Horton}, {Hosseinzadeh}, {Hu}, {Hunkeler}, {Ivezi{\'c}}, {Jain}, {Jenness}, {Kanarek}, {Kendrew}, {Kern}, {Kerzendorf}, {Khvalko}, {King}, {Kirkby}, {Kulkarni},
  {Kumar}, {Lee}, {Lenz}, {Littlefair}, {Ma}, {Macleod}, {Mastropietro}, {McCully}, {Montagnac}, {Morris}, {Mueller}, {Mumford}, {Muna}, {Murphy}, {Nelson}, {Nguyen}, {Ninan}, {N{\"o}the}, {Ogaz}, {Oh}, {Parejko}, {Parley}, {Pascual}, {Patil}, {Patil}, {Plunkett}, {Prochaska}, {Rastogi}, {Reddy Janga}, {Sabater}, {Sakurikar}, {Seifert}, {Sherbert}, {Sherwood-Taylor}, {Shih}, {Sick}, {Silbiger}, {Singanamalla}, {Singer}, {Sladen}, {Sooley}, {Sornarajah}, {Streicher}, {Teuben}, {Thomas}, {Tremblay}, {Turner}, {Terr{\'o}n}, {van Kerkwijk}, {de la Vega}, {Watkins}, {Weaver}, {Whitmore}, {Woillez}, {Zabalza}, \& {Astropy Contributors}}]{Astropy18}
{Astropy Collaboration}, {Price-Whelan}, A.~M., {Sip{\H{o}}cz}, B.~M., {et~al.} 2018, \aj, 156, 123, \dodoi{10.3847/1538-3881/aabc4f}

\bibitem[{{Bailer-Jones} {et~al.}(2021){Bailer-Jones}, {Rybizki}, {Fouesneau}, {Demleitner}, \& {Andrae}}]{Bailer-Jones}
{Bailer-Jones}, C.~A.~L., {Rybizki}, J., {Fouesneau}, M., {Demleitner}, M., \& {Andrae}, R. 2021, \aj, 161, 147, \dodoi{10.3847/1538-3881/abd806}

\bibitem[{{Balick} \& {Frank}(2002)}]{Balick02}
{Balick}, B., \& {Frank}, A. 2002, \araa, 40, 439, \dodoi{10.1146/annurev.astro.40.060401.093849}

\bibitem[{{Bhattacharjee} {et~al.}(2025){Bhattacharjee}, {Kulkarni}, {Kong}, {Tam}, {Bond}, {El-Badry}, {Caiazzo}, {Chornay}, {Graham}, {Rodriguez}, {Zeimann}, {Fremling}, {Drake}, {Werner}, {Rodriguez}, {Prince}, {Laher}, {Chen}, \& {Riddle}}]{Bhattacharjee24}
{Bhattacharjee}, S., {Kulkarni}, S.~R., {Kong}, A. K.~H., {et~al.} 2025, \pasp, 137, 024201, \dodoi{10.1088/1538-3873/ada702}

\bibitem[{{Boffin} \& {Jones}(2019)}]{Boffin19}
{Boffin}, H. M.~J., \& {Jones}, D. 2019, {The Importance of Binaries in the Formation and Evolution of Planetary Nebulae}, \dodoi{10.1007/978-3-030-25059-1}

\bibitem[{{Bollen} {et~al.}(2017){Bollen}, {Van Winckel}, \& {Kamath}}]{Bollen17}
{Bollen}, D., {Van Winckel}, H., \& {Kamath}, D. 2017, \aap, 607, A60, \dodoi{10.1051/0004-6361/201731493}

\bibitem[{{Bond} {et~al.}(2024){Bond}, {Chaturvedi}, {Ciardullo}, {Werner}, {Zeimann}, \& {Siegel}}]{Bond24egb6}
{Bond}, H.~E., {Chaturvedi}, A.~S., {Ciardullo}, R., {et~al.} 2024, \apj, 970, 164, \dodoi{10.3847/1538-4357/ad4f84}

\bibitem[{{Bond} \& {Livio}(1990)}]{Bond90Binary}
{Bond}, H.~E., \& {Livio}, M. 1990, \apj, 355, 568, \dodoi{10.1086/168789}

\bibitem[{{Bond} \& {Meakes}(1990)}]{Bond90Pulsator}
{Bond}, H.~E., \& {Meakes}, M.~G. 1990, \aj, 100, 788, \dodoi{10.1086/115560}

\bibitem[{{Bond} {et~al.}(2023){Bond}, {Werner}, {Jacoby}, \& {Zeimann}}]{Bond23}
{Bond}, H.~E., {Werner}, K., {Jacoby}, G.~H., \& {Zeimann}, G.~R. 2023, \mnras, 521, 668, \dodoi{10.1093/mnras/stad524}

\bibitem[{{Bond} \& {Zeimann}(2024)}]{Bond24}
{Bond}, H.~E., \& {Zeimann}, G.~R. 2024, \apj, 967, 122, \dodoi{10.3847/1538-4357/ad3df9}

\bibitem[{Budaj {et~al.}(2025)Budaj, Bernhard, Jones, \& Munday}]{Budaj25}
Budaj, J., Bernhard, K., Jones, D., \& Munday, J. 2025, Nature Astronomy, \dodoi{10.1038/s41550-024-02446-x}

\bibitem[{{Chen} {et~al.}(2025){Chen}, {Fang}, {Chen}, \& {Liu}}]{Chen25}
{Chen}, P., {Fang}, X., {Chen}, X., \& {Liu}, J. 2025, arXiv e-prints, arXiv:2501.06056.
\newblock \doarXiv{2501.06056}

\bibitem[{{Chen} {et~al.}(2017){Chen}, {Frank}, {Blackman}, {Nordhaus}, \& {Carroll-Nellenback}}]{Chen17}
{Chen}, Z., {Frank}, A., {Blackman}, E.~G., {Nordhaus}, J., \& {Carroll-Nellenback}, J. 2017, \mnras, 468, 4465, \dodoi{10.1093/mnras/stx680}

\bibitem[{{Choi} {et~al.}(2016){Choi}, {Dotter}, {Conroy}, {Cantiello}, {Paxton}, \& {Johnson}}]{Choi16}
{Choi}, J., {Dotter}, A., {Conroy}, C., {et~al.} 2016, \apj, 823, 102, \dodoi{10.3847/0004-637X/823/2/102}

\bibitem[{{Chonis} {et~al.}(2014){Chonis}, {Hill}, {Lee}, {Tuttle}, \& {Vattiat}}]{Chonis14}
{Chonis}, T.~S., {Hill}, G.~J., {Lee}, H., {Tuttle}, S.~E., \& {Vattiat}, B.~L. 2014, in Society of Photo-Optical Instrumentation Engineers (SPIE) Conference Series, Vol. 9147, Ground-based and Airborne Instrumentation for Astronomy V, ed. S.~K. {Ramsay}, I.~S. {McLean}, \& H.~{Takami}, 91470A, \dodoi{10.1117/12.2056005}

\bibitem[{{Chonis} {et~al.}(2016){Chonis}, {Hill}, {Lee}, {Tuttle}, {Vattiat}, {Drory}, {Indahl}, {Peterson}, \& {Ramsey}}]{Chonis16}
{Chonis}, T.~S., {Hill}, G.~J., {Lee}, H., {et~al.} 2016, in Society of Photo-Optical Instrumentation Engineers (SPIE) Conference Series, Vol. 9908, Ground-based and Airborne Instrumentation for Astronomy VI, ed. C.~J. {Evans}, L.~{Simard}, \& H.~{Takami}, 99084C, \dodoi{10.1117/12.2232209}

\bibitem[{{Chornay} \& {Walton}(2021)}]{Chornay21Distance}
{Chornay}, N., \& {Walton}, N.~A. 2021, \aap, 656, A110, \dodoi{10.1051/0004-6361/202142008}

\bibitem[{{Ciardullo} \& {Bond}(1996)}]{Ciardullo96}
{Ciardullo}, R., \& {Bond}, H.~E. 1996, \aj, 111, 2332, \dodoi{10.1086/117967}

\bibitem[{{Clayton} \& {De Marco}(1997)}]{Clayton97}
{Clayton}, G.~C., \& {De Marco}, O. 1997, \aj, 114, 2679, \dodoi{10.1086/118678}

\bibitem[{{C{\'o}rsico} {et~al.}(2021){C{\'o}rsico}, {Uzundag}, {Kepler}, {Althaus}, {Silvotti}, {Baran}, {Vu{\v{c}}kovi{\'c}}, {Werner}, {Bell}, \& {Higgins}}]{Corsico21}
{C{\'o}rsico}, A.~H., {Uzundag}, M., {Kepler}, S.~O., {et~al.} 2021, \aap, 645, A117, \dodoi{10.1051/0004-6361/202039202}

\bibitem[{{Croxall} \& {Pogge}(2019)}]{Croxall+2019}
{Croxall}, K.~V., \& {Pogge}, R.~W. 2019, {rwpogge/modsIDL: modsIDL Binocular Release}, v1.0, Zenodo,  Zenodo, \dodoi{10.5281/zenodo.2561424}

\bibitem[{{Culpan} {et~al.}(2022){Culpan}, {Geier}, {Reindl}, {Pelisoli}, {Gentile Fusillo}, \& {Vorontseva}}]{Culpan22}
{Culpan}, R., {Geier}, S., {Reindl}, N., {et~al.} 2022, \aap, 662, A40, \dodoi{10.1051/0004-6361/202243337}

\bibitem[{{De Marco}(2009)}]{demarco09}
{De Marco}, O. 2009, \pasp, 121, 316, \dodoi{10.1086/597765}

\bibitem[{{Decleir} {et~al.}(2022){Decleir}, {Gordon}, {Andrews}, {Clayton}, {Cushing}, {Misselt}, {Pendleton}, {Rayner}, {Vacca}, \& {Whittet}}]{Decleir22}
{Decleir}, M., {Gordon}, K.~D., {Andrews}, J.~E., {et~al.} 2022, \apj, 930, 15, \dodoi{10.3847/1538-4357/ac5dbe}

\bibitem[{{Dotter}(2016)}]{Dotter16}
{Dotter}, A. 2016, \apjs, 222, 8, \dodoi{10.3847/0067-0049/222/1/8}

\bibitem[{{El-Badry} {et~al.}(2019){El-Badry}, {Rix}, {Tian}, {Duch{\^e}ne}, \& {Moe}}]{Elbadry19}
{El-Badry}, K., {Rix}, H.-W., {Tian}, H., {Duch{\^e}ne}, G., \& {Moe}, M. 2019, \mnras, 489, 5822, \dodoi{10.1093/mnras/stz2480}

\bibitem[{{Faltov{\'a}} {et~al.}(2021){Faltov{\'a}}, {Kallov{\'a}}, {Pri{\v{s}}egen}, {Stan{\v{e}}k}, {Sup{\'\i}kov{\'a}}, {Xia}, {Bernhard}, {H{\"u}mmerich}, \& {Paunzen}}]{Faltova21}
{Faltov{\'a}}, N., {Kallov{\'a}}, K., {Pri{\v{s}}egen}, M., {et~al.} 2021, \aap, 656, A125, \dodoi{10.1051/0004-6361/202141534}

\bibitem[{{Feline} {et~al.}(2005){Feline}, {Dhillon}, {Marsh}, {Watson}, \& {Littlefair}}]{Feline05}
{Feline}, W.~J., {Dhillon}, V.~S., {Marsh}, T.~R., {Watson}, C.~A., \& {Littlefair}, S.~P. 2005, \mnras, 364, 1158, \dodoi{10.1111/j.1365-2966.2005.09668.x}

\bibitem[{{Fitzpatrick} {et~al.}(2019{\natexlab{a}}){Fitzpatrick}, {Massa}, {Gordon}, {Bohlin}, \& {Clayton}}]{Fritzpatrick19}
{Fitzpatrick}, E.~L., {Massa}, D., {Gordon}, K.~D., {Bohlin}, R., \& {Clayton}, G.~C. 2019{\natexlab{a}}, \apj, 886, 108, \dodoi{10.3847/1538-4357/ab4c3a}

\bibitem[{{Fitzpatrick} {et~al.}(2019{\natexlab{b}}){Fitzpatrick}, {Massa}, {Gordon}, {Bohlin}, \& {Clayton}}]{Fitzpatrick+2019}
---. 2019{\natexlab{b}}, \apj, 886, 108, \dodoi{10.3847/1538-4357/ab4c3a}

\bibitem[{{Froebrich} {et~al.}(2023){Froebrich}, {Hillenbrand}, {Herbert}, {De}, {Eisl{\"o}ffel}, {Campbell-White}, {Kahar}, {Hambsch}, {Urtly}, {Popowicz}, {Bernacki}, {Malcher}, {Lasota}, {Fiolka}, {Jozwik-Wabik}, {Dubois}, {Logie}, {Rau}, {Phillips}, {Fleming}, {Gonzalez Farf{\'a}n}, {Sold{\'a}n Alfaro}, {Nelson}, {Futcher}, {Rolfe}, {Campbell}, {Vale}, {Devine}, {Mo{\'z}dzierski}, {Miko{\l}ajczyk}, {Eggenstein}, {Rodriguez}, {Walton}, {Vanaverbeke}, {Merrikin}, {{\"O}{\u{g}}men}, {Perez}, {Aimar}, {Piehler}, {Dover}, {Patel}, {Miller}, {Finch}, {Hankins}, {Moore}, {Travouillon}, \& {Szczepanski}}]{Froebrich23}
{Froebrich}, D., {Hillenbrand}, L.~A., {Herbert}, C., {et~al.} 2023, \mnras, 520, 5413, \dodoi{10.1093/mnras/stad407}

\bibitem[{{Garc{\'\i}a-Hern{\'a}ndez} \& {G{\'o}rny}(2014)}]{HernandezGarcia14}
{Garc{\'\i}a-Hern{\'a}ndez}, D.~A., \& {G{\'o}rny}, S.~K. 2014, \aap, 567, A12, \dodoi{10.1051/0004-6361/201423620}

\bibitem[{{Gezer} {et~al.}(2015){Gezer}, {Van Winckel}, {Bozkurt}, {De Smedt}, {Kamath}, {Hillen}, \& {Manick}}]{Gezer15}
{Gezer}, I., {Van Winckel}, H., {Bozkurt}, Z., {et~al.} 2015, \mnras, 453, 133, \dodoi{10.1093/mnras/stv1627}

\bibitem[{{Ginsburg} {et~al.}(2019){Ginsburg}, {Sip{\H{o}}cz}, {Brasseur}, {Cowperthwaite}, {Craig}, {Deil}, {Guillochon}, {Guzman}, {Liedtke}, {Lian Lim}, {Lockhart}, {Mommert}, {Morris}, {Norman}, {Parikh}, {Persson}, {Robitaille}, {Segovia}, {Singer}, {Tollerud}, {de Val-Borro}, {Valtchanov}, {Woillez}, {Astroquery Collaboration}, \& {a subset of astropy Collaboration}}]{astroquery19}
{Ginsburg}, A., {Sip{\H{o}}cz}, B.~M., {Brasseur}, C.~E., {et~al.} 2019, \aj, 157, 98, \dodoi{10.3847/1538-3881/aafc33}

\bibitem[{{G{\'o}mez-Gonz{\'a}lez} {et~al.}(2022){G{\'o}mez-Gonz{\'a}lez}, {Rubio}, {Toal{\'a}}, {Guerrero}, {Sabin}, {Todt}, {G{\'o}mez-Llanos}, {Ramos-Larios}, \& {Mayya}}]{Gomez22}
{G{\'o}mez-Gonz{\'a}lez}, V.~M.~A., {Rubio}, G., {Toal{\'a}}, J.~A., {et~al.} 2022, \mnras, 509, 974, \dodoi{10.1093/mnras/stab3042}

\bibitem[{{G{\'o}mez-Mu{\~n}oz} {et~al.}(2023){G{\'o}mez-Mu{\~n}oz}, {Bianchi}, \& {Manchado}}]{Munoz23}
{G{\'o}mez-Mu{\~n}oz}, M.~A., {Bianchi}, L., \& {Manchado}, A. 2023, \apjs, 266, 34, \dodoi{10.3847/1538-4365/acca77}

\bibitem[{{Gonzalez} {et~al.}(1998){Gonzalez}, {Lambert}, {Wallerstein}, {Rao}, {Smith}, \& {McCarthy}}]{Gonzalez98}
{Gonzalez}, G., {Lambert}, D.~L., {Wallerstein}, G., {et~al.} 1998, \apjs, 114, 133, \dodoi{10.1086/313068}

\bibitem[{{Gordon}(2024{\natexlab{a}})}]{dust_extinction}
{Gordon}, K. 2024{\natexlab{a}}, {dust\_extinction: Interstellar Dust Extinction Models}, v1.5,  Zenodo, \dodoi{10.5281/zenodo.4658887}

\bibitem[{{Gordon}(2024{\natexlab{b}})}]{Gordon24_package_paper}
---. 2024{\natexlab{b}}, The Journal of Open Source Software, 9, 7023, \dodoi{10.21105/joss.07023}

\bibitem[{{Gordon} {et~al.}(2009){Gordon}, {Cartledge}, \& {Clayton}}]{Gordon09}
{Gordon}, K.~D., {Cartledge}, S., \& {Clayton}, G.~C. 2009, \apj, 705, 1320, \dodoi{10.1088/0004-637X/705/2/1320}

\bibitem[{{Gordon} {et~al.}(2023){Gordon}, {Clayton}, {Decleir}, {Fitzpatrick}, {Massa}, {Misselt}, \& {Tollerud}}]{Karl23}
{Gordon}, K.~D., {Clayton}, G.~C., {Decleir}, M., {et~al.} 2023, \apj, 950, 86, \dodoi{10.3847/1538-4357/accb59}

\bibitem[{{Gordon} {et~al.}(2021){Gordon}, {Misselt}, {Bouwman}, {Clayton}, {Decleir}, {Hines}, {Pendleton}, {Rieke}, {Smith}, \& {Whittet}}]{Gordon21}
{Gordon}, K.~D., {Misselt}, K.~A., {Bouwman}, J., {et~al.} 2021, \apj, 916, 33, \dodoi{10.3847/1538-4357/ac00b7}

\bibitem[{{Graham} {et~al.}(2019){Graham}, {Kulkarni}, {Bellm}, {Adams}, {Barbarino}, {Blagorodnova}, {Bodewits}, {Bolin}, {Brady}, {Cenko}, {Chang}, {Coughlin}, {De}, {Eadie}, {Farnham}, {Feindt}, {Franckowiak}, {Fremling}, {Gezari}, {Ghosh}, {Goldstein}, {Golkhou}, {Goobar}, {Ho}, {Huppenkothen}, {Ivezi{\'c}}, {Jones}, {Juric}, {Kaplan}, {Kasliwal}, {Kelley}, {Kupfer}, {Lee}, {Lin}, {Lunnan}, {Mahabal}, {Miller}, {Ngeow}, {Nugent}, {Ofek}, {Prince}, {Rauch}, {van Roestel}, {Schulze}, {Singer}, {Sollerman}, {Taddia}, {Yan}, {Ye}, {Yu}, {Barlow}, {Bauer}, {Beck}, {Belicki}, {Biswas}, {Brinnel}, {Brooke}, {Bue}, {Bulla}, {Burruss}, {Connolly}, {Cromer}, {Cunningham}, {Dekany}, {Delacroix}, {Desai}, {Duev}, {Feeney}, {Flynn}, {Frederick}, {Gal-Yam}, {Giomi}, {Groom}, {Hacopians}, {Hale}, {Helou}, {Henning}, {Hover}, {Hillenbrand}, {Howell}, {Hung}, {Imel}, {Ip}, {Jackson}, {Kaspi}, {Kaye}, {Kowalski}, {Kramer}, {Kuhn}, {Landry}, {Laher}, {Mao}, {Masci}, {Monkewitz}, {Murphy}, {Nordin}, {Patterson}, {Penprase},
  {Porter}, {Rebbapragada}, {Reiley}, {Riddle}, {Rigault}, {Rodriguez}, {Rusholme}, {van Santen}, {Shupe}, {Smith}, {Soumagnac}, {Stein}, {Surace}, {Szkody}, {Terek}, {Van Sistine}, {van Velzen}, {Vestrand}, {Walters}, {Ward}, {Zhang}, \& {Zolkower}}]{Graham19}
{Graham}, M.~J., {Kulkarni}, S.~R., {Bellm}, E.~C., {et~al.} 2019, \pasp, 131, 078001, \dodoi{10.1088/1538-3873/ab006c}

\bibitem[{{Gr{\"o}bel} {et~al.}(2017){Gr{\"o}bel}, {H{\"u}mmerich}, {Paunzen}, \& {Bernhard}}]{Grobel17}
{Gr{\"o}bel}, R., {H{\"u}mmerich}, S., {Paunzen}, E., \& {Bernhard}, K. 2017, \na, 50, 104, \dodoi{10.1016/j.newast.2016.07.012}

\bibitem[{{Guerrero} {et~al.}(2013){Guerrero}, {Miranda}, {Ramos-Larios}, \& {V{\'a}zquez}}]{Guerrero13}
{Guerrero}, M.~A., {Miranda}, L.~F., {Ramos-Larios}, G., \& {V{\'a}zquez}, R. 2013, \aap, 551, A53, \dodoi{10.1051/0004-6361/201220592}

\bibitem[{{Hajduk} {et~al.}(2014){Hajduk}, {van Hoof}, {Gesicki}, {Zijlstra}, {G{\'o}rny}, \& {G{\l}adkowski}}]{Hajduk14}
{Hajduk}, M., {van Hoof}, P.~A.~M., {Gesicki}, K., {et~al.} 2014, \aap, 567, A15, \dodoi{10.1051/0004-6361/201322742}

\bibitem[{{Hajduk} {et~al.}(2008){Hajduk}, {Zijlstra}, \& {Gesicki}}]{Hajduk08}
{Hajduk}, M., {Zijlstra}, A.~A., \& {Gesicki}, K. 2008, \aap, 490, L7, \dodoi{10.1051/0004-6361:200810492}

\bibitem[{{Handler} {et~al.}(1997){Handler}, {Mendez}, {Medupe}, {Costero}, {Birch}, {Alvarez}, {Sullivan}, {Kurtz}, {Herrero}, {Guerrero}, {Ciardullo}, \& {Breger}}]{Handler97}
{Handler}, G., {Mendez}, R.~H., {Medupe}, R., {et~al.} 1997, \aap, 320, 125

\bibitem[{Harris {et~al.}(2020)Harris, Millman, van~der Walt, Gommers, Virtanen, Cournapeau, Wieser, Taylor, Berg, Smith, Kern, Picus, Hoyer, van Kerkwijk, Brett, Haldane, del R{\'{i}}o, Wiebe, Peterson, G{\'{e}}rard-Marchant, Sheppard, Reddy, Weckesser, Abbasi, Gohlke, \& Oliphant}]{harris2020array}
Harris, C.~R., Millman, K.~J., van~der Walt, S.~J., {et~al.} 2020, Nature, 585, 357, \dodoi{10.1038/s41586-020-2649-2}

\bibitem[{{Hart} {et~al.}(2023){Hart}, {Shappee}, {Hey}, {Kochanek}, {Stanek}, {Lim}, {Dobbs}, {Tucker}, {Jayasinghe}, {Beacom}, {Boright}, {Holoien}, {Ong}, {Prieto}, {Thompson}, \& {Will}}]{Hart23}
{Hart}, K., {Shappee}, B.~J., {Hey}, D., {et~al.} 2023, arXiv e-prints, arXiv:2304.03791, \dodoi{10.48550/arXiv.2304.03791}

\bibitem[{{Heber} {et~al.}(2018){Heber}, {Irrgang}, \& {Schaffenroth}}]{Heber+2018}
{Heber}, U., {Irrgang}, A., \& {Schaffenroth}, J. 2018, Open Astronomy, 27, 35, \dodoi{10.1515/astro-2018-0008}

\bibitem[{{Hill} {et~al.}(2021){Hill}, {Lee}, {MacQueen}, {Kelz}, {Drory}, {Vattiat}, {Good}, {Ramsey}, {Kriel}, {Peterson}, {DePoy}, {Gebhardt}, {Marshall}, {Tuttle}, {Bauer}, {Chonis}, {Fabricius}, {Froning}, {H{\"a}user}, {Indahl}, {Jahn}, {Landriau}, {Leck}, {Montesano}, {Prochaska}, {Snigula}, {Zeimann}, {Bryant}, {Damm}, {Fowler}, {Janowiecki}, {Martin}, {Mrozinski}, {Odewahn}, {Rostopchin}, {Shetrone}, {Spencer}, {Mentuch Cooper}, {Armandroff}, {Bender}, {Dalton}, {Hopp}, {Komatsu}, {Nicklas}, {Ramsey}, {Roth}, {Schneider}, {Sneden}, \& {Steinmetz}}]{Hill21}
{Hill}, G.~J., {Lee}, H., {MacQueen}, P.~J., {et~al.} 2021, \aj, 162, 298, \dodoi{10.3847/1538-3881/ac2c02}

\bibitem[{Hunter(2007)}]{Hunter:2007}
Hunter, J.~D. 2007, Computing in Science \& Engineering, 9, 90, \dodoi{10.1109/MCSE.2007.55}

\bibitem[{{Husser} {et~al.}(2013){Husser}, {Wende-von Berg}, {Dreizler}, {Homeier}, {Reiners}, {Barman}, \& {Hauschildt}}]{Husser+2013}
{Husser}, T.~O., {Wende-von Berg}, S., {Dreizler}, S., {et~al.} 2013, \aap, 553, A6, \dodoi{10.1051/0004-6361/201219058}

\bibitem[{{Hyung} {et~al.}(2010){Hyung}, {Lee}, \& {Kim}}]{Hyung10}
{Hyung}, S., {Lee}, S.-J., \& {Kim}, M.-H. 2010, Journal of Korean Physical Society, 57, 514, \dodoi{10.3938/jkps.57.514}

\bibitem[{{Irrgang} {et~al.}(2021){Irrgang}, {Geier}, {Heber}, {Kupfer}, {El-Badry}, \& {Bloemen}}]{Irrgang+2021}
{Irrgang}, A., {Geier}, S., {Heber}, U., {et~al.} 2021, \aap, 650, A102, \dodoi{10.1051/0004-6361/202038757}

\bibitem[{{Ivezi{\'c}} {et~al.}(2019){Ivezi{\'c}}, {Kahn}, {Tyson}, {Abel}, {Acosta}, {Allsman}, {Alonso}, {AlSayyad}, {Anderson}, {Andrew}, {Angel}, {Angeli}, {Ansari}, {Antilogus}, {Araujo}, {Armstrong}, {Arndt}, {Astier}, {Aubourg}, {Auza}, {Axelrod}, {Bard}, {Barr}, {Barrau}, {Bartlett}, {Bauer}, {Bauman}, {Baumont}, {Bechtol}, {Bechtol}, {Becker}, {Becla}, {Beldica}, {Bellavia}, {Bianco}, {Biswas}, {Blanc}, {Blazek}, {Blandford}, {Bloom}, {Bogart}, {Bond}, {Booth}, {Borgland}, {Borne}, {Bosch}, {Boutigny}, {Brackett}, {Bradshaw}, {Brandt}, {Brown}, {Bullock}, {Burchat}, {Burke}, {Cagnoli}, {Calabrese}, {Callahan}, {Callen}, {Carlin}, {Carlson}, {Chandrasekharan}, {Charles-Emerson}, {Chesley}, {Cheu}, {Chiang}, {Chiang}, {Chirino}, {Chow}, {Ciardi}, {Claver}, {Cohen-Tanugi}, {Cockrum}, {Coles}, {Connolly}, {Cook}, {Cooray}, {Covey}, {Cribbs}, {Cui}, {Cutri}, {Daly}, {Daniel}, {Daruich}, {Daubard}, {Daues}, {Dawson}, {Delgado}, {Dellapenna}, {de Peyster}, {de Val-Borro}, {Digel}, {Doherty}, {Dubois},
  {Dubois-Felsmann}, {Durech}, {Economou}, {Eifler}, {Eracleous}, {Emmons}, {Fausti Neto}, {Ferguson}, {Figueroa}, {Fisher-Levine}, {Focke}, {Foss}, {Frank}, {Freemon}, {Gangler}, {Gawiser}, {Geary}, {Gee}, {Geha}, {Gessner}, {Gibson}, {Gilmore}, {Glanzman}, {Glick}, {Goldina}, {Goldstein}, {Goodenow}, {Graham}, {Gressler}, {Gris}, {Guy}, {Guyonnet}, {Haller}, {Harris}, {Hascall}, {Haupt}, {Hernandez}, {Herrmann}, {Hileman}, {Hoblitt}, {Hodgson}, {Hogan}, {Howard}, {Huang}, {Huffer}, {Ingraham}, {Innes}, {Jacoby}, {Jain}, {Jammes}, {Jee}, {Jenness}, {Jernigan}, {Jevremovi{\'c}}, {Johns}, {Johnson}, {Johnson}, {Jones}, {Juramy-Gilles}, {Juri{\'c}}, {Kalirai}, {Kallivayalil}, {Kalmbach}, {Kantor}, {Karst}, {Kasliwal}, {Kelly}, {Kessler}, {Kinnison}, {Kirkby}, {Knox}, {Kotov}, {Krabbendam}, {Krughoff}, {Kub{\'a}nek}, {Kuczewski}, {Kulkarni}, {Ku}, {Kurita}, {Lage}, {Lambert}, {Lange}, {Langton}, {Le Guillou}, {Levine}, {Liang}, {Lim}, {Lintott}, {Long}, {Lopez}, {Lotz}, {Lupton}, {Lust}, {MacArthur}, {Mahabal},
  {Mandelbaum}, {Markiewicz}, {Marsh}, {Marshall}, {Marshall}, {May}, {McKercher}, {McQueen}, {Meyers}, {Migliore}, {Miller}, {Mills}, {Miraval}, {Moeyens}, {Moolekamp}, {Monet}, {Moniez}, {Monkewitz}, {Montgomery}, {Morrison}, {Mueller}, {Muller}, {Mu{\~n}oz Arancibia}, {Neill}, {Newbry}, {Nief}, {Nomerotski}, {Nordby}, {O'Connor}, {Oliver}, {Olivier}, {Olsen}, {O'Mullane}, {Ortiz}, {Osier}, {Owen}, {Pain}, {Palecek}, {Parejko}, {Parsons}, {Pease}, {Peterson}, {Peterson}, {Petravick}, {Libby Petrick}, {Petry}, {Pierfederici}, {Pietrowicz}, {Pike}, {Pinto}, {Plante}, {Plate}, {Plutchak}, {Price}, {Prouza}, {Radeka}, {Rajagopal}, {Rasmussen}, {Regnault}, {Reil}, {Reiss}, {Reuter}, {Ridgway}, {Riot}, {Ritz}, {Robinson}, {Roby}, {Roodman}, {Rosing}, {Roucelle}, {Rumore}, {Russo}, {Saha}, {Sassolas}, {Schalk}, {Schellart}, {Schindler}, {Schmidt}, {Schneider}, {Schneider}, {Schoening}, {Schumacher}, {Schwamb}, {Sebag}, {Selvy}, {Sembroski}, {Seppala}, {Serio}, {Serrano}, {Shaw}, {Shipsey}, {Sick}, {Silvestri},
  {Slater}, {Smith}, {Smith}, {Sobhani}, {Soldahl}, {Storrie-Lombardi}, {Stover}, {Strauss}, {Street}, {Stubbs}, {Sullivan}, {Sweeney}, {Swinbank}, {Szalay}, {Takacs}, {Tether}, {Thaler}, {Thayer}, {Thomas}, {Thornton}, {Thukral}, {Tice}, {Trilling}, {Turri}, {Van Berg}, {Vanden Berk}, {Vetter}, {Virieux}, {Vucina}, {Wahl}, {Walkowicz}, {Walsh}, {Walter}, {Wang}, {Wang}, {Warner}, {Wiecha}, {Willman}, {Winters}, {Wittman}, {Wolff}, {Wood-Vasey}, {Wu}, {Xin}, {Yoachim}, \& {Zhan}}]{Ivezic19}
{Ivezi{\'c}}, {\v{Z}}., {Kahn}, S.~M., {Tyson}, J.~A., {et~al.} 2019, \apj, 873, 111, \dodoi{10.3847/1538-4357/ab042c}

\bibitem[{{Jacob} {et~al.}(2013){Jacob}, {Sch{\"o}nberner}, \& {Steffen}}]{Jacob13}
{Jacob}, R., {Sch{\"o}nberner}, D., \& {Steffen}, M. 2013, \aap, 558, A78, \dodoi{10.1051/0004-6361/201321532}

\bibitem[{{Jacoby} {et~al.}(2021){Jacoby}, {Hillwig}, {Jones}, {Martin}, {De Marco}, {Kronberger}, {Hurowitz}, {Crocker}, \& {Dey}}]{Jacoby21}
{Jacoby}, G.~H., {Hillwig}, T.~C., {Jones}, D., {et~al.} 2021, \mnras, 506, 5223, \dodoi{10.1093/mnras/stab2045}

\bibitem[{{Jeffery} \& {Hambsch}(2019)}]{Jeffery19}
{Jeffery}, C.~S., \& {Hambsch}, F.~J. 2019, \mnras, 487, 4128, \dodoi{10.1093/mnras/stz1600}

\bibitem[{{Jeffery} \& {Sch{\"o}nberner}(2006)}]{Jeffery06}
{Jeffery}, C.~S., \& {Sch{\"o}nberner}, D. 2006, \aap, 459, 885, \dodoi{10.1051/0004-6361:20047075}

\bibitem[{{Jones} \& {Boffin}(2017)}]{jones17}
{Jones}, D., \& {Boffin}, H. M.~J. 2017, Nature Astronomy, 1, 0117, \dodoi{10.1038/s41550-017-0117}

\bibitem[{{Jones} {et~al.}(2017){Jones}, {Van Winckel}, {Aller}, {Exter}, \& {De Marco}}]{Jones17lotrngc}
{Jones}, D., {Van Winckel}, H., {Aller}, A., {Exter}, K., \& {De Marco}, O. 2017, \aap, 600, L9, \dodoi{10.1051/0004-6361/201730700}

\bibitem[{{K{\H{o}}v{\'a}ri} {et~al.}(2019){K{\H{o}}v{\'a}ri}, {Strassmeier}, {Ol{\'a}h}, {Kriskovics}, {Vida}, {Carroll}, {Granzer}, {Ilyin}, {Jurcsik}, {K{\H{o}}v{\'a}ri}, \& {Weber}}]{Kovari+2019}
{K{\H{o}}v{\'a}ri}, Z., {Strassmeier}, K.~G., {Ol{\'a}h}, K., {et~al.} 2019, \aap, 624, A83, \dodoi{10.1051/0004-6361/201834810}

\bibitem[{{Kiss} \& {B{\'o}di}(2017)}]{Kiss17}
{Kiss}, L.~L., \& {B{\'o}di}, A. 2017, \aap, 608, A99, \dodoi{10.1051/0004-6361/201731876}

\bibitem[{{Kiss} {et~al.}(2006){Kiss}, {Szab{\'o}}, \& {Bedding}}]{Kiss06}
{Kiss}, L.~L., {Szab{\'o}}, G.~M., \& {Bedding}, T.~R. 2006, \mnras, 372, 1721, \dodoi{10.1111/j.1365-2966.2006.10973.x}

\bibitem[{{Kostyakova} \& {Arkhipova}(2009)}]{Kostyakova09}
{Kostyakova}, E.~B., \& {Arkhipova}, V.~P. 2009, Astronomy Reports, 53, 1155, \dodoi{10.1134/S1063772909120087}

\bibitem[{{Lawlor}(2023)}]{Lawlor23}
{Lawlor}, T.~M. 2023, \mnras, 519, 5373, \dodoi{10.1093/mnras/stad042}

\bibitem[{{Lee} {et~al.}(2001){Lee}, {Kang}, \& {Byun}}]{Lee01}
{Lee}, H.-W., {Kang}, Y.-W., \& {Byun}, Y.-I. 2001, \apjl, 551, L121, \dodoi{10.1086/319830}

\bibitem[{{Liebert} {et~al.}(2013){Liebert}, {Bond}, {Dufour}, {Ciardullo}, {Meakes}, {Renzini}, \& {Gianninas}}]{Liebert13}
{Liebert}, J., {Bond}, H.~E., {Dufour}, P., {et~al.} 2013, \apj, 769, 32, \dodoi{10.1088/0004-637X/769/1/32}

\bibitem[{{Lomb}(1976)}]{Lomb76}
{Lomb}, N.~R. 1976, \apss, 39, 447, \dodoi{10.1007/BF00648343}

\bibitem[{{Manick} {et~al.}(2021){Manick}, {Miszalski}, {Kamath}, {Whitelock}, {Van Winckel}, {Hrivnak}, {Barlow}, \& {Mohamed}}]{Manick21}
{Manick}, R., {Miszalski}, B., {Kamath}, D., {et~al.} 2021, \mnras, 508, 2226, \dodoi{10.1093/mnras/stab2428}

\bibitem[{{Marcolino} \& {de Ara{\'u}jo}(2003)}]{Marcolino+2003}
{Marcolino}, W.~L.~F., \& {de Ara{\'u}jo}, F.~X. 2003, \aj, 126, 887, \dodoi{10.1086/375908}

\bibitem[{{Mart{\'\i}nez} {et~al.}(2022){Mart{\'\i}nez}, {Mauas}, \& {Buccino}}]{Martinez22}
{Mart{\'\i}nez}, C.~I., {Mauas}, P.~J.~D., \& {Buccino}, A.~P. 2022, \mnras, 512, 4835, \dodoi{10.1093/mnras/stac755}

\bibitem[{{Masci} {et~al.}(2019){Masci}, {Laher}, {Rusholme}, {Shupe}, {Groom}, {Surace}, {Jackson}, {Monkewitz}, {Beck}, \& {Flynn}}]{Masci19}
{Masci}, F.~J., {Laher}, R.~R., {Rusholme}, B., {et~al.} 2019, \pasp, 131, 018003, \dodoi{10.1088/1538-3873/aae8ac}

\bibitem[{{Miller Bertolami}(2016)}]{M3B2016}
{Miller Bertolami}, M.~M. 2016, \aap, 588, A25, \dodoi{10.1051/0004-6361/201526577}

\bibitem[{{Miszalski} {et~al.}(2009{\natexlab{a}}){Miszalski}, {Acker}, {Moffat}, {Parker}, \& {Udalski}}]{Miszalski09a}
{Miszalski}, B., {Acker}, A., {Moffat}, A.~F.~J., {Parker}, Q.~A., \& {Udalski}, A. 2009{\natexlab{a}}, \aap, 496, 813, \dodoi{10.1051/0004-6361/200811380}

\bibitem[{{Miszalski} {et~al.}(2009{\natexlab{b}}){Miszalski}, {Acker}, {Parker}, \& {Moffat}}]{Miszalski09b}
{Miszalski}, B., {Acker}, A., {Parker}, Q.~A., \& {Moffat}, A.~F.~J. 2009{\natexlab{b}}, \aap, 505, 249, \dodoi{10.1051/0004-6361/200912176}

\bibitem[{{Miszalski} {et~al.}(2011{\natexlab{a}}){Miszalski}, {Corradi}, {Boffin}, {Jones}, {Sabin}, {Santander-Garc{\'\i}a}, {Rodr{\'\i}guez-Gil}, \& {Rubio-D{\'\i}ez}}]{Miszalski11}
{Miszalski}, B., {Corradi}, R.~L.~M., {Boffin}, H.~M.~J., {et~al.} 2011{\natexlab{a}}, \mnras, 413, 1264, \dodoi{10.1111/j.1365-2966.2011.18212.x}

\bibitem[{{Miszalski} {et~al.}(2011{\natexlab{b}}){Miszalski}, {Miko{\l}ajewska}, {K{\"o}ppen}, {Rauch}, {Acker}, {Cohen}, {Frew}, {Moffat}, {Parker}, {Jones}, \& {Udalski}}]{Miszalski11pnm229}
{Miszalski}, B., {Miko{\l}ajewska}, J., {K{\"o}ppen}, J., {et~al.} 2011{\natexlab{b}}, \aap, 528, A39, \dodoi{10.1051/0004-6361/201015469}

\bibitem[{{Moreno-Ib{\'a}{\~n}ez} {et~al.}(2016){Moreno-Ib{\'a}{\~n}ez}, {Villaver}, {Shaw}, \& {Stanghellini}}]{MorenoIbanez16}
{Moreno-Ib{\'a}{\~n}ez}, M., {Villaver}, E., {Shaw}, R.~A., \& {Stanghellini}, L. 2016, \aap, 593, A29, \dodoi{10.1051/0004-6361/201628191}

\bibitem[{{Muzerolle} {et~al.}(2013){Muzerolle}, {Furlan}, {Flaherty}, {Balog}, \& {Gutermuth}}]{Muzerolle13}
{Muzerolle}, J., {Furlan}, E., {Flaherty}, K., {Balog}, Z., \& {Gutermuth}, R. 2013, \nat, 493, 378, \dodoi{10.1038/nature11746}

\bibitem[{{Nicholls} {et~al.}(2009){Nicholls}, {Wood}, {Cioni}, \& {Soszy{\'n}ski}}]{Nicholls09}
{Nicholls}, C.~P., {Wood}, P.~R., {Cioni}, M. R.~L., \& {Soszy{\'n}ski}, I. 2009, \mnras, 399, 2063, \dodoi{10.1111/j.1365-2966.2009.15401.x}

\bibitem[{{Oke} \& {Gunn}(1982)}]{Oke82}
{Oke}, J.~B., \& {Gunn}, J.~E. 1982, \pasp, 94, 586, \dodoi{10.1086/131027}

\bibitem[{{Oliveira da Rosa} {et~al.}(2024){Oliveira da Rosa}, {Kepler}, {Soethe}, {Romero}, \& {Bell}}]{Rosa24}
{Oliveira da Rosa}, G., {Kepler}, S.~O., {Soethe}, L.~T.~T., {Romero}, A.~D., \& {Bell}, K.~J. 2024, \apj, 974, 314, \dodoi{10.3847/1538-4357/ad6987}

\bibitem[{{Osterbrock} \& {Ferland}(2006)}]{Osterbrock04}
{Osterbrock}, D.~E., \& {Ferland}, G.~J. 2006, {Astrophysics of gaseous nebulae and active galactic nuclei}

\bibitem[{pandas~development team(2020)}]{reback2020pandas}
pandas~development team, T. 2020, pandas-dev/pandas: Pandas, latest,  Zenodo, \dodoi{10.5281/zenodo.3509134}

\bibitem[{{Parker} {et~al.}(2016){Parker}, {Boji{\v{c}}i{\'c}}, \& {Frew}}]{Parker16}
{Parker}, Q.~A., {Boji{\v{c}}i{\'c}}, I.~S., \& {Frew}, D.~J. 2016, in Journal of Physics Conference Series, Vol. 728, Journal of Physics Conference Series (IOP), 032008, \dodoi{10.1088/1742-6596/728/3/032008}

\bibitem[{{Pawlak}(2021)}]{Pawlak21}
{Pawlak}, M. 2021, \aap, 649, A110, \dodoi{10.1051/0004-6361/202038642}

\bibitem[{{Paxton} {et~al.}(2011){Paxton}, {Bildsten}, {Dotter}, {Herwig}, {Lesaffre}, \& {Timmes}}]{Paxton11}
{Paxton}, B., {Bildsten}, L., {Dotter}, A., {et~al.} 2011, \apjs, 192, 3, \dodoi{10.1088/0067-0049/192/1/3}

\bibitem[{{Paxton} {et~al.}(2013){Paxton}, {Cantiello}, {Arras}, {Bildsten}, {Brown}, {Dotter}, {Mankovich}, {Montgomery}, {Stello}, {Timmes}, \& {Townsend}}]{Paxton13}
{Paxton}, B., {Cantiello}, M., {Arras}, P., {et~al.} 2013, \apjs, 208, 4, \dodoi{10.1088/0067-0049/208/1/4}

\bibitem[{{Paxton} {et~al.}(2015){Paxton}, {Marchant}, {Schwab}, {Bauer}, {Bildsten}, {Cantiello}, {Dessart}, {Farmer}, {Hu}, {Langer}, {Townsend}, {Townsley}, \& {Timmes}}]{Paxton15}
{Paxton}, B., {Marchant}, P., {Schwab}, J., {et~al.} 2015, \apjs, 220, 15, \dodoi{10.1088/0067-0049/220/1/15}

\bibitem[{{Pereyra} {et~al.}(2013){Pereyra}, {Richer}, \& {L{\'o}pez}}]{Pereyra13}
{Pereyra}, M., {Richer}, M.~G., \& {L{\'o}pez}, J.~A. 2013, \apj, 771, 114, \dodoi{10.1088/0004-637X/771/2/114}

\bibitem[{{Pogge}(2019)}]{Pogge2019}
{Pogge}, R. 2019, {rwpogge/modsCCDRed: v2.0.1}, 2.0.1, Zenodo,  Zenodo, \dodoi{10.5281/zenodo.2647501}

\bibitem[{{Pogge} {et~al.}(2010){Pogge}, {Atwood}, {Brewer}, {Byard}, {Derwent}, {Gonzalez}, {Martini}, {Mason}, {O'Brien}, {Osmer}, {Pappalardo}, {Steinbrecher}, {Teiga}, \& {Zhelem}}]{Pogge+2010}
{Pogge}, R.~W., {Atwood}, B., {Brewer}, D.~F., {et~al.} 2010, in Society of Photo-Optical Instrumentation Engineers (SPIE) Conference Series, Vol. 7735, Ground-based and Airborne Instrumentation for Astronomy III, ed. I.~S. {McLean}, S.~K. {Ramsay}, \& H.~{Takami}, 77350A, \dodoi{10.1117/12.857215}

\bibitem[{{Prochaska} {et~al.}(2020{\natexlab{a}}){Prochaska}, {Hennawi}, {Westfall}, {Cooke}, {Wang}, {Hsyu}, {Davies}, {Farina}, \& {Pelliccia}}]{PypeIt_1}
{Prochaska}, J., {Hennawi}, J., {Westfall}, K., {et~al.} 2020{\natexlab{a}}, The Journal of Open Source Software, 5, 2308, \dodoi{10.21105/joss.02308}

\bibitem[{{Prochaska} {et~al.}(2020{\natexlab{b}}){Prochaska}, Hennawi, Cooke, Westfall, Wang, EmAstro, tiffanyhsyu, Wasserman, Villaume, marijana777, Schindler, Young, Simha, Wilde, Tejos, Isbell, Flörs, Sandford, Vasović, Betts, \& Holden}]{PypeIt_2}
{Prochaska}, J., Hennawi, J., Cooke, R., {et~al.} 2020{\natexlab{b}}, pypeit/PypeIt: Release 1.0.0, v1.0.0,  Zenodo, \dodoi{10.5281/zenodo.3743493}

\bibitem[{{Ramsey} {et~al.}(1998){Ramsey}, {Adams}, {Barnes}, {Booth}, {Cornell}, {Fowler}, {Gaffney}, {Glaspey}, {Good}, {Hill}, {Kelton}, {Krabbendam}, {Long}, {MacQueen}, {Ray}, {Ricklefs}, {Sage}, {Sebring}, {Spiesman}, \& {Steiner}}]{Ramsey98}
{Ramsey}, L.~W., {Adams}, M.~T., {Barnes}, T.~G., {et~al.} 1998, in Society of Photo-Optical Instrumentation Engineers (SPIE) Conference Series, Vol. 3352, Advanced Technology Optical/IR Telescopes VI, ed. L.~M. {Stepp}, 34--42, \dodoi{10.1117/12.319287}

\bibitem[{{Reindl} {et~al.}(2017){Reindl}, {Rauch}, {Miller Bertolami}, {Todt}, \& {Werner}}]{Reindl+2017}
{Reindl}, N., {Rauch}, T., {Miller Bertolami}, M.~M., {Todt}, H., \& {Werner}, K. 2017, \mnras, 464, L51, \dodoi{10.1093/mnrasl/slw175}

\bibitem[{{Reindl} {et~al.}(2014){Reindl}, {Rauch}, {Parthasarathy}, {Werner}, {Kruk}, {Hamann}, {Sander}, \& {Todt}}]{Reindl+2014a}
{Reindl}, N., {Rauch}, T., {Parthasarathy}, M., {et~al.} 2014, \aap, 565, A40, \dodoi{10.1051/0004-6361/201323189}

\bibitem[{{Reindl} {et~al.}(2023){Reindl}, {Islami}, {Werner}, {Kepler}, {Pritzkuleit}, {Dawson}, {Dorsch}, {Istrate}, {Pelisoli}, {Geier}, {Uzundag}, {Provencal}, \& {Justham}}]{Reindl+2023}
{Reindl}, N., {Islami}, R., {Werner}, K., {et~al.} 2023, \aap, 677, A29, \dodoi{10.1051/0004-6361/202346865}

\bibitem[{{Rosenbush} \& {Efimov}(2015)}]{Rosenbush15}
{Rosenbush}, A.~{\'E}., \& {Efimov}, Y.~S. 2015, Astrophysics, 58, 46, \dodoi{10.1007/s10511-015-9365-x}

\bibitem[{{Scargle}(1982)}]{Scargle82}
{Scargle}, J.~D. 1982, \apj, 263, 835, \dodoi{10.1086/160554}

\bibitem[{{Schaefer} \& {Edwards}(2015)}]{Schaefer15}
{Schaefer}, B.~E., \& {Edwards}, Z.~I. 2015, \apj, 812, 133, \dodoi{10.1088/0004-637X/812/2/133}

\bibitem[{{Shappee} {et~al.}(2014){Shappee}, {Prieto}, {Grupe}, {Kochanek}, {Stanek}, {De Rosa}, {Mathur}, {Zu}, {Peterson}, {Pogge}, {Komossa}, {Im}, {Jencson}, {Holoien}, {Basu}, {Beacom}, {Szczygie{\l}}, {Brimacombe}, {Adams}, {Campillay}, {Choi}, {Contreras}, {Dietrich}, {Dubberley}, {Elphick}, {Foale}, {Giustini}, {Gonzalez}, {Hawkins}, {Howell}, {Hsiao}, {Koss}, {Leighly}, {Morrell}, {Mudd}, {Mullins}, {Nugent}, {Parrent}, {Phillips}, {Pojmanski}, {Rosing}, {Ross}, {Sand}, {Terndrup}, {Valenti}, {Walker}, \& {Yoon}}]{Shappee14}
{Shappee}, B.~J., {Prieto}, J.~L., {Grupe}, D., {et~al.} 2014, \apj, 788, 48, \dodoi{10.1088/0004-637X/788/1/48}

\bibitem[{{Shingles} {et~al.}(2021){Shingles}, {Smith}, {Young}, {Smartt}, {Tonry}, {Denneau}, {Heinze}, {Weiland}, {Flewelling}, {Stalder}, {Clocchiatti}, {F{\"o}rster}, {Pignata}, {Rest}, {Anderson}, {Stubbs}, \& {Erasmus}}]{Shingles21}
{Shingles}, L., {Smith}, K.~W., {Young}, D.~R., {et~al.} 2021, Transient Name Server AstroNote, 7, 1

\bibitem[{{Soszy{\'n}ski} {et~al.}(2021){Soszy{\'n}ski}, {Olechowska}, {Ratajczak}, {Iwanek}, {Skowron}, {Mr{\'o}z}, {Pietrukowicz}, {Udalski}, {Szyma{\'n}ski}, {Skowron}, {Gromadzki}, {Poleski}, {Koz{\l}owski}, {Wrona}, {Ulaczyk}, \& {Rybicki}}]{Soszynski21}
{Soszy{\'n}ski}, I., {Olechowska}, A., {Ratajczak}, M., {et~al.} 2021, \apjl, 911, L22, \dodoi{10.3847/2041-8213/abf3c9}

\bibitem[{{Stanghellini} {et~al.}(2016){Stanghellini}, {Shaw}, \& {Villaver}}]{Stanghellini16}
{Stanghellini}, L., {Shaw}, R.~A., \& {Villaver}, E. 2016, \apj, 830, 33, \dodoi{10.3847/0004-637X/830/1/33}

\bibitem[{{Tyndall} {et~al.}(2013){Tyndall}, {Jones}, {Boffin}, {Miszalski}, {Faedi}, {Lloyd}, {Boumis}, {L{\'o}pez}, {Martell}, {Pollacco}, \& {Santander-Garc{\'\i}a}}]{Tyndall13}
{Tyndall}, A.~A., {Jones}, D., {Boffin}, H.~M.~J., {et~al.} 2013, \mnras, 436, 2082, \dodoi{10.1093/mnras/stt1713}

\bibitem[{{van Genderen} \& {Gautschy}(1995)}]{Genderen95}
{van Genderen}, A.~M., \& {Gautschy}, A. 1995, \aap, 294, 453

\bibitem[{{Van Winckel} {et~al.}(2014){Van Winckel}, {Jorissen}, {Exter}, {Raskin}, {Prins}, {Perez Padilla}, {Merges}, \& {Pessemier}}]{VanWinckel14}
{Van Winckel}, H., {Jorissen}, A., {Exter}, K., {et~al.} 2014, \aap, 563, L10, \dodoi{10.1051/0004-6361/201423650}

\bibitem[{{Van Winckel} {et~al.}(1999){Van Winckel}, {Waelkens}, {Fernie}, \& {Waters}}]{VanWinckel99}
{Van Winckel}, H., {Waelkens}, C., {Fernie}, J.~D., \& {Waters}, L.~B.~F.~M. 1999, \aap, 343, 202

\bibitem[{{VanderPlas}(2018)}]{VanderPlas18}
{VanderPlas}, J.~T. 2018, \apjs, 236, 16, \dodoi{10.3847/1538-4365/aab766}

\bibitem[{Virtanen {et~al.}(2020)Virtanen, Gommers, Oliphant, Haberland, Reddy, Cournapeau, Burovski, Peterson, Weckesser, Bright, {van der Walt}, Brett, Wilson, Millman, Mayorov, Nelson, Jones, Kern, Larson, Carey, Polat, Feng, Moore, {VanderPlas}, Laxalde, Perktold, Cimrman, Henriksen, Quintero, Harris, Archibald, Ribeiro, Pedregosa, {van Mulbregt}, \& {SciPy 1.0 Contributors}}]{2020SciPy-NMeth}
Virtanen, P., Gommers, R., Oliphant, T.~E., {et~al.} 2020, Nature Methods, 17, 261, \dodoi{10.1038/s41592-019-0686-2}

\bibitem[{von Neumann(1941)}]{VonNeumann41}
von Neumann, J. 1941, The Annals of Mathematical Statistics, 12, 367 , \dodoi{10.1214/aoms/1177731677}

\bibitem[{{Weidmann} {et~al.}(2018){Weidmann}, {Gamen}, {Mast}, {Fari{\~n}a}, {Gimeno}, {Schmidt}, {Ashley}, {Peralta de Arriba}, {Sowicka}, \& {Ordonez-Etxeberria}}]{Weidmann18}
{Weidmann}, W., {Gamen}, R., {Mast}, D., {et~al.} 2018, \aap, 614, A135, \dodoi{10.1051/0004-6361/201731805}

\bibitem[{{Werner} {et~al.}(2003){Werner}, {Deetjen}, {Dreizler}, {Nagel}, {Rauch}, \& {Schuh}}]{2003ASPC..288...31W}
{Werner}, K., {Deetjen}, J.~L., {Dreizler}, S., {et~al.} 2003, in Astronomical Society of the Pacific Conference Series, Vol. 288, Stellar Atmosphere Modeling, ed. I.~{Hubeny}, D.~{Mihalas}, \& K.~{Werner}, 31

\bibitem[{{Werner} {et~al.}(2019){Werner}, {Rauch}, \& {Reindl}}]{Werner19}
{Werner}, K., {Rauch}, T., \& {Reindl}, N. 2019, \mnras, 483, 5291, \dodoi{10.1093/mnras/sty3408}

\bibitem[{{Werner} {et~al.}(2020){Werner}, {Reindl}, {L{\"o}bling}, {Pelisoli}, {Schaffenroth}, {Rebassa-Mansergas}, {Irawati}, \& {Ren}}]{Werner+2020}
{Werner}, K., {Reindl}, N., {L{\"o}bling}, L., {et~al.} 2020, \aap, 642, A228, \dodoi{10.1051/0004-6361/202038574}

\bibitem[{{Wright} {et~al.}(2005){Wright}, {Corradi}, \& {Perinotto}}]{Wright05}
{Wright}, S.~A., {Corradi}, R.~L.~M., \& {Perinotto}, M. 2005, \aap, 436, 967, \dodoi{10.1051/0004-6361:20052666}

\end{thebibliography}
\bibliographystyle{aasjournal}



\end{document}